\documentclass[publication,lncs,normal,final]{format/paper}

\newcommand{\getsub}[1]{\BeforeSubString{|}{#1}}

\newcommand{\gettxt}[1]{\BehindSubString{|}{#1}}

\newcommand{\showtxtsub}[1]{\ensuremath{\gettxt{#1}_{\getsub{#1}}}}

\newcommand{\txtstyname}[1]{\textsc{#1}}

\newcommand{\txtname}[2][]
	{{\ifx&#1&\txtstyname{#2}\else\txtstyname{#2}\txtsub{#1}\fi}}

\newcommand{\txtsub}[1]{$_{\text{\textnormal{#1}}}$}

\newcommand{\mthstyname}[1]{\mathcal{#1}}

\newcommand{\mthname}[2][]
	{{\ensuremath{\ifx&#1&\mthstyname{#2}\else\mthstyname{#2}\showtxtsub{#1}\fi}}}

\newcommand{\mthstystr}[1]{\mathfrak{#1}}

\newcommand{\mthstr}[2][]
	{{\ensuremath{\ifx&#1&\mthstystr{#2}\else\mthstystr{#2}\showtxtsub{#1}\fi}}}

\newcommand{\mthstycls}[1]{\mathbb{#1}}

\newcommand{\mthcls}[2][]
	{{\ensuremath{\ifx&#1&\mthstycls{#2}\else\mthstycls{#2}\showtxtsub{#1}\fi}}}

\newcommand{\mthstyfam}[1]{\mathscr{#1}}

\newcommand{\mthfam}[2][]
	{{\ensuremath{\ifx&#1&\mthstyfam{#2}\else\mthstyfam{#2}\showtxtsub{#1}\fi}}}

\newcommand{\mthstyset}[1]{\mathrm{#1}}

\newcommand{\mthset}[2][]
	{{\ensuremath{\ifx&#1&\mthstyset{#2}\else\mthstyset{#2}\showtxtsub{#1}\fi}}}

\newcommand{\mthstyfun}[1]{\mathsf{#1}}

\newcommand{\mthfun}[2][]
	{{\ensuremath{\ifx&#1&\mthstyfun{#2}\else\mthstyfun{#2}\showtxtsub{#1}\fi}}}

\newcommand{\mthstyrel}[1]{\mathit{#1}}

\newcommand{\mthrel}[2][]
	{{\ensuremath{\ifx&#1&\mthstyrel{#2}\else\mthstyrel{#2}\showtxtsub{#1}\fi}}}

\newcommand{\mthstysym}[1]{\mathsf{#1}}

\newcommand{\mthsym}[2][]
	{{\ensuremath{\ifx&#1&\mthstysym{#2}\else\mthstysym{#2}\showtxtsub{#1}\fi}}}

\newcommand{\mthstynot}[1]{\mathtt{#1}}

\newcommand{\mthnot}[2][]
	{{\ensuremath{\ifx&#1&\mthstynot{#2}\else\mthstynot{#2}\showtxtsub{#1}\fi}}}

\newcommand{\mthstyelm}[1]{#1}

\newcommand{\mthelm}[2][]
	{{\ensuremath{\ifx&#1&\mthstyelm{#2}\else\mthstyelm{#2}\showtxtsub{#1}\fi}}}

\newcommand{\AName}[1][]{\mthname[#1]{A}}

\newcommand{\GName}[1][]{\mthname[#1]{G}}

\newcommand{\NName}[1][]{\mthname[#1]{N}}

\newcommand{\TName}[1][]{\mthname[#1]{T}}
\newcommand{\UName}[1][]{\mthname[#1]{U}}

\newcommand{\ASet}[1][]{\mthset[#1]{A}}

\newcommand{\CSet}[1][]{\mthset[#1]{C}}
\newcommand{\DSet}[1][]{\mthset[#1]{D}}
\newcommand{\ESet}[1][]{\mthset[#1]{E}}
\newcommand{\FSet}[1][]{\mthset[#1]{F}}

\newcommand{\LSet}[1][]{\mthset[#1]{L}}
\newcommand{\MSet}[1][]{\mthset[#1]{M}}

\newcommand{\PSet}[1][]{\mthset[#1]{P}}
\newcommand{\QSet}[1][]{\mthset[#1]{Q}}
\newcommand{\RSet}[1][]{\mthset[#1]{R}}
\newcommand{\SSet}[1][]{\mthset[#1]{S}}
\newcommand{\TSet}[1][]{\mthset[#1]{T}}
\newcommand{\USet}[1][]{\mthset[#1]{U}}
\newcommand{\VSet}[1][]{\mthset[#1]{V}}
\newcommand{\WSet}[1][]{\mthset[#1]{W}}
\newcommand{\XSet}[1][]{\mthset[#1]{X}}
\newcommand{\YSet}[1][]{\mthset[#1]{Y}}
\newcommand{\ZSet}[1][]{\mthset[#1]{Z}}

\newcommand{\AFun}[1][]{\mthfun[#1]{A}}

\newcommand{\DFun}[1][]{\mthfun[#1]{D}}

\newcommand{\LFun}[1][]{\mthfun[#1]{L}}

\newcommand{\TFun}[1][]{\mthfun[#1]{T}}

\newcommand{\WFun}[1][]{\mthfun[#1]{W}}

\newcommand{\bFun}[1][]{\mthfun[#1]{b}}

\newcommand{\dFun}[1][]{\mthfun[#1]{d}}

\newcommand{\fFun}[1][]{\mthfun[#1]{f}}
\newcommand{\gFun}[1][]{\mthfun[#1]{g}}
\newcommand{\hFun}[1][]{\mthfun[#1]{h}}

\newcommand{\mFun}[1][]{\mthfun[#1]{m}}

\newcommand{\uFun}[1][]{\mthfun[#1]{u}}
\newcommand{\vFun}[1][]{\mthfun[#1]{v}}
\newcommand{\wFun}[1][]{\mthfun[#1]{w}}

\newcommand{\zFun}[1][]{\mthfun[#1]{z}}

\newcommand{\RRel}[1][]{\mthrel[#1]{R}}
\newcommand{\SRel}[1][]{\mthrel[#1]{S}}

\newcommand{\PSym}[1][]{\mthsym[#1]{P}}

\newcommand{\aSym}[1][]{\mthsym[#1]{a}}
\newcommand{\bSym}[1][]{\mthsym[#1]{b}}

\newcommand{\gSym}[1][]{\mthsym[#1]{g}}

\newcommand{\pSym}[1][]{\mthsym[#1]{p}}
\newcommand{\qSym}[1][]{\mthsym[#1]{q}}
\newcommand{\rSym}[1][]{\mthsym[#1]{r}}
\newcommand{\sSym}[1][]{\mthsym[#1]{s}}

\newcommand{\xSym}[1][]{\mthsym[#1]{x}}
\newcommand{\ySym}[1][]{\mthsym[#1]{y}}
\newcommand{\zSym}[1][]{\mthsym[#1]{z}}

\newcommand{\BNot}[1][]{\mthnot[#1]{B}}

\newcommand{\PNot}[1][]{\mthnot[#1]{P}}

\newcommand{\EElm}[1][]{\mthelm[#1]{E}}

\newcommand{\aElm}[1][]{\mthelm[#1]{a}}

\newcommand{\cElm}[1][]{\mthelm[#1]{c}}
\newcommand{\dElm}[1][]{\mthelm[#1]{d}}
\newcommand{\eElm}[1][]{\mthelm[#1]{e}}

\newcommand{\lElm}[1][]{\mthelm[#1]{l}}

\newcommand{\nElm}[1][]{\mthelm[#1]{n}}

\newcommand{\pElm}[1][]{\mthelm[#1]{p}}
\newcommand{\qElm}[1][]{\mthelm[#1]{q}}

\newcommand{\sElm}[1][]{\mthelm[#1]{s}}
\newcommand{\tElm}[1][]{\mthelm[#1]{t}}

\newcommand{\wElm}[1][]{\mthelm[#1]{w}}
\newcommand{\xElm}[1][]{\mthelm[#1]{x}}
\newcommand{\yElm}[1][]{\mthelm[#1]{y}}
\newcommand{\zElm}[1][]{\mthelm[#1]{z}}

\newcommand{\set}[2]
	{\{ \arga{#1} \allowbreak : \allowbreak \arga{#2} \}}

\newcommand{\card}[1]
	{\vert \arga{#1} \vert}

\newcommand{\pow}[1]
	{2^{\arga{#1}}}

\newcommand{\class}[2]
	{(\arga{#1} \allowbreak /\!\! \allowbreak \arga{#2})}

\newcommand{\defeq}
	{\triangleq}

\newcommand{\der}[2][]
	{{\widehat{\arga{#2}}_{#1}}}
\newcommand{\adj}[2][]
	{{\widetilde{\arga{#2}}_{#1}}}

\newcommand{\flip}[2][]
	{{\widehat{\arga{#2}}_{#1}}}
\newcommand{\dual}[2][]
	{{\overline{\arga{#2}}_{#1}}}

\newcommand{\arga}[1]
	{#1}
\newcommand{\argb}[2]
	{#1, \allowbreak #2}
\newcommand{\argc}[3]
	{#1, \allowbreak #2, \allowbreak #3}
\newcommand{\argd}[4]
	{#1, \allowbreak #2, \allowbreak #3, \allowbreak #4}
\newcommand{\arge}[5]
	{#1, \allowbreak #2, \allowbreak #3, \allowbreak #4, \allowbreak #5}
\newcommand{\argf}[6]
	{#1, \allowbreak #2, \allowbreak #3, \allowbreak #4, \allowbreak #5,
	\allowbreak #6}
\newcommand{\argg}[7]
	{#1, \allowbreak #2, \allowbreak #3, \allowbreak #4, \allowbreak #5,
	\allowbreak #6, \allowbreak #7}
\newcommand{\argh}[8]
	{#1, \allowbreak #2, \allowbreak #3, \allowbreak #4, \allowbreak #5,
	\allowbreak #6, \allowbreak #7, \allowbreak #8}

\newcommand{\tupleb}[2]
	{\langle \argb{#1}{#2} \rangle}
\newcommand{\tuplec}[3]
	{\langle \argc{#1}{#2}{#3} \rangle}
\newcommand{\tupled}[4]
	{\langle \argd{#1}{#2}{#3}{#4} \rangle}
\newcommand{\tuplee}[5]
	{\langle \arge{#1}{#2}{#3}{#4}{#5} \rangle}
\newcommand{\tuplef}[6]
	{\langle \argf{#1}{#2}{#3}{#4}{#5}{#6} \rangle}
\newcommand{\tupleg}[7]
	{\langle \argg{#1}{#2}{#3}{#4}{#5}{#6}{#7} \rangle}
\newcommand{\tupleh}[8]
	{\langle \argh{#1}{#2}{#3}{#4}{#5}{#6}{#7}{#8} \rangle}

\newcommand{\SetN}
	{\mthcls{N}}

\newcommand{\SetNI}
	{\der{\SetN}}

\newcommand{\numcc}[2]
	{[\argb{#1}{#2}]}
\newcommand{\numco}[2]
	{[\argb{#1}{#2}[\:}
\newcommand{\numoc}[2]
	{\:]\argb{#1}{#2}]}
\newcommand{\numoo}[2]
	{\:]\argb{#1}{#2}[\:}

\newcommand{\emptyfun}{\varnothing}

\newcommand{\pto}{\rightharpoonup}

\newcommand{\dom}[1]{\mthfun{dom}(\arga{#1})}

\newcommand{\cod}[1]{\mthfun{cod}(\arga{#1})}

\newcommand{\rng}[1]{\mthfun{rng}(\arga{#1})}

\newcommand{\cmp}{\circ}

\newcommand{\rst}{\upharpoonright}

\newcommand{\AOmicron}[1]
	{\mthnot{O}(\arga{#1})}

\newcommand{\pfx}[1]
	{\mthfun{pfx}(\arga{#1})}

\newcommand{\fst}[1]
	{\mthfun{fst}(\arga{#1})}
\newcommand{\lst}[1]
	{\mthfun{lst}(\arga{#1})}

\newcommand{\FOL}[1][]{\txtname{Fol\small\ifx&#1&\else\textnormal{[#1]}\fi}}

\newcommand{\MSOL}[1][]{\txtname{Msol\small\ifx&#1&\else\textnormal{[#1]}\fi}}

\newcommand{\MTL}[1][]{\txtname{Mtl\small\ifx&#1&\else\textnormal{[#1]}\fi}}

\newcommand{\MPL}[1][]{\txtname{Mpl\small\ifx&#1&\else\textnormal{[#1]}\fi}}

\newcommand{\PML}[1][]{\txtname{Pml\small\ifx&#1&\else\textnormal{[#1]}\fi}}

\newcommand{\MuCalculus}[1][]
	{\txtname{$\mu$Calculus\small\ifx&#1&\else\textnormal{[#1]}\fi}}

\newcommand{\AMuCalculus}{\txtname{A}\MuCalculus}

\newcommand{\LTL}[1][]{\txtname{Ltl\small\ifx&#1&\else\textnormal{[#1]}\fi}}

\newcommand{\QPTL}[1][]{\txtname{QPtl\small\ifx&#1&\else\textnormal{[#1]}\fi}}

\newcommand{\X}{{\mthsym{X}\:}}

\newcommand{\U}{{\mthsym{U}\:}}

\newcommand{\R}{{\mthsym{R}\:}}

\renewcommand{\S}{{\mthsym{S}\:}}

\newcommand{\F}{{\mthsym{F}\:}}

\newcommand{\G}{{\mthsym{G}\:}}

\renewcommand{\P}{{\mthsym{P}\:}}

\newcommand{\CTL}[1][]{\txtname{Ctl\small\ifx&#1&\else\textnormal{[#1]}\fi}}

\newcommand{\CTLP}[1][]
	{\txtname{Ctl\!$^{+}$\small\ifx&#1&\else\textnormal{[#1]}\fi}}

\newcommand{\CTLS}[1][]
	{\txtname{Ctl\!$^{\ast}$\small\ifx&#1&\else\textnormal{[#1]}\fi}}

\providecommand{\E}{\mthsym{E}}

\providecommand{\A}{\mthsym{A}}

\newcommand{\ATL}[1][]{\txtname{Atl\small\ifx&#1&\else\textnormal{[#1]}\fi}}

\newcommand{\ATLP}[1][]
	{\txtname{Atl\!$^{+}$\small\ifx&#1&\else\textnormal{[#1]}\fi}}

\newcommand{\ATLS}[1][]
	{\txtname{Atl\!$^{\ast}$\small\ifx&#1&\else\textnormal{[#1]}\fi}}

\newcommand{\SL}[1][]{\txtname{Sl\small\ifx&#1&\else\textnormal{[#1]}\fi}}

\newcommand{\OGSL}[1][]{\SL[\txtname{1g}\ifx&#1&\else\textnormal,~#1\fi]}

\newcommand{\BGSL}[1][]{\SL[\txtname{bg}\ifx&#1&\else\textnormal,~#1\fi]}

\newcommand{\NGSL}[1][]{\SL[\txtname{ng}\ifx&#1&\else\textnormal,~#1\fi]}

\newcommand{\CHPSL}{\txtname{CHP}-\SL}

\providecommand{\emodels}{\models_{\txtname{E}}}

\newcommand{\CaRet}[1][]{\txtname{CaRet\small\ifx&#1&\else\textnormal{[#1]}\fi}}

\providecommand{\Tt}{\mthstr{t}}

\providecommand{\Ff}{\mthstr{f}}

\renewcommand{\implies}
	{\Rightarrow}

\newcommand{\GL}{\txtname{Gl}}

\newcommand{\CL}{\txtname{Cl}}

\newcommand{\APSet}
	{\mthset{A\!P}}

\newcommand{\VarSet}
	{\mthset{Var}}

\newcommand{\AsgSet}[1][]{\mthset[#1]{Asg}}

\newcommand{\asgFun}[1][]{\mthfun[#1]{\chi}}

\newcommand{\BndSet}[1][]{\mthset[#1]{Bnd}}

\newcommand{\bndFun}[1][]{\mthfun[#1]{\zeta}}

\newcommand{\fun}[3][]{\mthfun[#1]{#2}(\arga{#3})}

\newcommand{\sub}[2][]{\fun[#1]{sub}{#2}}
\newcommand{\snt}[2][]{\fun[#1]{snt}{#2}}

\newcommand{\psnt}[2][]{\fun[#1]{psnt}{#2}}

\newcommand{\free}[2][]{\fun[#1]{free}{#2}}

\newcommand{\Exs}[1]{\langle \arga{#1} \rangle}
\newcommand{\All}[1]{\lbrack \arga{#1} \rbrack}

\newcommand{\EExs}[1]{{\Exs{\!\Exs{#1}\!}}}
\newcommand{\AAll}[1]{{\All{\:\!\!\All{#1}\:\!\!}}}

\providecommand{\LabSet}[1][]{\mthset[#1]{\Sigma}}

\providecommand{\DirSet}[1][]{\mthset[#1]{\Delta}}

\providecommand{\TSet}[1][]{\mthset[#1]{T}}

\providecommand{\vFun}[1][]{\mthfun[#1]{v}}

\newcommand{\LTTuple}[4]
	{
	\ifx&#1&
		\ifx&#2&
			\tupleb{#3}{#4}
		\else
			\tuplec{#2}{#3}{#4}
		\fi
	\else
		\ifx&#2&
			\tuplec{#1}{#3}{#4}
		\else
			\tupled{#1}{#2}{#3}{#4}
		\fi
	\fi
	}

\newcommand{\LTDef}[3][]
	{
	\LTTuple {#2} {#3} {\TSet[#1]} {\vFun[#1]}
	}

\newcommand{\LTStruct}[1][]
	{
	\LTDef [#1] {} {}
	}

\providecommand{\NdESet}[1][]{\mthset[#1]{N_{e}}}

\providecommand{\NdOSet}[1][]{\mthset[#1]{N_{o}}}

\providecommand{\EdgRel}[1][]{\mthrel[#1]{E}}

\newcommand{\TPATuple}[4]
	{
	\ifx&#4&
		\tuplec{#1}{#2}{#3}
	\else
		\tupled{#1}{#2}{#3}{#4}
	\fi
	}

\newcommand{\TPADef}[2][]
	{
	\TPATuple {\NdESet[#2]} {\NdOSet[#2]} {\EdgRel[#2]} {#1}
	}

\newcommand{\TPAStruct}[1][ {\nElm[0]} ]
	{
	\TPADef [#1] {}
	}

\providecommand{\APSet}[1][]{\mthset[#1]{AP}}

\providecommand{\WSet}[1][]{\mthset[#1]{W}}

\providecommand{\RRel}[1][]{\mthrel[#1]{R}}

\providecommand{\LFun}[1][]{\mthfun[#1]{L}}

\newcommand{\KSTuple}[5]
	{
	\ifx&#5&
		\tupled{#1}{#2}{#3}{#4}
	\else
		\tuplee{#1}{#2}{#3}{#4}{#5}
	\fi
	}

\newcommand{\KSDef}[2][]
	{
	\KSTuple {\APSet} {\WSet[#2]} {\RRel[#2]} {\LFun[#2]} {#1}
	}

\newcommand{\KSStruct}[1][ {\wElm[0]} ]
	{
	\KSDef [#1] {}
	}

\providecommand{\PthSet}[1][]{\mthset[#1]{Pth}}

\providecommand{\TrkSet}[1][]{\mthset[#1]{Trk}}

\providecommand{\unwFun}[1][]{\mthfun[#1]{unw}}

\newcommand{\CGS}{\txtname{Cgs}}

\newcommand{\CGT}{\txtname{Cgt}}

\newcommand{\DT}{\txtname{Dt}}

\providecommand{\APSet}[1][]{\mthset[#1]{AP}}

\providecommand{\AgSet}[1][]{\mthset[#1]{Ag}}

\providecommand{\AcSet}[1][]{\mthset[#1]{Ac}}

\providecommand{\StSet}[1][]{\mthset[#1]{St}}

\providecommand{\labFun}[1][]{\mthfun[#1]{\lambda}}

\providecommand{\chcFun}[1][]{\mthfun[#1]{\xi}}

\providecommand{\trnFun}[1][]{\mthfun[#1]{\tau}}

\providecommand{\DecSet}[1][]{\mthset[#1]{Dc}}

\providecommand{\decFun}[1][]{\mthfun[#1]{d}}

\newcommand{\CGSTuple}[7]
	{
	\ifx&#7&
		\tuplef{#1}{#2}{#3}{#4}{#5}{#6}
	\else
		\tupleg{#1}{#2}{#3}{#4}{#5}{#6}{#7}
	\fi
	}

\newcommand{\ECGSTuple}[8]
	{
	\ifx&#8&
		\tupleg{#1}{#2}{#3}{#4}{#5}{#6}{#7}
	\else
		\tupleh{#1}{#2}{#3}{#4}{#5}{#6}{#7}{#8}
	\fi
	}

\newcommand{\CGSDef}[2][]
	{
	\CGSTuple {\APSet} {\AgSet} {\AcSet[#2]} {\StSet[#2]} {\labFun[#2]}
		{\trnFun[#2]} {#1}
	}

\newcommand{\ECGSDef}[2][]
	{
	\ECGSTuple {\APSet} {\AgSet} {\AcSet[#2]} {\StSet[#2]} {\labFun[#2]}
		{\chcFun[#2]} {\trnFun[#2]} {#1}
	}

\newcommand{\CGSStruct}[1][ {\sElm[0]} ]
	{
	\CGSDef [#1] {}
	}

\newcommand{\ECGSStruct}[1][ {\sElm[0]} ]
	{
	\ECGSDef [#1] {}
	}

\providecommand{\StrSet}[1][]{\mthset[#1]{Str}}

\providecommand{\strFun}[1][]{\mthfun[#1]{f}}

\providecommand{\TrkSet}[1][]{\mthset[#1]{Trk}}

\providecommand{\trkElm}[1][]{\mthelm[#1]{\rho}}

\providecommand{\PthSet}[1][]{\mthset[#1]{Pth}}

\providecommand{\pthElm}[1][]{\mthelm[#1]{\pi}}

\providecommand{\PlnSet}[1][]{\mthset[#1]{Pln}}

\providecommand{\plnElm}[1][]{\mthelm[#1]{\kappa}}

\providecommand{\TPlnSet}[1][]{\mthset[#1]{TPln}}

\providecommand{\PPlnSet}[1][]{\mthset[#1]{PPln}}

\providecommand{\playElm}[1][]{\mthelm[#1]{\pi}}

\providecommand{\playFun}[1][]{\mthfun[#1]{play}}

\providecommand{\unwFun}[1][]{\mthfun[#1]{unw}}

\newcommand{\QPSet}[1][]{\mthset[#1]{Qnt}}

\newcommand{\qpSym}[1][]{\mthsym[#1]{\wp}}

\newcommand{\qpElm}[1][]{\mthelm[#1]{\wp}}

\newcommand{\QPVSet}[1][]{\mthset[#1]{V}}

\newcommand{\QPEVSet}[2][]{{\EExs{#2}}}

\newcommand{\QPAVSet}[2][]{{\AAll{#2}}}

\newcommand{\qpordRel}[1][]{\mthrel[#1]{<}}

\newcommand{\QPDepSet}[1][]{\mthset[#1]{Dep}}

\newcommand{\BPSet}[1][]{\mthset[#1]{Bnd}}

\newcommand{\bpFun}[1][]{\mthfun[#1]{\zeta}}

\newcommand{\bpSym}[1][]{\mthsym[#1]{\flat}}

\newcommand{\bpElm}[1][]{\mthelm[#1]{\flat}}

\newcommand{\BPColSet}[1][]{\mthset[#1]{Col}}

\providecommand{\ValSet}[1][]{\mthset[#1]{Val}}

\providecommand{\valFun}[1][]{\mthfun[#1]{v}}

\newcommand{\SpcSet}[1][]{\mthset[#1]{DM}}

\newcommand{\ESpcSet}[1][]{\mthset[#1]{EDM}}

\newcommand{\spcFun}[1][]{\mthfun[#1]{\theta}}

\newcommand{\SigSet}[1][]{\mthset[#1]{Sig}}

\newcommand{\LSigSet}[1][]{\mthset[#1]{LSig}}

\newcommand{\sigElm}[1][]{\mthelm[#1]{\sigma}}

\newcommand{\SpcMapSet}[1][]{\mthset[#1]{SigDep}}

\newcommand{\LSpcMapSet}[1][]{\mthset[#1]{LSigDep}}

\newcommand{\spcmapFun}[1][]{\mthfun[#1]{\wFun}}

\newcommand{\headFun}[1][]{\mthfun[#1]{head}}

\newcommand{\bodyFun}[1][]{\mthfun[#1]{body}}

\providecommand{\DSet}[1][]{\mthset[#1]{D}}

\newcommand{\DSTuple}[4]
	{
	\ifx&#4&
		\tuplec{#1}{#2}{#3}
	\else
		\tupled{#1}{#2}{#3}{#4}
	\fi
	}

\newcommand{\PCPTuple}[4]
	{
	\ifx&#4&
		\tuplec{#1}{#2}{#3}
	\else
		\tupled{#1}{#2}{#3}{#4}
	\fi
	}

\newcommand{\NBW}{\txtname{Nbw}}

\newcommand{\UCW}{\txtname{Ucw}}

\providecommand{\SymSet}[1][]{\mthset[#1]{\Sigma}}

\providecommand{\QSet}[1][]{\mthset[#1]{Q}}

\providecommand{\PSet}[1][]{\mthset[#1]{P}}

\providecommand{\atFun}[1][]{\mthfun[#1]{\delta}}

\newcommand{\WATuple}[5]
	{
	\ifx&#5&
		\tupled{#1}{#2}{#3}{#4}
	\else
		\tuplee{#1}{#2}{#3}{#4}{#5}
	\fi
	}

\newcommand{\WMTuple}[7]
	{
	\ifx&#7&
		\tuplef{#1}{#2}{#3}{#4}{#5}{#6}
	\else
		\tupleg{#1}{#2}{#3}{#4}{#5}{#6}{#7}
	\fi
	}

\newcommand{\NTA}{\txtname{Nta}}
\newcommand{\UTA}{\txtname{Uta}}
\newcommand{\ATA}{\txtname{Ata}}

\newcommand{\UCT}{\txtname{Uct}}
\newcommand{\ACT}{\txtname{Act}}

\newcommand{\NPT}{\txtname{Npt}}
\newcommand{\UPT}{\txtname{Upt}}
\newcommand{\APT}{\txtname{Apt}}

\providecommand{\SymSet}[1][]{\mthset[#1]{\Sigma}}

\providecommand{\RDirSet}[1][]{\mthset[#1]{\Delta}}

\providecommand{\QSet}[1][]{\mthset[#1]{Q}}

\providecommand{\PSet}[1][]{\mthset[#1]{P}}

\providecommand{\atFun}[1][]{\mthfun[#1]{\delta}}

\newcommand{\TATuple}[6]
	{
	\ifx&#2&
		\ifx&#6&
			\tupled{#1}{#3}{#4}{#5}
		\else
			\tuplee{#1}{#3}{#4}{#5}{#6}
		\fi
	\else
		\ifx&#6&
			\tuplee{#1}{#2}{#3}{#4}{#5}
		\else
			\tuplef{#1}{#2}{#3}{#4}{#5}{#6}
		\fi
	\fi
	}

\newcommand{\TMTuple}[8]
	{
	\ifx&#3&
		\ifx&#8&
			\tuplef{#1}{#2}{#4}{#5}{#6}{#7}
		\else
			\tupleg{#1}{#2}{#4}{#5}{#6}{#7}{#8}
		\fi
	\else
		\ifx&#8&
			\tupleg{#1}{#2}{#3}{#4}{#5}{#6}{#7}
		\else
			\tupleh{#1}{#2}{#3}{#4}{#5}{#6}{#7}{#8}
		\fi
	\fi
	}

\newcommand{\TADef}[4][q_{0}]
	{
	\TATuple {\SymSet} {#2} {\QSet[#4]} {\atFun[#4]} {#1} {#3}
	}

\newcommand{\TAStruct}[2][q_{0}]
	{
	\TADef [#1] {#2} {\aleph} {}
	}

\newcommand{\ATAStruct}[1][q_{0}]
	{
	\TAStruct [#1] {\RDirSet}
	}

\providecommand{\PBoolSet}[1][]{\BNot[#1]^{+}}

\providecommand{\LangSet}[1][]{\mthset[#1]{L}}

\providecommand{\infFun}[1][]{\mthfun[#1]{inf}}

\newcommand{\ExpTime}{\txtname{ExpTime}}

\newcommand{\ExpTimeC}{\ExpTime-\CComp}

\newcommand{\SComp}[2]{$\Sigma_{\arga{#1}}^{\arga{#2}}$}
\newcommand{\SCompH}[2]{\SComp{#1}{#2}-\HComp}

\newcommand{\HComp}{\txtname{hard}}

\newcommand{\CComp}{\txtname{complete}}

\renewcommand{\epsilon}{\varepsilon}

\usetikzlibrary{arrows,shapes,backgrounds}

\tikzstyle{every node} = [draw = black, fill = white, thin]
\tikzstyle{every edge} += [thick]

\tikzstyle{plain0} = [draw = none, fill = none]
\tikzstyle{plain1} = [draw = none]

\tikzstyle{player} = [circle]
\tikzstyle{player0} = [circle]
\tikzstyle{player1} = [regular polygon, regular polygon sides = 4]
\tikzstyle{player2} = [diamond]
\tikzstyle{player3} = [regular polygon, regular polygon sides = 5]

\newcommand{\toignore}[1]{}

\begin{document}

	\ztitle
		{A Decidable Fragment of Strategy Logic}
		{A Decidable Fragment of Strategy Logic}
	\zauthor
		{Fabio Mogavero$^{1}$, Aniello Murano$^{1}$, Giuseppe Perelli$^{1}$, and
		Moshe Y. Vardi$^{2}$}
		{F. Mogavero, A. Murano, G. Perelli, and M.Y. Vardi}
	\zemail
		{\{mogavero, murano\}@na.infn.it \hspace{1ex} perelli.gi@gmail.com
		\hspace{1ex} vardi@cs.rice.edu}
	\zurl
		{}
	\zinstitute
		{$^{1}$Universit\'a degli Studi di Napoli "Federico II", Napoli, Italy.
		$^{2}$Rice University, Houston, Texas, USA.}
	\zkeywords
		{}
	\zsubjclass
		{}

	\zmaketitle



\begin{abstract}

	\emph{Strategy Logic} (\SL, for short) has been recently introduced by
	Mogavero, Murano, and Vardi as a useful formalism for reasoning explicitly
	about strategies, as first-order objects, in multi-agent concurrent games.
	This logic turns to be very powerful, subsuming all major previously studied
	modal logics for strategic reasoning, including \ATL, \ATLS, and the like.
	Unfortunately, due to its expressiveness, \SL\ has a non-elementarily
	decidable model-checking problem and a highly undecidable satisfiability
	problem, specifically, \SCompH{1}{1}.
	In order to obtain a decidable sublogic, we introduce and study here
	\emph{One-Goal Strategy Logic} (\OGSL, for short).
	This logic is a syntactic fragment of \SL, strictly subsuming \ATLS, which
	encompasses formulas in prenex normal form having a single temporal goal at a
	time, for every strategy quantification of agents.
	\OGSL\ is known to have an elementarily decidable model-checking problem.
	Here we prove that, unlike \SL, it has the bounded tree-model property and its
	satisfiability problem is decidable in 2\ExpTime, thus not harder than the one
	for \ATLS.

\end{abstract}




\begin{section}{Introduction}

	In open-system verification~\cite{CGP02,KVW01}, an important area of research
	is the study of modal logics for strategic reasoning in the setting of
	multi-agent games~\cite{AHK02,JH04,Pau02}.
	An important contribution in this field has been the development of
	\emph{Alternating-Time Temporal Logic} (\ATLS, for short), introduced by
	Alur, Henzinger, and Kupferman~\cite{AHK02}.
	\ATLS\ allows reasoning about strategic behavior of agents with temporal
	goals.
	Formally, it is obtained as a generalization of the branching-time temporal
	logic \CTLS~\cite{EH86}, where the path quantifiers \emph{there exists}
	``$\E$'' and \emph{for all} ``$\A$'' are replaced with strategic modalities
	of the form ``$\EExs{\ASet}$'' and ``$\AAll{\ASet}$'', for a set $\ASet$ of
	\emph{agents}.
	Such strategic modalities are used to express cooperation and competition
	among agents in order to achieve certain temporal goals.
	In particular, these modalities express selective quantifications over those
	paths that are the results of infinite games between a coalition and its
	complement.
	\ATLS\ formulas are interpreted over \emph{concurrent game structures} (\CGS,
	for short)~\cite{AHK02}, which model interacting processes.
	Given a \CGS\ $\GName$ and a set $\ASet$ of agents, the \ATLS\ formula
	$\EExs{\ASet} \psi$ holds at a state $\sElm$ of $\GName$ if there is a set of
	strategies for the agents in $\ASet$ such that, no matter which strategy is
	executed by the agents not in $\ASet$, the resulting outcome of the
	interaction in $\GName$ satisfies $\psi$ at $\sElm$.
	Several decision problems have been investigated about \ATLS; both its
	model-checking and satisfiability problems are decidable in
	2\ExpTime~\cite{Sch08}.
	The complexity of the latter is just like the one for \CTLS~\cite{EJ88,EJ99}.
	\\ \indent
	Despite its powerful expressiveness, \ATLS\ suffers from the strong limitation
	that strategies are treated only implicitly through modalities that refer to
	games between competing coalitions.
	To overcome this problem, Chatterjee, Henzinger, and Piterman introduced
	\emph{Strategy Logic} (\CHPSL, for short)~\cite{CHP07}, a logic that treats
	strategies in \emph{two-player turn-based games} as \emph{first-order
	objects}.
	The explicit treatment of strategies in this logic allows the expression of
	many properties not expressible in \ATLS.
	Although the model-checking problem of \CHPSL\ is known to be decidable, with
	a non-elementary upper bound, it is not known if the satisfiability problem is
	decidable~\cite{CHP10}.
	While the basic idea exploited in~\cite{CHP10} of explicitly quantify over
	strategies is powerful and useful~\cite{FKL10}, \CHPSL\ still suffers from
	various limitations.
	In particular, it is limited to two-player turn-based games.
	Furthermore, \CHPSL\ does not allow different players to share the same
	strategy, suggesting that strategies have yet to become truly first-class
	objects in this logic.
	For example, it is impossible to describe the classic strategy-stealing
	argument of combinatorial games such as Chess, Go, Hex, and the
	like~\cite{ANW07}.
	\\ \indent
	These considerations led us to introduce and investigate a new \emph{Strategy
	Logic}, denoted \SL, as a more general framework than \CHPSL, for explicit
	reasoning about strategies in multi-agent concurrent games~\cite{MMV10b}.
	Syntactically, \SL\ extends the linear-time temporal-logic \LTL~\cite{Pnu77}
	by means of \emph{strategy quantifiers}, the existential $\EExs{\xElm}$ and
	the universal $\AAll{\xElm}$, as well as \emph{agent binding} $(\aElm,
	\xElm)$, where $\aElm$ is an agent and $\xElm$ a variable.
	Intuitively, these elements can be read as \emph{``there exists a strategy
	$\xElm$''}, \emph{``for all strategies $\xElm$''}, and \emph{``bind agent
	$\aElm$ to the strategy associated with $\xElm$''}, respectively.
	For example, in a \CGS\ $\GName$ with agents $\alpha$, $\beta$, and $\gamma$,
	consider the property ``$\alpha$ and $\beta$ have a common strategy to avoid a
	failure''.
	This property can be expressed by the \SL\ formula $\EExs{\xSym} \AAll{\ySym}
	(\alpha, \xSym) (\beta, \xSym) (\gamma, \ySym) (\G \neg \mathit{fail})$.
	The variable $\xSym$ is used to select a strategy for the agents $\alpha$ and
	$\beta$, while $\ySym$ is used to select another one for agent $\gamma$ such
	that their composition, after the binding, results in a play where
	$\mathit{fail}$ is never met.
	Further examples, motivations, and results can be found in a technical
	report~\cite{MMPV11}.
	\\ \indent
	The price that one has to pay for the expressiveness of \SL\ w.r.t.\ \ATLS\ is
	the lack of important model-theoretic properties and an increased complexity
	of related decision problems.
	In particular, in~\cite{MMV10b}, it was shown that \SL\ does not have the
	bounded-tree model property and the related satisfiability problem is
	\mbox{\emph{highly undecidable}, precisely, \SCompH{1}{1}.}
	\\ \indent
	The contrast between the undecidability of the satisfiability problem for \SL\
	and the elementary decidability of the same problem for \ATLS, provides
	motivation for an investigation of decidable fragments of \SL\ that subsume
	\ATLS.
	In particular, we would like to understand why \SL\ is computationally more
	difficult than \ATLS.
	\\ \indent
	We introduce here the syntactic fragment \emph{One-Goal Strategy Logic}
	(\OGSL, for short), which encompasses formulas in a special prenex normal form
	having a single temporal goal at a time.
	This means that every temporal formula $\psi$ is prefixed with a
	quantification-binding prefix that quantifies over a tuple of strategies and
	bind strategies to all agents.
	With \OGSL\ one can express, for example, visibility constraints on
	strategies among agents, i.e., only some agents from a coalition have
	knowledge of the strategies taken by those in the opponent coalition.
	Also, one can describe the fact that, in the Hex game, the
	strategy-stealing argument does not let the player who adopts it to win.
	Observe that both the above properties cannot be expressed neither in \ATLS\
	nor in \CHPSL.
	\\ \indent
	In a technical report~\cite{MMPV11}, we showed that \OGSL\ is strictly more
	expressive that \ATLS, yet its model-checking problem is 2\ExpTimeC, just like
	the one for \ATLS, while the same problem for \SL\ is non-elementarily
	decidable.
	Our main result here is that the satisfiability problem for \OGSL\ is also
	2\ExpTimeC.
	Thus, in spite of its expressiveness, \OGSL\ has the same computational
	properties of \ATLS, which suggests that the one-goal restriction is the key
	to the elementary complexity of the latter logic too.
	\\ \indent
	To achieve our main result, we use a fundamental property of the semantics of
	\OGSL\, called \emph{elementariness}, which allows us to simplify reasoning
	about strategies by reducing it to a set of reasonings about actions.
	This intrinsic characteristic of \OGSL\ means that, to choose an existential
	strategy, we do not need to know the entire structure of
	universally-quantified strategies, as it is the case for \SL, but only their
	values on the histories of interest.
	Technically, to formally describe this property, we make use of the machinery
	of \emph{dependence maps}, which is introduced to define a Skolemization
	procedure for \SL, inspired by the one in first-order logic.
	Using elementariness, we show that \OGSL\ satisfies the \emph{bounded
	tree-model property}.
	This allows us to efficiently make use of a \emph{tree automata-theoretic
	approach}~\cite{Var96,VW86a} to solve the satisfiability problem.
	Given a formula $\varphi$, we build an \emph{alternating co-B\"uchi tree
	automaton}~\cite{KVW00,MS95}, whose size is only exponential in the size of
	$\varphi$, accepting all bounded-branching tree models of the formula.
	Then, together with the complexity of automata-nonemptiness checking, we get
	that the satisfiability procedure for \OGSL\ is 2\ExpTime.
	We believe that our proof techniques are of independent interest and
	applicable to other logics as well.
	\\ \indent
	\emph{Related works.}
		Several works have focused on extensions of \ATLS\ to incorporate more
		powerful strategic constructs.
		Among them, we recall the \emph{Alternating-Time \MuCalculus} (\AMuCalculus,
		for short)~\cite{AHK02}, \emph{Game Logic} (\GL, for short)~\cite{AHK02},
		\emph{Quantified Decision Modality \MuCalculus} (\textsc{q}D$\mu$, for
		short)~\cite{Pin07}, \emph{Coordination Logic} (\CL, for short)~\cite{FS10},
		and some other extensions considered in~\cite{CLM10},~\cite{MMV10a},
		and~\cite{WHY11}.
		\AMuCalculus\ and \textsc{q}D$\mu$ are intrinsically different from \OGSL\
		(as well as from \CHPSL\ and \ATLS) as they are obtained by extending the
		propositional $\mu$-calculus~\cite{Koz83} with strategic modalities.
		\CL\ is similar to \textsc{q}D$\mu$, but with \LTL\ temporal operators
		instead of explicit fixpoint constructors.
		\GL\ and \CHPSL\ are orthogonal to \OGSL.
		Indeed, they both use more than a temporal goal, \GL\ has quantifier
		alternation fixed to one, and \CHPSL\ only works for two agents.
	\\ \indent
	The paper is almost self contained; all proofs are reported in the appendixes.
	In Appendix~\ref{app:mthnot}, we recall standard mathematical notation and
	some basic definitions that are used in the paper.
	Additional details on \OGSL\ can be found in the technical
	report~\cite{MMPV11}.

\end{section}




\newcommand{\figexmsv}
	{
	\begin{wrapfigure}[11]{r}{0.350\textwidth}
		\vspace{-37.5pt}
		\begin{center}
			\footnotesize
			\mbox{\scalebox{0.80}[0.80]{
			\begin{tikzpicture}
				[node distance = 3cm, bend angle = 15, shorten >= 2pt, shorten <= 2pt]
				\node [player]
							(S0)
							{$\stackrel{\sSym[0]}{\emptyset}$};
				\node [player]
							(S1)
							[below left of = S0]
							{$\stackrel{\sSym[1]}{\pSym}$};
				\node [player]
							(S2)
							[below of = S0]
							{$\stackrel{\sSym[2]}{\pSym, \qSym}$};
				\node [player]
							(S3)
							[below right of = S0]
							{$\stackrel{\sSym[3]}{\qSym}$};
				\path[-stealth']
					(S0)	edge	[bend left]
											node [pos = 0.65, plain1] {$00$}
											(S1)
								edge	[bend left]
											node [plain1] {$01$}
											(S2)
								edge	[bend left]
											node [plain1] {$10$}
											(S3)
								edge	[loop above]
											node [plain1] {$11$}
											()
					(S1)	edge	[bend left]
											node [plain1] {$**$}
											(S0)
					(S2)	edge	[bend left]
											node [plain1] {$**$}
											(S0)
					(S3)	edge	[bend left]
											node [pos = 0.35, plain1] {$**$}
											(S0)
					;
			\end{tikzpicture}
			}}
			\caption{\label{fig:exm:sv} A \CGS\ $\GName$.}
		\end{center}
		\vspace{-15pt}
	\end{wrapfigure}
	}

\newcommand{\figexmsa}
	{
	\begin{wrapfigure}[9]{l}{0.250\textwidth}
		\vspace{-30pt}
		\begin{center}
			\footnotesize
			\mbox{\scalebox{0.80}[0.80]{
			\begin{tikzpicture}
				[node distance = 2cm, bend angle = 10, shorten >= 2pt, shorten <= 2pt]
				\node [player]
							(S0)
							{$\stackrel{\sSym[0]}{\emptyset}$};
				\node [player]
							(S1)
							[below left of = S0]
							{$\stackrel{\sSym[1]}{\pSym}$};
				\node [player]
							(S2)
							[below right of = S0]
							{$\stackrel{\sSym[2]}{\emptyset}$};
				\path[-stealth']
					(S0)	edge	[bend right]
											node [plain1] {$*0*$}
											(S1)
								edge	[bend left]
											node [plain1] {$*1*$}
											(S2)
					(S1)	edge	[loop below]
											node [plain1] {$***$}
											()
					(S2)	edge	[loop below]
											node [plain1] {$***$}
											()
					;
			\end{tikzpicture}
			}}
			\caption{\label{fig:exm:sa} The \CGS\ $\GName[SA]$.}
		\end{center}
		\vspace{-15pt}
	\end{wrapfigure}
	}

\newcommand{\figexmcd}
	{
	\begin{wrapfigure}[9]{r}{0.275\textwidth}
		\vspace{-30pt}
		\begin{center}
			\footnotesize
			\mbox{\scalebox{0.80}[0.80]{
			\begin{tikzpicture}
				[node distance = 2.25cm, bend angle = 10, shorten >= 2pt, shorten <=
				2pt]
				\node [player]
							(S0)
							{$\stackrel{\sSym[0]}{\emptyset}$};
				\node [player]
							(S1)
							[below left of = S0]
							{$\stackrel{\sSym[1]}{\pSym}$};
				\node [player]
							(S2)
							[below right of = S0]
							{$\stackrel{\sSym[2]}{\emptyset}$};
				\path[-stealth']
					(S0)	edge	[bend right]
											node [plain1] {$\!0\!*\!0, \!1\!*\!1\!$}
											(S1)
								edge	[bend left]
											node [plain1] {$\!0\!*\!1, \!1\!*\!0\!$}
											(S2)
					(S1)	edge	[loop below]
											node [plain1] {$***$}
											()
					(S2)	edge	[loop below]
											node [plain1] {$***$}
											()
					;
			\end{tikzpicture}
			}}
			\caption{\label{fig:exm:cd} The \CGS\ $\GName[CD]$.}
		\end{center}
		\vspace{-15pt}
	\end{wrapfigure}
	}




\begin{section}{Preliminaries}
	\label{sec:preliminaries}

	A \emph{concurrent game structure} (\emph{\CGS}, for short)~\cite{AHK02} is a
	tuple $\GName \defeq \CGSStruct$, where $\APSet$ and $\AgSet$ are finite
	non-empty sets of \emph{atomic propositions} and \emph{agents}, $\AcSet$ and
	$\StSet$ are enumerable non-empty sets of \emph{actions} and \emph{states},
	$\sElm[0] \in \StSet$ is a designated \emph{initial state}, and $\labFun :
	\StSet \to \pow{\APSet}$ is a \emph{labeling function} that maps each state to
	the set of atomic propositions true in that state.
	Let $\DecSet \defeq \AcSet^{\AgSet}$ be the set of \emph{decisions}, i.e.,
	functions from $\AgSet$ to $\AcSet$ representing the choices of an action for
	each agent.
	Then, $\trnFun : \StSet \times \DecSet \to \StSet$ is a \emph{transition
	function} mapping a pair of a state and a decision to a state.
	If the set of actions is finite, i.e., $b = \card{\AcSet} < \omega$, we say
	that $\GName$ is $b$-bounded, or simply bounded.
	If both the sets of \mbox{actions and states are finite, we say that $\GName$
	is finite.}
	\\ \indent
	A \emph{track} (resp., \emph{path}) in a \CGS\ $\GName$ is a finite (resp., an
	infinite) sequence of states $\trkElm \in \StSet^{*}$ (resp., $\pthElm \in
	\StSet^{\omega}$) such that, for all $i \in \numco{0}{\card{\trkElm} - 1}$
	(resp., $i \in \SetN$), there exists a decision $\decFun \in \DecSet$ such
	that $(\trkElm)_{i + 1} = \trnFun((\trkElm)_{i}, \decFun)$ (resp.,
	$(\pthElm)_{i + 1} = \trnFun((\pthElm)_{i}, \decFun)$).
	A track $\trkElm$ is \emph{non-trivial} if $\card{\trkElm} > 0$, i.e.,
	$\trkElm \neq \epsilon$.
	$\TrkSet \subseteq \StSet^{+}$ (resp., $\PthSet \subseteq \StSet^{\omega}$)
	denotes the set of all non-trivial tracks (resp., paths).
	Moreover, $\TrkSet(\sElm) \defeq \set{ \trkElm \in \TrkSet }{ \fst{\trkElm} =
	\sElm }$ (resp., $\PthSet(\sElm) \defeq \set{ \pthElm \in \PthSet }{
	\fst{\pthElm} = \sElm }$) indicates the subsets of tracks (resp., paths)
	starting at a state $\sElm \in \StSet$.
	\\ \indent
	A \emph{strategy} is a partial function $\strFun : \TrkSet \pto \AcSet$ that
	maps each non-trivial track in its domain to an action.
	For a state $\sElm \in \StSet$, a strategy $\strFun$ is said
	\emph{$\sElm$-total} if it is defined on all tracks starting in $\sElm$, i.e.,
	$\dom{\strFun} = \TrkSet(\sElm)$.
	$\StrSet \defeq \TrkSet \pto \AcSet$ (resp., $\StrSet(\sElm) \defeq
	\TrkSet(\sElm) \to \AcSet$) denotes the set of all (resp., $\sElm$-total)
	strategies.
	For all tracks $\rho \in \TrkSet$, by $(\strFun)_{\trkElm} \in \StrSet$ we
	denote the \emph{translation} of $\strFun$ along $\trkElm$, i.e., the strategy
	with $\dom{(\strFun)_{\trkElm}} \defeq \set{ \lst{\trkElm} \cdot \trkElm' }{
	\trkElm \cdot \trkElm' \in \dom{\strFun} }$ such that
	$(\strFun)_{\trkElm}(\lst{\trkElm} \cdot \trkElm') \defeq \strFun(\trkElm
	\cdot \trkElm')$, for all $\trkElm \cdot \trkElm' \in \dom{\strFun}$.
	\\ \indent
	Let $\VarSet$ be a fixed set of \emph{variables}.
	An \emph{assignment} is a partial function $\asgFun : \VarSet \cup \AgSet \pto
	\StrSet$ mapping variables and agents in its domain to a strategy.
	An assignment $\asgFun$ is \emph{complete} if it is defined on all agents,
	i.e., $\AgSet \subseteq \dom{\asgFun}$.
	For a state $\sElm \in \StSet$, it is said that $\asgFun$ is
	\emph{$\sElm$-total} if all strategies $\asgFun(\lElm)$ are $\sElm$-total,
	for $\lElm \in \dom{\asgFun}$.
	$\AsgSet \defeq \VarSet \cup \AgSet \pto \StrSet$ (resp., $\AsgSet(\sElm)
	\defeq \VarSet \cup \AgSet \pto \StrSet(\sElm)$) denotes the set of all
	(resp., $\sElm$-total) assignments.
	Moreover, $\AsgSet(\XSet) \defeq \XSet \to \StrSet$ (resp.,
	$\AsgSet(\XSet, \sElm) \defeq \XSet \to \StrSet(\sElm)$) indicates the
	subset of \emph{$\XSet$-defined} (resp., $\sElm$-total) assignments, i.e.,
	(resp., $\sElm$-total) assignments defined on the set $\XSet \subseteq
	\VarSet \cup \AgSet$.
	For all tracks $\rho \in \TrkSet$, by $(\asgFun)_{\trkElm} \in
	\AsgSet(\lst{\trkElm})$ we denote the \emph{translation} of $\asgFun$ along
	$\trkElm$, i.e., the $\lst{\trkElm}$-total assignment with
	$\dom{(\asgFun)_{\trkElm}} \defeq \dom{\asgFun}$, such that
	$(\asgFun)_{\trkElm}(\lElm) \defeq (\asgFun(\lElm))_{\trkElm}$, for all $\lElm
	\in \dom{\asgFun}$.
	For all elements $\lElm \in \VarSet \cup \AgSet$, by $\asgFun[][\lElm \mapsto
	\strFun] \in \AsgSet$ we denote the new assignment defined on
	$\dom{\asgFun[][\lElm \mapsto \strFun]} \defeq \dom{\asgFun} \cup \{ \lElm \}$
	that returns $\strFun$ on $\lElm$ and $\asgFun$ otherwise, i.e.,
	$\asgFun[][\lElm \!\mapsto\! \strFun](\lElm) \!\defeq\! \strFun$ and
	$\asgFun[][\lElm \!\mapsto\! \strFun](\lElm') \!\defeq\! \asgFun(\lElm')$, for
	all $\lElm' \!\in\! \dom{\asgFun} \!\setminus\! \{ \lElm \}$.
	\\ \indent
	A path $\playElm \in \PthSet(\sElm)$ starting at a state $\sElm \in \StSet$ is
	a \emph{play} w.r.t.\ a complete $\sElm$-total assignment $\asgFun \in
	\AsgSet(\sElm)$ (\emph{$(\asgFun, \sElm)$-play}, for short) if, for all $i \in
	\SetN$, it holds that $(\playElm)_{i + 1} = \trnFun((\playElm)_{i}, \decFun)$,
	where $\decFun(\aElm) \defeq \asgFun(\aElm)((\playElm)_{\leq i})$, for each
	$\aElm \in \AgSet$.
	The partial function $\playFun : \AsgSet \times \StSet \pto \PthSet$, with
	$\dom{\playFun} \defeq \set{ (\asgFun, \sElm) }{ \AgSet \subseteq
	\dom{\asgFun} \land \asgFun \in \AsgSet(\sElm) \land \sElm \in \StSet }$,
	returns the $(\asgFun, \sElm)$-play $\playFun(\asgFun, \sElm) \in
	\PthSet(\sElm)$, for all $(\asgFun, \sElm)$ in its domain.
	\\ \indent
	For a state $\sElm \in \StSet$ and a complete $\sElm$-total assignment
	$\asgFun \in \AsgSet(\sElm)$, the \emph{$i$-th global translation} of
	$(\asgFun, \sElm)$, with $i \in \SetN$, is the pair of a complete assignment
	and a state $(\asgFun, \sElm)^{i} \defeq ((\asgFun)_{(\playElm)_{\leq i}},
	(\playElm)_{i})$, where $\playElm = \playFun(\asgFun, \sElm)$.

	From now on, we use the name of a \CGS\ as a subscript to extract the
	components from its tuple-structure.
	Accordingly, if $\GName = \CGSStruct$, we have $\AcSet[\GName] = \AcSet$,
	$\labFun[\GName] = \labFun$, $\sElm[0\GName] = \sSym[0]$, and so on.
	Also, we use the same notational concept to make explicit to which \CGS\ the
	sets $\DecSet$, $\TrkSet$, $\PthSet$, etc. are related to.
	Note that, we omit the subscripts if the structure can be unambiguously
	individuated from the context.

\end{section}




\begin{section}{One-Goal Strategy Logic}
	\label{sec:ogsl}

	In this section, we introduce syntax and semantics of One-Goal Strategy Logic
	(\OGSL, for short), as a syntactic fragment of \SL, which we also report here
	for technical reasons.
	For more about \OGSL, see~\cite{MMPV11}.

	\begin{paragraph}{\SL\ Syntax}

		\emph{\SL} syntactically extends \LTL\ by means of two \emph{strategy
		quantifiers}, existential $\EExs{\xElm}$ and universal $\AAll{\xElm}$, and
		\emph{agent binding} $(\aElm, \xElm)$, where $\aElm$ is an agent and $\xElm$
		is a variable.
		Intuitively, these elements can be read, respectively, as \emph{``there
		exists a strategy $\xElm$''}, \emph{``for all strategies $\xElm$''}, and
		\emph{``bind agent $\aElm$ to the strategy associated with the variable
		$\xElm$''}.
		\mbox{The formal syntax of \SL\ follows.}
		\begin{definition}[\SL\ Syntax]
			\label{def:sl(syntax)}
			\SL\ \emph{formulas} are built inductively from the sets of atomic
			propositions $\APSet$, variables $\VarSet$, and agents $\AgSet$, by using
			the following grammar, where $\pElm \in \APSet$, $\xElm \in \VarSet$, and
			$\aElm \in \AgSet$:
			\begin{center}
				$\varphi ::= \pElm \mid \neg \varphi \mid \varphi \wedge \varphi \mid
				\varphi \vee \varphi \mid \X \varphi \mid \varphi \:\U \varphi \mid
				\varphi \:\R \varphi \mid \EExs{\xElm} \varphi \mid \AAll{\xElm} \varphi
				\mid (\aElm, \xElm) \varphi$.
			\end{center}
		\end{definition}

		By $\mthfun{sub}(\varphi)$ we denote the set of all \emph{subformulas} of
		the \SL\ formula $\varphi$.
		For instance, with $\varphi = \EExs{\xSym} (\alpha, \xSym) (\F \pSym)$, we
		have that $\sub{\varphi} = \{ \varphi, (\alpha, \xSym) (\F \pSym), (\F
		\pSym), \pSym, \Tt \}$.
		By $\free{\varphi}$ we indicate the set of \emph{free agents/variables}
		of $\varphi$ defined as the subset of $\AgSet \cup \VarSet$ containing
		\emph{(i)} all the agents for which there is no variable application before
		the occurrence of a temporal operator and \emph{(ii)} all the variables for
		which there is an application but no quantification.
		For example, let $\varphi = \EExs{\xElm} (\alpha, \xElm)(\beta, \yElm)(\F
		\pElm)$ be the formula on agents $\AgSet = \{ \alpha, \beta, \gamma \}$.
		Then, we have $\free{\varphi} = \{ \gamma, \yElm \}$, since $\gamma$ is an
		agent without any application before $\F \pElm$ and $\yElm$ has no
		quantification at all.
		A formula $\varphi$ without free agents (resp., variables), i.e., with
		$\free{\varphi} \cap \AgSet = \emptyset$ (resp., $\free{\varphi} \cap
		\VarSet = \emptyset)$, is named \emph{agent-closed} (resp.,
		\emph{variable-closed}).
		If $\varphi$ is both agent- and variable-closed, it is named
		\emph{sentence}.
		By $\mthfun{snt}(\varphi)$ we denote the set of all sentences that are
		subformulas of $\varphi$.

	\end{paragraph}

	\begin{paragraph}{\SL\ Semantics}

		As for \ATLS, we define the semantics of \SL\ w.r.t. concurrent game
		structures.
		For a \CGS\ $\GName$, a state $\sElm$, and an $\sElm$-total assignment
		$\asgFun$ with $\free{\varphi} \subseteq \dom{\asgFun}$, we write $\GName,
		\asgFun, \sElm \models \varphi$ to indicate that the formula $\varphi$ holds
		at $\sElm$ under the assignment $\asgFun$.
		The semantics of \SL\ formulas involving $\pElm$, $\neg$, $\wedge$, and
		$\vee$ is defined as usual in \LTL\ and we omit it here (see~\cite{MMPV11},
		for the full definition).
		The semantics of the remaining part, which involves quantifications,
		bindings, and temporal operators follows.
		\begin{definition}[\SL\ Semantics]
			\label{def:sl(semantics)}
			Given a \CGS\ $\GName$, for all \SL\ formulas $\varphi$, states $\sElm \in
			\StSet$, and $\sElm$-total assignments $\asgFun \in \AsgSet(\sElm)$ with
			$\free{\varphi} \subseteq \dom{\asgFun}$, the relation $\GName, \asgFun,
			\sElm \models \varphi$ is inductively defined as follows.
			\begin{enumerate}
				\item\label{def:sl(semantics:eqnt)}
					$\GName, \asgFun, \sElm \models \EExs{\xElm} \varphi$ iff there
					exists an $\sElm$-total strategy $\strFun \in \StrSet(\sElm)$ such
					that $\GName, \asgFun[][\xElm \mapsto \strFun], \sElm \models
					\varphi$;
				\item\label{def:sl(semantics:aqnt)}
					$\GName, \asgFun, \sElm \models \AAll{\xElm} \varphi$ iff for all
					$\sElm$-total strategies $\strFun \in \StrSet(\sElm)$ it holds
					that $\GName, \asgFun[][\xElm \mapsto \strFun], \sElm \models
					\varphi$.
			\end{enumerate}
			Moreover, if $\free{\varphi} \cup \{\xElm \} \subseteq \dom{\asgFun}
			\cup \{\aElm \}$ for an agent $\aElm \in \AgSet$, it holds that:
			\begin{enumerate}
				\setcounter{enumi}{2}
				\item\label{def:sl(semantics:bnd)}
					$\GName, \asgFun, \sElm \models (\aElm, \xElm) \varphi$ iff $\GName,
					\asgFun[][\aElm \mapsto \asgFun(\xElm)], \sElm \models \varphi$.
			\end{enumerate}
			Finally, if $\asgFun$ is also complete, it holds that:
			\begin{enumerate}
				\setcounter{enumi}{3}
				\item\label{def:sl(semantics:next)}
					$\GName, \asgFun, \sElm \models \X \varphi$ if $\GName, (\asgFun,
					\sElm)^{1} \models \varphi$;
				\item\label{def:sl(semantics:until)}
					$\GName, \asgFun, \sElm \models \varphi_{1} \U \varphi_{2}$ if there
					is an index $i \in \SetN$~with $k \!\leq\! i$ such that $\GName,
					(\asgFun, \sElm)^{i} \!\models\! \varphi_{2}$ and, for all indexes
					$j \!\in\! \SetN$ with $k \!\leq j \!<\! i$, it holds that $\GName,
					(\asgFun, \sElm)^{j} \!\models\! \varphi_{1}$;
				\item\label{def:sl(semantics:release)}
					$\GName, \asgFun, \sElm \models \varphi_{1} \R \varphi_{2}$ if, for
					all indexes $i \in \SetN$~with $k \!\leq\! i$, it holds that $\GName,
					(\asgFun, \sElm)^{i} \!\models\! \varphi_{2}$ or there~is~an~in\-dex
					$j \!\in\! \SetN$ with $k \!\leq\! j \!<\! i$ such that \mbox{$\GName,
					(\asgFun, \sElm)^{j} \!\models\! \varphi_{1}$.}
			\end{enumerate}
		\end{definition}
		\noindent
		Intuitively, at Items~\ref{def:sl(semantics:eqnt)}
		and~\ref{def:sl(semantics:aqnt)}, respectively, we evaluate the existential
		$\EExs{\xElm}$ and universal $\AAll{\xElm}$ quantifiers over strategies, by
		associating them to the variable $\xElm$.
		Moreover, at Item~\ref{def:sl(semantics:bnd)}, by means of an agent binding
		$(\aElm, \xElm)$, we commit the agent $\aElm$ to a strategy associated with
		the variable $\xElm$.
		It is evident that the \LTL\ semantics is simply embedded into the \SL\ one.

		A \CGS\ $\GName$ is a \emph{model} of an \SL\ sentence $\varphi$, denoted by
		$\GName \models \varphi$, iff $\GName , \emptyset, \sElm_{0} \models
		\varphi$, where $\emptyset$ is the empty assignment.
		Moreover, $\varphi$ is \emph{satisfiable} iff there is a model for it.
		Given two \CGS s $\GName_{1}$, $\GName_{2}$ and a sentence $\varphi$, we
		say that $\varphi$ is \emph{invariant} under $\GName_{1}$ and $\GName_{2}$
		iff it holds that: $\GName_{1} \models \varphi$ iff $\GName_{2} \models
		\varphi$.
		Finally, given two \SL\ formulas $\varphi_{1}$ and $\varphi_{2}$ with
		$\free{\varphi_{1}} = \free{\varphi_{2}}$, we say that $\varphi_{1}$
		\emph{implies} $\varphi_{2}$, in symbols $\varphi_{1} \implies \varphi_{2}$,
		if, for all \CGS s $\GName$, states $\sElm \in \StSet$, and
		$\free{\varphi_{1}}$-defined $\sElm$-total assignments $\asgFun \in
		\AsgSet(\free{\varphi_{1}}, \sElm)$, it holds that if $\GName, \asgFun,
		\sElm \models \varphi_{1}$ then $\GName, \asgFun, \sElm \models
		\varphi_{2}$.
		Accordingly, we say that $\varphi_{1}$ is \emph{equivalent} to
		$\varphi_{2}$, in symbols $\varphi_{1} \equiv \varphi_{2}$, if $\varphi_{1}
		\implies \varphi_{2}$ and $\varphi_{2} \implies \varphi_{1}$.

		\figexmsv
		As an example, consider the \SL\ sentence $\varphi \!=\! \EExs{\xSym} \!
		\AAll{\ySym} \allowbreak \EExs{\zSym} ((\alpha, \xSym) (\beta, \ySym) (\X
		\pSym) \!\wedge\! (\alpha, \ySym) (\beta, \zSym) (\X \qSym))$.
		Note that both agents $\alpha$ and $\beta$ use the strategy associated with
		$\ySym$ to achieve simultaneously the \LTL\ goals $\X \pSym$ and $\X \qSym$,
		respectively.
		A model for $\varphi$ is the \CGS\ $\GName \defeq \CGSTuple {\{ \pSym, \qSym
		\}} {\{ \alpha, \beta\}} {\{ 0, 1 \}} {\{ \sSym[0], \sSym[1], \sSym[2],
		\sSym[3] \}} {\labFun} {\trnFun} {\sSym[0]}$, where $\labFun(\sSym[0])
		\defeq \emptyset$, $\labFun(\sSym[1]) \defeq \{ \pSym \}$,
		$\labFun(\sSym[2]) \defeq \{ \pSym, \qSym \}$, $\labFun(\sSym[3]) \defeq \{
		\qSym \}$, $\trnFun(\sSym[0]$, $ (0, 0)) \defeq \sSym[1]$,
		$\trnFun(\sSym[0], (0, 1)) \defeq \sSym[2]$, $\trnFun(\sSym[0], (1, 0))
		\defeq \sSym[3]$, and all the remaining transitions go to $\sSym[0]$.
		See the representation of $\GName$ depicted in Figure~\ref{fig:exm:sv}, in
		which vertexes are states of the game and labels on edges represent
		decisions of agents or sets of them, where the symbol $*$ is used in place
		of every possible action.
		Clearly, $\GName \models \varphi$ by letting, on $\sSym[0]$, the variables
		$\xSym$ to chose action $0$ (the formula $(\alpha, \xSym) (\beta, \ySym) (\X
		\pSym)$ is satisfied for any choice of $\ySym$, since we can move from
		$\sSym[0]$ to either $\sSym[1]$ or $\sSym[2]$, both labeled with $\pSym$)
		and $\zSym$ to choose action $1$ when $\ySym$ has action $0$ and, vice
		versa, $0$ when $\ySym$ has $1$ (in both cases, the formula $(\alpha, \ySym)
		(\beta, \zSym) (\X \qSym)$ is satisfied, since one can move from $\sSym[0]$
		to either $\sSym[2]$ or $\sSym[3]$, both labeled with $\qSym$).

	\end{paragraph}

	\begin{paragraph}{\OGSL\ Syntax}

		To formalize the syntactic fragment \OGSL\ of \SL, we need first to define
		the concepts of \emph{quantification} and \emph{binding prefixes}.
		\begin{definition}[Prefixes]
			\label{def:prf}
			A \emph{quantification prefix} over a set $\VSet \!\subseteq\! \VarSet$ of
			variables is a finite word $\qpElm \!\in\! \set{ \EExs{\xElm},
			\AAll{\xElm} }{ \xElm \in \VSet }^{\card{\VSet}}$ of length $\card{\VSet}$
			such that each variable $\xElm \in \VSet$ occurs just once in
			$\qpElm$.
			A \emph{binding prefix} over a set $\VSet \subseteq \VarSet$ of variables
			is a finite word $\bpElm \in \set{ (\aElm, \xElm) }{ \aElm \in \AgSet
			\land \xElm \in \VSet }^{\card{\AgSet}}$ of length $\card{\AgSet}$ such
			that each agent $\aElm \in \AgSet$ occurs just once in $\bpElm$.
			Finally, $\QPSet(\VSet) \subseteq \set{ \EExs{\xElm}, \AAll{\xElm} }{
			\xElm \in \VSet }^{\card{\VSet}}$ and $\BPSet(\VSet) \subseteq \set{
			(\aElm, \xElm) }{ \aElm \in \AgSet \land \xElm \in \VSet
			}^{\card{\AgSet}}$ denote, respectively, the sets of all quantification
			and binding prefixes over variables in $\VSet$.
		\end{definition}

		We can now define the syntactic fragment we want to analyze.
		The idea is to force each group of agent bindings, represented by a binding
		prefix, to be coupled with a quantification prefix.
		\begin{definition}[\OGSL\ Syntax]
			\label{def:ogsl(syntax)}
			\OGSL\ formulas are built inductively from the sets of atomic propositions
			$\APSet$, quantification prefixes $\QPSet(\VSet)$, for $\VSet \subseteq
			\VarSet$, and binding prefixes $\BPSet(\VarSet)$, by using the following
			grammar, with $\pElm \in \APSet$, $\qpElm \in \cup_{\VSet \subseteq
			\VarSet} \QPSet(\VSet)$, and $\bpElm \in \BPSet(\VarSet)$:
			\begin{center}
				$\varphi ::= \pElm \mid \neg \varphi \mid \varphi \wedge \varphi \mid
				\varphi \vee \varphi \mid \X \varphi \mid \varphi \:\U \varphi \mid
				\varphi \:\R \varphi \mid \qpElm \bpElm \varphi$,
			\end{center}
			with $\qpElm \in \QPSet(\free{\bpElm \varphi})$, in the formation rule
			$\qpElm \bpElm \varphi$.
		\end{definition}

		In the following, for a \emph{goal} we mean an \SL\ agent-closed formula of
		the kind $\bpElm \psi$, where $\psi$ is variable-closed and $\bpElm \in
		\BPSet(\free{\psi})$.
		Note that, since $\bpElm \varphi$ is a goal, it is agent-closed, so,
		$\free{\bpElm \varphi} \subseteq \VarSet$.
		Moreover, an \OGSL\ sentence $\varphi$ is \emph{principal} if it is of the
		form $\varphi = \qpElm \bpElm \psi$, where $\bpElm \psi$ is a goal and
		$\qpElm \in \QPSet(\free{\bpElm \psi})$.
		By $\psnt{\varphi} \subseteq \snt{\varphi}$ we denote the set of
		\emph{principal subsentences} of the \OGSL\ formula $\varphi$.

		As an example, let $\varphi_{1} = \qpSym \bpSym[1] \psi_{1}$ and
		$\varphi_{2} = \qpSym (\bpSym[1] \psi_{1} \wedge \bpSym[2] \psi_{2})$, where
		$\qpSym = \AAll{\xSym} \EExs{\ySym} \AAll{\zSym}$, $\bpSym[1] = (\alpha,
		\xSym) (\beta, \ySym) (\gamma, \zSym)$, $\bpSym[2] = (\alpha,
		\ySym) (\beta, \zSym) (\gamma, \ySym)$, $\psi_{1} = \X \pSym$, and $\psi_{2}
		= \X \qSym$.
		Then, it is evident that $\varphi_{1} \in \OGSL$ but $\varphi_{2} \not\in
		\OGSL$, since the quantification prefix $\qpSym$ of the latter does not have
		in its scope a unique goal.

		It is fundamental to observe that the formula $\varphi_{1}$ of the above
		example cannot be expressed in \ATLS, as proved in~\cite{MMPV11} and
		reported in the following theorem, since its $2$-quantifier alternation
		cannot be encompassed in the $1$-alternation \ATLS\ modalities.
		On the contrary, each \ATLS\ formula of the type $\EExs{\ASet} \psi$, where
		$\ASet = \{\alpha_{1}, \ldots, \alpha_{n}\} \subseteq \AgSet = \{\alpha_{1},
		\ldots, \alpha_{n}, \beta_{1}, \ldots, \beta_{m}\}$ can be expressed in
		\OGSL\ as follows: $\EExs{\xSym[1]} \cdots \EExs{\xElm[n]} \AAll{\ySym[1]}
		\allowbreak \cdots \allowbreak \AAll{\ySym[m]} (\alpha_{1}, \xSym[1]) \cdots
		(\alpha_{n}, \xSym[n]) (\beta_{1}, \ySym[1]) \cdots (\beta_{m}, \ySym[m])
		\psi$.
		\begin{theorem}
			\OGSL\ is strictly more expressive than \ATLS.
		\end{theorem}

		We now give two examples in which we show the importance of the ability to
		write specifications with alternation of quantifiers greater than $1$ along
		with strategy sharing.

		\begin{example}[Escape from Alcatraz\footnote{We thank Luigi Sauro for
			having pointed out this example.}]
			Consider the situation in which an Alcatraz prisoner tries to escape from
			jail with the help of an external accomplice of him, by helicopter.
			Due to his panoramic point of view, assume that the accomplice has the
			full visibility on the behaviors of guards, while the prisoner does
			not have the same ability.
			Therefore, the latter has to put in practice an escape strategy that,
			independently from guards moves, can be supported by his accomplice to
			escape.
			We can formalize such an intricate situation by means of an \OGSL\
			sentence as follows.
			First, let $\GName[A]$ be a \CGS\ modeling the possible situations in
			which the agents ``$\pSym$'' prisoner, ``$\gSym$'' guards, and ``$\aSym$''
			accomplice can reside, together with all related possible moves.
			Then, we can verify the existence of an escape strategy by checking the
			assertion $\GName[A] \models \EExs{\xSym} \AAll{\ySym} \EExs{\zSym}
			(\pSym, \xSym) (\gSym, \ySym) (\aSym, \zSym) (\F \mthfun[\PSym]{free})$.
		\end{example}

		\begin{example}[Stealing-Strategy in Hex]
			Hex is a two-player game, red vs blue, in which each player in
			turn places a stone of his color on a single empty hexagonal cell of the
			rhomboidal playing board having opposite sides equally colored, either red
			or blue.
			The goal of each player is to be the first to form a path connecting the
			opposing sides of the board marked by his color.
			It is easy to prove that the stealing-strategy argument does not lead to a
			winning strategy in Hex, i.e., if the player that moves second copies the
			moves of the opponent, he surely loses the play.
			It is possible to formalize this fact in \OGSL\ as follows.
			First model Hex with a \CGS\ $\GName[H]$ whose states represent a possible
			possible configurations reached during a play between ``$\rSym$'' red and
			``$\bSym$'' blue.
			Then, express the negation of the stealing-strategy argument by asserting
			that $\GName[H] \models \EExs{\xSym} (\rSym, \xSym) (\bSym, \xSym) (\F
			\mthfun[\rSym]{cnc})$.
			Intuitively, this sentence says that agent $\rSym$ has a strategy that,
			once it is copied (binded) by $\bSym$ it allows the former to win, i.e.,
			to be the first to connect the related red edges ($\F
			\mthfun[\rSym]{cnc}$).
		\end{example}

	\end{paragraph}

\end{section}




\begin{section}{Strategy Quantifications}
	\label{sec:strqnt}

	We now define the concept of \emph{dependence map}.
	The key idea is that every quantification prefix occurring in an \SL\ formula
	can be represented by a suitable choice of a dependence map over strategies.
	Such a result is at the base of the definition of the \emph{elementariness}
	property and allows us to prove that \OGSL\ is elementarily satisfiable, i.e.,
	we can simplify a reasoning about strategies by reducing it to a set of local
	reasonings about actions~\cite{MMPV11}.

	\begin{paragraph}{Dependence map}

		First, we introduce some notation regarding quantification prefixes.
		Let $\qpElm \in \QPSet(\VSet)$ be a quantification prefix over a set
		$\QPVSet(\qpElm) \defeq \VSet \subseteq \VarSet$ of variables.
		By $\QPEVSet{\qpElm} \defeq \set{ \xElm \in \VSet }{ \exists i \in
		\numco{0}{\card{\qpElm}} .\: (\qpElm)_{i} = \EExs{\xElm} }$ and
		$\QPAVSet{\qpElm} \defeq \VSet \setminus \QPEVSet{\qpElm}$ we denote,
		respectively, the sets of \emph{existential} and \emph{universal variables}
		quantified in $\qpElm$.
		For two variables $\xElm, \yElm \in \VSet$, we say that $\xElm$
		\emph{precedes} $\yElm$ in $\qpElm$, in symbols $\xElm \qpordRel[\qpElm]
		\yElm$, if $\xElm$ occurs before $\yElm$ in $\qpElm$.
		Moreover, by $\QPDepSet(\qpElm) \defeq \set{ (\xElm, \yElm) \in \VSet \times
		\VSet }{ \xElm \in	\QPAVSet{\qpElm}, \yElm \in \QPEVSet{\qpElm} \land \xElm
		\qpordRel[\qpElm] \yElm}$ we denote the set of \emph{dependence pairs},
		i.e., a dependence relation, on which we derive the parameterized version
		$\QPDepSet(\qpElm, \yElm) \defeq \set{ \xElm \in \VSet }{ (\xElm, \yElm) \in
		\QPDepSet(\qpElm)}$ containing all variables from which $\yElm$ depends.
		Also, we use $\dual{\qpElm} \in \QPSet(\VSet)$ to indicate the
		quantification derived from $\qpElm$ by \emph{dualizing} each quantifier
		contained in it, i.e., for all $i \in \numco{0}{\card{\qpElm}}\!$, it holds
		that $(\dual{\qpElm})_{i} = \EExs{\xElm}$ iff $(\qpElm)_{i} = \AAll{\xElm}$,
		with $\xElm \in \VSet$.
		Clearly, $\QPEVSet{\dual{\qpElm}} = \QPAVSet{\qpElm}$ and
		$\QPAVSet{\dual{\qpElm}} = \QPEVSet{\qpElm}$.
		Finally, we define the notion of \emph{valuation} of variables over a
		generic set $\DSet$ as a partial function $\valFun : \VarSet \pto \DSet$
		mapping every variable in its domain to an element in $\DSet$.
		By $\ValSet[\DSet](\VSet) \defeq \VSet \to \DSet$ we denote the set of all
		valuation functions over $\DSet$ defined on $\VSet \subseteq \VarSet$.

		We now give the semantics for quantification prefixes via the following
		definition of \emph{dependence map}.
		\begin{definition}[Dependence Maps]
			\label{def:qntspc}
			Let $\qpElm \in \QPSet(\VSet)$ be a quantification prefix over a set of
			variables $\VSet \subseteq \VarSet$, and $\DSet$ a set.
			Then, a \emph{dependence map} for $\qpElm$ over $\DSet$ is a function
			$\spcFun : \ValSet[\DSet](\QPAVSet{\qpElm}) \to \ValSet[\DSet](\VSet)$
			satisfying the following properties:
			\begin{enumerate}
				\item
					\label{def:qntspc(aqnt)}
					$\spcFun(\valFun)_{\rst \QPAVSet{\qpElm}} \!=\! \valFun$, for all
					$\valFun \in \ValSet[\DSet](\QPAVSet{\qpElm})$;
				\item
					\label{def:qntspc(eqnt)}
					$\spcFun(\valFun[1])(\xElm) \!=\! \spcFun(\valFun[2])(\xElm)$, for all
					$\valFun[1], \valFun[2] \in \ValSet[\DSet](\QPAVSet{\qpElm})$ and
					$\xElm \in \QPEVSet{\qpElm}$ such that $\valFun[1]_{\rst
					\QPDepSet(\qpElm, \xElm)} \!=\! \valFun[2]_{\rst \QPDepSet(\qpElm,
					\xElm)}$.
			\end{enumerate}
			$\SpcSet[\DSet](\qpElm)$ denotes the set of all dependence maps for
			$\qpElm$ over $\DSet$.
		\end{definition}
		\noindent
		Intuitively, Item~\ref{def:qntspc(aqnt)} asserts that $\spcFun$ takes the
		same values of its argument w.r.t.\ the universal variables in $\qpElm$ and
		Item~\ref{def:qntspc(eqnt)} ensures that the value of $\spcFun$ w.r.t.\ an
		existential variable $\xElm$ in $\qpElm$ does not depend on variables not in
		$\QPDepSet(\qpElm, \xElm)$.
		To get better insight into this definition, a dependence map $\spcFun$ for
		$\qpElm$ can be considered as a set of \emph{Skolem functions} that, given a
		value for each variable in $\VSet$ that is universally quantified in
		$\qpElm$, returns a possible value for all the existential variables in
		$\qpElm$, in a way that is consistent w.r.t.\ the order of quantifications.

		We now state a fundamental theorem that describes how to eliminate
		strategy quantifications of an \SL\ formula via a choice of a dependence map
		over strategies.
		This procedure, easily proved to be correct by induction on the structure of
		the formula in~\cite{MMPV11}, can be seen as the equivalent of the
		\emph{Skolemization} in first order logic~\cite{Hod93}.
		\begin{theorem}[\SL\ Strategy Quantification]
			\label{thm:sl(strqnt)}
			Let $\GName$ be a \CGS\ and $\varphi = \qpElm \psi$ an \SL\ sentence,
			where $\psi$ is agent-closed and $\qpElm \in \QPSet(\free{\psi})$.
			Then, $\GName \models \varphi$ iff there exists a dependence map
			$\spcFun \in \SpcSet[ {\StrSet(\sElm[0])} ](\qpElm)$ such that $\GName,
			\spcFun(\asgFun), \sElm[0] \models \psi$, for all $\asgFun \in
			\AsgSet(\QPAVSet{\qpElm}, \sElm[0])$.
		\end{theorem}

		The above theorem substantially characterizes the \SL\ semantics by means of
		the concept of dependence map.
		In particular, it shows that if a formula is satisfiable then it is always
		possible to find a suitable dependence map returning the existential
		strategies in response to the universal ones.
		Such a characterization lends itself to define alternative
		semantics of \SL, based
		on the choice of a subset of dependence maps that meet a certain given
		property.
		We do this on the aim of identifying semantic fragments of \SL\ having
		better model properties and easier decision problems.
		With more details, given a \CGS\ $\GName$, one of its states $\sElm$, and a
		property $\PNot$, we say that a sentence $\qpElm \psi$ is
		$\PNot$-satisfiable, in symbols $\GName \models_{\PNot} \qpElm \psi$, if
		there exists a dependence map $\spcFun$ meeting $\PNot$ such that, for all
		assignment $\asgFun \in \AsgSet(\QPAVSet{\qpElm}, \sElm)$, it holds that
		$\GName, \spcFun(\asgFun), \sElm \models \psi$.
		Alternative semantics identified by a property $\PNot$ are even more
		interesting if there exists a syntactic fragment corresponding to it, i.e.,
		each satisfiable sentence of such a fragment is $\PNot$-satisfiable and vice
		versa.
		In the following, we put in practice this idea in order to show that \OGSL\
		has the same complexity of \ATLS w.r.t.\ the satisfiability problem.

	\end{paragraph}

	\begin{paragraph}{Elementary quantifications}

		According to the above description, we now introduce a suitable property
		of dependence maps, called elementariness, together with the related
		alternative semantics.
		Then, in Theorem~\ref{thm:ogsl(elm)}, we state that \OGSL\ has the
		elementariness property, i.e., each \OGSL\ sentence is satisfiable iff it is
		elementary satisfiable.

		Intuitively, a dependence map $\spcFun \in \SpcSet[\TSet \to \DSet](\qpElm)$
		over functions from a set $\TSet$ to a set $\DSet$ is elementary if it can
		be split into a set of dependence maps over $\DSet$, one for each element of
		$\TSet$, represented by a function $\adj{\spcFun}: \TSet \to
		\SpcSet[\DSet](\qpElm)$.
		This idea allows us to enormously simplify the reasoning about strategy
		quantifications, since we can reduce them to a set of quantifications over
		actions, one for each track in their domains.

		Note that sets $\DSet$ and $\TSet$, as well as $\USet$ and $\VSet$ used in
		the following, are generic and in our framework they may refer to actions
		and strategies ($\DSet$), tracks ($\TSet$), and variables ($\USet$ and
		$\VSet$).
		In particular, observe that functions from $\TSet$ to $\DSet$ represent
		strategies.
		We prefer to use abstract name, as the properties we describe hold
		generally.

		To formally develop the above idea, we have first to introduce the generic
		concept of \emph{adjoint function}.
		From now on, we denote by $\flip{\gFun}: \YSet \to (\XSet \to \ZSet)$ the
		operation of \emph{flipping} of a generic function $\gFun : \XSet \to (\YSet
		\to \ZSet)$, i.e., the transformation of $\gFun$ by swapping the order of
		its arguments.
		Such a flipping is well-grounded due to the following chain of isomorphisms:
		$\XSet \to (\YSet \to \ZSet) \cong (\XSet \times \YSet) \to \ZSet \cong
		(\YSet \times \XSet) \to \ZSet \cong \YSet \to (\XSet \to \ZSet)$.
		\begin{definition}[Adjoint Functions]
			\label{def:adjfun}
			Let $\DSet$, $\TSet$, $\USet$, and $\VSet$ be four sets, and
			$\mFun : (\TSet \to \DSet)^{\USet} \to (\TSet \to \DSet)^{\VSet}$ and
			$\adj{\mFun} : \TSet \to (\DSet^{\USet} \to \DSet^{\VSet})$ two functions.
			Then, $\adj{\mFun}$ is the \emph{adjoint} of $\mFun$ if
			$\adj{\mFun}(\tElm)(\flip{\gFun}(\tElm))(\xElm) =
			\mFun(\gFun)(\xElm)(\tElm)$, for all $\gFun \in (\TSet \to
			\DSet)^{\USet}$, $\xElm \in \VSet$, and $\tElm \in \TSet$.
		\end{definition}
		\noindent
		Intuitively, a function $\mFun$ transforming a map of kind $(\TSet \to
		\DSet)^{\USet}$ into a new map of kind $(\TSet \to \DSet)^{\VSet}$ has an
		adjoint $\adj{\mFun}$ if such a transformation can be done pointwisely
		w.r.t.\ the set $\TSet$, i.e., we can put out as a common domain the set
		$\TSet$ and then transform a map of kind $\DSet^{\USet}$ in a map of kind
		$\DSet^{\VSet}$.
		Observe that, if a function has an adjoint, this is unique.
		Similarly, from an adjoint function it is possible to determine the original
		function unambiguously.
		Thus, it is established a one-to-one correspondence between functions
		admitting an adjoint and the adjoint itself.

		The formal meaning of the elementariness of a dependence map over
		generic functions follows.
		\begin{definition}[Elementary Dependence Maps]
			\label{def:elmspc}
			Let $\qpElm \in \QPSet(\VSet)$ be a quantification prefix over a set
			$\VSet \subseteq \VarSet$ of variables, $\DSet$ and $\TSet$ two sets, and
			$\spcFun \in \SpcSet[\TSet \to \DSet](\qpElm)$ a dependence map
			for $\qpElm$ over $\TSet \to \DSet$.
			Then, $\spcFun$ is \emph{elementary} if it admits an adjoint function.
			$\ESpcSet[\TSet \to \DSet](\qpElm)$ denotes the set of all
			elementary dependence maps for $\qpElm$ over $\TSet \to \DSet$.
		\end{definition}

		As mentioned above, we now introduce the important variant of \OGSL\
		semantics based on the property of elementariness of dependence maps over
		strategies.
		We refer to the related satisfiability concept as \emph{elementary
		satisfiability}, in symbols $\emodels$.


		The new semantics of \OGSL\ formulas involving atomic propositions, Boolean
		connectives, temporal operators, and agent bindings is defined as for the
		classic one, where the modeling relation $\models$ is substituted with
		$\emodels$, and we omit to report it here.
		In the following definition, we only describe the part concerning the
		quantification prefixes.
		Observe that by $\bpFun[\bpElm] : \AgSet \to \VarSet$, for $\bpElm \in
		\BndSet(\VarSet)$, we denote the function associating to each agent the
		variable of its binding in $\bpElm$.
		\begin{definition}[\OGSL\ Elementary Semantics]
			\label{def:ogsl(elementarysemantics)}
			Let $\GName$ be a \CGS, $\sElm \in \StSet$ one of its states, and $\qpElm
			\bpElm \psi$ an \OGSL\ principal sentence.
			Then $\GName, \emptyfun, \sElm \emodels \qpElm \bpElm \psi$ iff there is
			an elementary dependence map $\spcFun \in \ESpcSet[
			{\StrSet(\sElm)} ](\qpElm)$ for $\qpElm$ over $\StrSet(\sElm)$ such that
			$\GName, \spcFun(\asgFun) \cmp \bpFun[\bpElm], \sElm \emodels \psi$, for
			all $\asgFun \in \AsgSet(\QPAVSet{\qpElm}, \sElm)$.
		\end{definition}
		\noindent
		It is immediate to see a strong similarity between the statement of
		Theorem~\ref{thm:sl(strqnt)} of \SL\ strategy quantification and the
		previous definition.
		The only crucial difference resides in the choice of the kind of dependence
		map.
		Moreover, observe that, differently from the classic semantics, the
		quantifications in a prefix are not treated individually but as an atomic
		block.
		This is due to the necessity of having a strict correlation between the
		point-wise structure of the quantified strategies.

		Finally, we state the following fundamental theorem which is a key step in
		the proof of the bounded model property and decidability of the
		satisfiability for \OGSL, whose correctness has been proved
		in~\cite{MMPV11}.
		The idea behind the proof of the elementariness property resides in the
		strong similarity between the statement of Theorem~\ref{thm:sl(strqnt)} of
		\SL\ strategy quantification and the definition of the winning condition in
		a classic turn-based two-player game.
		Indeed, on one hand, we say that a sentence is satisfiable iff ``there
		exists a dependence map such that, for all all assignments, it
		holds that ...''.
		On the other hand, we say that the first player wins a game iff ``there
		exists a strategy for him such that, for all strategies of the other player,
		it holds that ...''.
		The gap between these two formulations is solved in \OGSL\ by using the
		concept of elementary quantification.
		So, we build a two-player turn-based game in which the two players are
		viewed one as a dependence map and the other as a valuation over universal
		quantified variables, both over actions, such that the formula is satisfied
		iff the first player wins the game.
		This construction is a deep technical evolution of the proof method used for
		the dualization of alternating automata on infinite objects~\cite{MS87}.
		Precisely, it uses Martin's Determinacy Theorem~\cite{Mar75} on the
		auxiliary turn-based game to prove that, if there is no dependence map of a
		given prefix that satisfies the given property, there is a dependence map of
		the dual prefix satisfying its negation.
		\begin{theorem}[\OGSL\ Elementariness]
			\label{thm:ogsl(elm)}
			Let $\GName$ be a \CGS\ and $\varphi$ an \OGSL\ sentence.
			Then, $\GName \!\models\! \varphi$ iff $\GName \!\emodels\! \varphi$.
		\end{theorem}

		In order to understand what elementariness means from a syntactic point of
		view, note that in \OGSL\ it holds that $\qpElm \bpElm \X \psi \equiv \qpElm
		\bpElm \X \qpElm \bpElm\psi$, i.e., we can requantify the strategies to
		satisfy the inner subformula $\psi$.
		This equivalence is a generalization of what is well know to hold for \CTLS:
		$\E \X \psi \equiv \E \X \E \psi$.
		Moreover, note that, as reported in \cite{MMPV11}, elementariness does not
		hold for more expressive fragments of \SL, such as \BGSL.

	\end{paragraph}

\end{section}




\begin{section}{Bounded Dependence Maps}
	\label{sec:bndspc}

	Here we prove a boundedness property for dependence maps crucial to get, in
	Section~\ref{sec:modprp}, the \emph{bounded tree-model property} for \OGSL,
	which is a preliminary step towards our decidability proof for the logic.

	As already mentioned, on reasoning about the satisfiability of an \OGSL\
	sentence, one can simplify the process, via elementariness, by splitting a
	dependence map over strategies in a set of dependence maps over actions.
	Thus, to gain the bounded model property, it is worth understanding how to
	build dependence maps over a predetermined finite set of actions, while
	preserving the satisfiability of the sentence of interest.

	The main difficulty here is that, the verification process of a sentence
	$\varphi$ over an (unbounded) \CGT\ $\TName$ may require some of its
	subsentences, perhaps in contradiction among them, to be checked on disjoint
	subtrees of $\TName$.
	For example, consider the formula $\varphi = \phi_{1} \wedge \phi_{2}$,
	where $\phi_{1} = \qpElm[1] \bpElm \X \pSym$ and $\phi_{2} = \qpElm[2]
	\bpElm \X \neg \pSym$ with $\bpElm = (\alpha, \xSym) (\beta, \ySym) (\gamma,
	\zSym)$.
	It is evident that, if $\TName \models \varphi$, the two strategy
	quantifications made via the prefixes $\qpElm[1]$ and $\qpElm[2]$ have to
	select two disjoint subtrees of $\TName$ on which verify the temporal
	properties $\X \pSym$ and $\X \neg \pSym$, respectively.
	So, a correct pruning of $\TName$ in a bounded tree-model has to keep the
	satisfiability of the subsentences $\phi_{1}$ and $\phi_{2}$ separated, by
	avoiding the collapse of the relative subtrees, which can be ensured via the
	use of an appropriate number of actions.

	By means of characterizing properties named \emph{overlapping} (see
	Definitions~\ref{def:intsig} and~\ref{def:spcmap}) on quantification-binding
	prefixes and sets of dependence maps, called \emph{signatures} (see
	Definition~\ref{def:sig}) and \emph{signature dependences} (see
	Definition~\ref{def:spcmap}), respectively, we ensure that the set of
	required actions is finite.
	Practically, we prove that sentences with overlapping signatures necessarily
	share a common subtree, independently from the number of actions in the
	model (see Corollary~\ref{cor:intspc}).
	Conversely, sentences with non-overlapping signatures may need different
	subtrees.
	So, a model must have a sufficient big set of actions, which we prove to
	be finite anyway (see Theorem~\ref{thm:nonintspc}).
	Note that, in the previous example, $\varphi$ to be satisfiable needs to
	have non-overlapping signatures, since otherwise there is at least a shared
	outcome on which verify the incompatible temporal properties $\X \pSym$ and
	$\X \neg \pSym$.

	We now give few more details on the idea behind the properties described
	above.
	Suppose to have a set of quantification prefixes $\QSet \subseteq
	\QPSet(\VSet)$ over a set of variables $\VSet$.
	We ask whether there is a relation among the elements of $\QSet$ that forces
	a set of related dependence maps to intersect their ranges in at least one
	valuation of variables.
	\figexmsa
	For instance, consider in the previous example the prefixes to be
	set as follows: $\qpElm[1] \defeq \AAll{\xSym} \EExs{\ySym} \EExs{\zSym}$
	and $\qpElm[2] \defeq \AAll{\zSym} \EExs{\ySym} \AAll{\xSym}$.
	Then, we want to know whether an arbitrary pair of dependence maps
	$\spcFun[1] \in \SpcSet[\DSet](\qpElm[1])$ and $\spcFun[2] \in
	\SpcSet[\DSet](\qpElm[2])$ has intersecting ranges, for a set $\DSet$.
	In this case, since $\ySym$ is existentially quantified in both prefixes, we
	can build $\spcFun[1]$ and $\spcFun[2]$ in such a way they choose different
	elements of $\DSet$ on $\ySym$, when they do the same choices on the
	other variables, supposed that $\card{\DSet} > 1$.
	Thus, if the prefixes share at least an existential variable, it is
	possible to find related dependence maps that are non-overlapping.
	Indeed, in this case, the formula $\varphi$ is satisfied on the \CGS\
	$\GName[SA]$ of Figure \ref{fig:exm:sa}, since we can allow $\ySym$ on
	$\sElm[0]$ to chose $0$ for $\qpElm[1]$ and $1$ for $\qpElm[2]$.

	\figexmcd
	Now, let consider the following prefixes: $\qpElm[1] \defeq \AAll{\xSym}
	\EExs{\zSym} \AAll{\ySym}$ and $\qpElm[2] \defeq \AAll{\zSym} \AAll{\ySym}
	\EExs{\xSym}$.
	Although, in this case, each variable is existentially quantified at most
	once, we have that $\xSym$ and $\zSym$ mutually depend in the different
	prefixes.
	So, there is a cyclic dependence that can make two related non-overlapping
	dependence maps.
	Indeed, suppose to have $\DSet = \{ 0, 1\}$.
	Then, we can choose $\spcFun[1] \in \SpcSet[\DSet](\qpElm[1])$ and
	$\spcFun[2] \in \SpcSet[\DSet](\qpElm[2])$ in the way that, for all
	valuations $\valFun[1] \in \dom{\spcFun[1]}$ and $\valFun[2] \in
	\dom{\spcFun[2]}$, it holds that $\spcFun[1](\valFun[1])(\zSym) \defeq
	\valFun[1](\xSym)$ and
	$\spcFun[2](\valFun[2])(\xSym) \defeq 1 - \valFun[2](\zSym)$.
	Thus, $\spcFun[1]$ and $\spcFun[2]$ do not intersect their ranges.
	Indeed, with the considered prefixes, the formula $\varphi$ is satisfied on
	the \CGS\ $\GName[CD]$ of \mbox{Figure \ref{fig:exm:cd}, by using the
	dependence maps described above.}

	Finally, consider a set of prefixes in which there is neither a shared
	existential quantified variable nor a cyclic dependence, such as the
	following: $\qpElm[1] \!\defeq\! \AAll{\xSym} \AAll{\ySym} \EExs{\zSym}$,
	$\qpElm[2] \!\defeq\! \EExs{\ySym} \AAll{\xSym} \AAll{\zSym}$, and
	$\qpElm[3] \!\defeq\! \AAll{\ySym} \EExs{\xSym} \AAll{\zSym}$.
	We now show that an arbitrary choice of dependence maps $\spcFun[1] \!\in\!
	\SpcSet[\DSet](\qpElm[1])$, $\spcFun[2] \!\in\! \SpcSet[\DSet](\qpElm[2])$,
	and $\spcFun[3] \!\in\! \SpcSet[\DSet](\qpElm[3])$ must have intersecting
	ranges, for every set $\DSet$.
	Indeed, since $\ySym$ in $\qpElm[2]$ does not depend from any other
	variable, there is a value $\dElm[\ySym] \!\in\! \DSet$ such that, for all
	$\valFun[2] \!\in\! \dom{\spcFun[2]}$, it holds that
	$\spcFun[2](\valFun[2])(\ySym) \!=\! \dElm[\ySym]$.
	Now, since $\xSym$ in $\qpElm[3]$ depends only on $\ySym$, there is a
	value $\dElm[\xSym] \!\in\! \DSet$ such that, for all $\valFun[3] \!\in\!
	\dom{\spcFun[3]}$ with $\valFun[3](\ySym) \!=\! \dElm[\ySym]$, it holds that
	$\spcFun[3](\valFun[3])(\xSym) \!=\! \dElm[\xSym]$.
	Finally, we can determine the value $\dElm[\zSym] \!\in\! \DSet$ of $\zSym$
	in $\qpElm[1]$ since $\xSym$ and $\ySym$ are fixed.
	So, for all $\valFun[1] \!\in\! \dom{\spcFun[1]}$ with $\valFun[1](\xSym)
	\!=\! \dElm[\xSym]$ and $\valFun[1](\ySym) \!=\! \dElm[\ySym]$, it holds
	that $\spcFun[1](\valFun[1])(\zSym) \!=\! \dElm[\zSym]$.
	Thus, the valuation $\valFun \!\in\! \ValSet[\DSet](\VSet)$, with
	$\valFun(\xSym) \!=\! \dElm[\xSym]$, $\valFun(\ySym) \!=\! \dElm[\ySym]$,
	and $\valFun(\zSym) \!=\! \dElm[\zSym]$, is such that $\valFun \in
	\rng{\spcFun[1]} \cap \rng{\spcFun[2]} \cap \rng{\spcFun[3]}$.
	Note that we run this procedure since we can find at each step an
	existential variable that depends only on universal variables
	previously determined.

	In order to formally define the above procedure, we need to introduce some
	preliminary definitions.
	As first thing, we generalize the described construction by taking into
	account not only quantification prefixes but binding prefixes too.
	This is due to the fact that different principal subsentences of the
	specification can share the same quantification prefix by having different
	binding prefixes.
	Moreover, we need to introduce a tool that gives us a way to
	differentiate the check of the satisfiability of a given sentence in
	different parts of the model, since it can use different actions when starts
	the check from different states.
	For this reason, we introduce the concepts of \emph{signature} and
	\emph{labeled signature}.
	The first is used to arrange opportunely prefixes with bindings, represented
	in a more general form through the use of a generic support set $\ESet$,
	while the second allows us to label signatures, by means of a set $\LSet$,
	to maintain an information on different instances of the same sentence.
	\begin{definition}[Signatures]
		\label{def:sig}
		A \emph{signature} on a set $\ESet$ is a pair $\sigElm \defeq (\qpElm,
		\bFun) \in \QPSet(\VSet) \times \VSet^{\ESet}$ of a quantification prefix
		$\qpElm$ over $\VSet$ and a surjective function $\bFun$ from $\ESet$ to
		$\VSet$, for a given set of variables $\VSet \subseteq \VarSet$.
		A \emph{labeled signature} on $\ESet$ w.r.t.\ a set $\LSet$ is a pair
		$(\sigElm, \lElm) \in (\QPSet(\VSet) \times \VSet^{\ESet}) \times \LSet$
		of a signature $\sigElm$ on $\ESet$ and a labeling $\lElm$ in $\LSet$.
		The sets $\SigSet(\ESet) \defeq \bigcup_{\VSet \subseteq \VarSet}
		\QPSet(\VSet) \!\times\! \VSet^{\ESet}$ and $\LSigSet(\ESet, \LSet) \defeq
		\SigSet(\ESet) \times \LSet$ contain, respectively, all \emph{signatures}
		on $\ESet$ and \emph{labeled signatures} on $\ESet$ w.r.t.\ $\LSet$.
	\end{definition}

	We now extend the concepts of existential quantification and functional
	dependence from prefixes to signatures.
	By $\QPEVSet{\sigElm} \defeq \set{ \eElm \in \ESet }{ \bFun(\eElm) \in
	\QPEVSet{\qpElm} }$, $\QPDepSet(\sigElm) \defeq \set{ (\eElm', \eElm'') \in
	\ESet \times \ESet }{ (\bFun(\eElm'), \bFun(\eElm'')) \in \QPDepSet(\qpElm)
	}$, and $\BPColSet(\sigElm) \defeq \set{ (\eElm', \eElm'') \in \ESet \times
	\ESet }{ \bFun(\eElm') = \bFun(\eElm'') \in \QPAVSet{\qpElm} }$, with
	$\sigElm = (\qpElm, \bFun) \in \SigSet(\ESet)$, we denote the set of
	existential elements, and the relation sets of functional
	dependent and collapsing elements, respectively.
	Moreover, for a set $\SSet \subseteq \SigSet(\ESet)$ of signatures, we
	define $\BPColSet(\SSet) \defeq (\bigcup_{\sigElm \in \SSet}
	\BPColSet(\sigElm))^{+}$ as the transitive relation set of collapsing
	elements and $\QPEVSet{\SSet} \defeq \bigcup_{\sigElm \in \SSet}
	\QPEVSet{\SSet, \sigElm}$, with $\QPEVSet{\SSet, \sigElm} \defeq \set{
	\eElm \in \QPEVSet{\sigElm} }{ \exists \sigElm' \in \SSet, \eElm' =
	(\qpElm', \bFun') \in \QPEVSet{\sigElm'} \:.\: (\sigElm \neq \sigElm' \lor
	\bFun(\eElm) \neq \bFun'(\eElm')) \land (\eElm, \eElm') \in
	\BPColSet(\SSet)}$, as the set of elements that are existential in two
	signatures, either directly or via a collapsing chain.
	Finally, by $\QPDepSet'(\sigElm) \defeq \set{(\eElm', \eElm'') \in \ESet
	\times \ESet}{ \exists \eElm''' \in \ESet \:.\: (\eElm', \eElm''') \in
	\BPColSet(\SSet) \land (\eElm''', \eElm'') \in \QPDepSet(\sigElm)}$ we
	indicate the relation set of functional dependent elements connected via a
	collapsing chain.

	As described above, if a set of prefixes has a \emph{cyclic dependence}
	between variables, we are sure to find a set of dependence maps,
	bijectively related to such prefixes, that do not share any total assignment
	in their codomains.
	Here, we formalize this concept of dependence by considering bindings too.
	In particular, the check of dependences is not done directly on variables,
	but by means of the associated elements of the support set $\ESet$.
	Note that, in the case of labeled signatures, we do not take into account
	the labeling component, since two instances of the same signature with
	different labeling cannot have a mutual dependent variable.

	To give the formal definition of cyclic dependence, we first provide the
	definition of \emph{$\SSet$-chain}.
	\begin{definition}[$\SSet$-Chain]
		\label{def:schain}
		An \emph{$\SSet$-chain} for a set of signatures $\SSet \subseteq
		\SigSet(\ESet)$ on $\ESet$ is a pair $(\vec{\eElm}, \vec{\sigElm}) \in
		\ESet^{k} \times \SSet^{k}$, with $k \in \numco{1}{\omega}$, for which the
		following hold:
		\begin{enumerate}
			\item \label{def:schain(lst)}
				$\lst{\vec{\eElm}} \in \QPAVSet{\lst{\vec{\sigElm}}}$;
			\item \label{def:schain(dep)}
				$((\vec{\eElm})_{i}, (\vec{\eElm})_{i+1}) \in
				\QPDepSet'((\vec{\sigElm})_{i})$, for all $i \in \numco{0}{k - 1}$;
			\item \label{def:schain(unq)}
				$(\vec{\sigElm})_{i} \neq (\vec{\sigElm})_{j}$, for all $i, j \in
				\numco{0}{k}$ with $i < j$.
		\end{enumerate}
	\end{definition}
	\noindent
	It is important to observe that, due to Item~\ref{def:schain(unq)}, each
	$\SSet$-chain cannot have length greater than $\card{\SSet}$.

	Now we can give the definition of \emph{cyclic dependence}.
	\begin{definition}[Cyclic Dependences]
		\label{def:cycdep}
		A \emph{cyclic dependence} for a set of signatures $\SSet \subseteq
		\SigSet(\ESet)$ on $\ESet$ is an \emph{$\SSet$-chain} $(\vec{\eElm},
		\vec{\sigElm})$ such that $(\lst{\vec{\eElm}}, \fst{\vec{\eElm}}) \in
		\QPDepSet'(\lst{\vec{\sigElm}})$.
		Moreover, it is a \emph{cyclic dependence} for a set of labeled signatures
		$\PSet \subseteq \LSigSet(\ESet, \LSet)$ on $\ESet$ w.r.t.\ $\LSet$ if it
		is a cyclic dependence for the set of signatures $\set{ \sigElm \in
		\SigSet(\ESet) }{ \exists \lElm \in \LSet \:.\: (\sigElm, \lElm) \in \PSet
		}$.
		The sets $\CSet(\SSet), \CSet(\PSet) \subseteq \ESet^{+} \times \SSet^{+}$
		contain, respectively, all cyclic dependences for signatures in $\SSet$
		and labeled signatures in $\PSet$.
	\end{definition}
	\noindent
	Observe that $\card{\CSet(\SSet)} \!\!\leq\!\! \card{\ESet}^{\card{\SSet}}
	\!\cdot\! \card{\SSet}!$, so, $\card{\CSet(\PSet)} \!\!\leq\!\!
	\card{\ESet}^{\card{\PSet}} \!\cdot\! \card{\PSet}!$.

	At this point, we can formally define the property of overlapping for
	signatures.
	According to the above description, this implies that dependence maps
	related to prefixes share at least one total variable valuation in their
	codomains.
	Thus, we say that a set of signatures is overlapping if they do not have
	common existential variables and there is no cyclic dependence.
	Observe that, if there are two different instances of the same signature
	having an existential variable, we can still construct a set of
	dependence maps that do not share any valuation, so we have to avoid
	this possibility too.
	\begin{definition}[Overlapping Signatures]
		\label{def:intsig}
		A set $\SSet \subseteq \SigSet(\ESet)$ of signatures on $\ESet$ is
		\emph{overlapping} if $\QPEVSet{\SSet} = \emptyset$ and $\CSet(\SSet) =
		\emptyset$.
		A set $\PSet \!\subseteq\! \LSigSet(\ESet, \LSet)$ of labeled signatures
		on $\ESet$ w.r.t.\ $\LSet$ is \emph{overlapping} if the derived set of
		signatures $\set{ \sigElm \in \SigSet(\ESet) }{ \exists \lElm \in \LSet
		\:.\: (\sigElm, \lElm) \in \PSet }$ is overlapping and, for all
		$(\sigElm, \lElm'), (\sigElm, \lElm'') \in \PSet$, if $\QPEVSet{\sigElm}
		\neq \emptyset$ then $\lElm' = \lElm''$.
	\end{definition}

	Finally, to manage the one-to-one connection between signatures and related
	dependence maps, it is useful to introduce the simple concept of signature
	dependence, which associates to every signature a related dependence map.
	We also define, as expected, the concept of overlapping for these
	functions, which intuitively states that the contained dependence maps have
	identical valuations of variables in their codomains, once they are composed
	with the related functions on the support set.
	\begin{definition}[Signature Dependences]
		\label{def:spcmap}
		A \emph{signature dependence} for a set of signatures $\SSet \subseteq
		\SigSet(\ESet)$ on $\ESet$ over $\DSet$ is a function
		$\spcmapFun : \SSet \to \cup_{(\qpElm, \bFun) \in \SSet}
		\SpcSet[\DSet](\qpElm)$ such that, for all $(\qpElm, \bFun) \in \SSet$, it
		holds that $\spcmapFun((\qpElm, \bFun)) \in \SpcSet[\DSet](\qpElm)$.
		A \emph{signature dependence} for a set of labeled signatures $\PSet
		\subseteq \LSigSet(\ESet, \LSet)$ on $\ESet$ w.r.t.\ $\LSet$ over $\DSet$
		is a function $\spcmapFun : \PSet \to \cup_{((\qpElm, \bFun), \lElm) \in
		\PSet} \SpcSet[\DSet](\qpElm)$ such that, for all $((\qpElm, \bFun),
		\lElm) \in \PSet$, it holds that $\spcmapFun(((\qpElm, \bFun), \lElm)) \in
		\SpcSet[\DSet](\qpElm)$.
		The sets $\SpcMapSet[\DSet](\SSet)$ and $\LSpcMapSet[\DSet](\PSet)$
		contain, respectively, all signature dependences for $\SSet$ and labeled
		signature dependences for $\PSet$ over $\DSet$.
		A signature dependence $\spcmapFun \in \SpcMapSet[\DSet](\SSet)$ is
		\emph{overlapping} if $\cap_{(\qpElm, \bFun) \in \SSet} \set{ \valFun
		\circ \bFun }{ \valFun \in \rng{\spcmapFun(\qpElm, \bFun)}} \neq
		\emptyset$.
		A labeled signature dependence $\spcmapFun \in \LSpcMapSet[\DSet](\PSet)$
		is \emph{overlapping} if $\cap_{((\qpElm, \bFun), \lElm) \in \PSet} \set{
		\valFun \circ \bFun }{ \valFun \in \rng{\spcmapFun((\qpElm, \bFun),
		\lElm)}} \neq \emptyset$.
	\end{definition}

	As explained above, signatures and signature dependences have a strict
	correlation w.r.t.\ the concept of overlapping.
	Indeed, the following result holds.
	The idea here is to find, at each step of the construction of the common
	valuation, a variable, called \emph{pivot}, that does not depend on other
	variables whose value is not already set.
	This is possible if there are no cyclic dependences and each variable is
	existential in at most one signature.
	\begin{theorem}[Overlapping Dependence Maps]
		\label{thm:intspc}
		Let $\SSet \subseteq \SigSet(\ESet)$ be a finite set of overlapping
		signatures on $\ESet$.
		Then, for all signature dependences $\spcmapFun \in
		\SpcMapSet[\DSet](\SSet)$ for $\SSet$ over a set $\DSet$, it holds that
		$\spcmapFun$ is overlapping.
	\end{theorem}

	This theorem can be easily lifted to labeled signatures, as stated in the
	following corollary.
	\begin{corollary}[Overlapping Dependence Maps]
		\label{cor:intspc}
		Let $\PSet \subseteq \LSigSet(\ESet, \allowbreak \LSet)$ be a finite set
		of overlapping labeled signatures on $\ESet$ w.r.t.\ $\LSet$.
		Then, for all labeled signature dependences $\spcmapFun \in
		\LSpcMapSet[\DSet](\PSet)$ for $\PSet$ over a set $\DSet$, it
		holds that $\spcmapFun$ is overlapping.
	\end{corollary}

	Finally, if the set $\DSet$ is sufficiently large, in the case of
	non-overlapping labeled signatures, we can find a signature dependence that
	is non-overlapping too, as reported in following theorem.
	The high-level combinatorial idea behind the proof is to assign to each
	existential variable, related to a given element of the support set in a
	signature, a value containing a univocal flag in $\PSet \times
	\QPVSet(\PSet)$, where $\QPVSet(\PSet) \defeq \bigcup_{((\qpElm, \bFun),
	\lElm) \in \PSet} \QPVSet(\qpElm)$, representing the signature itself.
	Thus, signatures sharing an existential element surely have related
	dependence maps that cannot share a common valuation.
	Moreover, for each cyclic dependence, we choose a particular element whose
	value is the inversion of that assigned to the element from which it
	depends, while all other elements preserve the related values.
	In this way, in a set of signature having cyclic dependences, there is one
	of them whose associated dependence maps have valuations that differ from
	those in the dependence maps of the other signatures, since it is the unique
	that has an inversion of the values.
	\begin{theorem}[Non-Overlapping Dependence Maps]
		\label{thm:nonintspc}
		Let $\PSet \subseteq \LSigSet(\ESet, \LSet)$ be a set of labeled
		signatures on $\ESet$ w.r.t.\ $\LSet$.
		Then, there exists a labeled signature dependence $\spcmapFun \in
		\LSpcMapSet[\DSet](\PSet)$ for $\PSet$ over $\DSet \!\defeq\! \PSet
		\!\times\! \QPVSet(\PSet) \!\times\! \{ 0, 1 \}^{\CSet(\PSet)}$ such
		that, for all $\PSet' \subseteq \PSet$, it holds that $\spcmapFun_{\rst
		\PSet'} \in \LSpcMapSet[\DSet](\PSet')$ is non-overlapping, if $\PSet'$
		is non-overlapping.
	\end{theorem}

\end{section}




\begin{section}{Model Properties}
	\label{sec:modprp}

	We now investigate basic model properties of \OGSL\ that turn out to be
	important on their own and useful to prove the decidability of the
	satisfiability problem.

	First, recall that the satisfiability problem for branching-time logics can be
	solved via tree automata, once a kind of bounded tree-model property holds.
	Indeed, by using it, one can build an automaton accepting all models of
	formulas, or their encoding.
	So, we first introduce the concepts of \emph{concurrent game tree},
	\emph{decision tree}, and \emph{decision-unwinding} and then show that \OGSL\
	is \emph{invariant under decision-unwinding}, which directly implies that it
	satisfies a \emph{unbounded tree-model property}.
	Finally, by using the techniques previously introduced, we further prove that
	the above property is actually a \emph{bounded tree-model property}.

	\begin{paragraph}{Tree-model property}

		We now introduce two particular kinds of \CGS\ whose structure is a directed
		tree.
		As already explained, we do this since the decidability procedure we give in
		the last section of the paper is based on alternating tree automata.
		\begin{definition}[Concurrent Game Trees]
			\label{def:cgt}
			A \emph{concurrent game tree} (\CGT, for short) is a \CGS\ $\TName \defeq
			\CGSStruct[\epsilon]$, where \emph{(i)} $\StSet \subseteq \DirSet^{*}$ is
			a $\DirSet$-tree for a given set $\DirSet$ of directions and \emph{(ii)}
			if $\tElm \cdot \eElm \in \StSet$ then there is a decision $\decFun \in
			\DecSet$ such that $\trnFun(\tElm, \decFun) = \tElm \cdot \eElm$, for all
			$\tElm \in \StSet$ and $\eElm \in \DirSet$.
			Furthermore, $\TName$ is a \emph{decision tree} (\DT, for short) if
			\emph{(i)} $\StSet = \DecSet^{*}$ and \emph{(ii)} if $\tElm \cdot \decFun
			\in \StSet$ then $\trnFun(\tElm, \decFun) = \tElm \cdot \decFun$, for all
			$\tElm \in \StSet$ and $\decFun \in \DecSet$.
		\end{definition}
		\noindent
		Intuitively, \CGT s are \CGS s with a tree-shaped transition relation and
		\DT s have, in addition, states uniquely determining the history of
		computation leading to them.

		At this point, we can define a generalization for \CGS s of the classic
		concept of \emph{unwinding} of labeled transition systems, namely
		decision-unwinding.
		Note that, in general and differently from \ATLS, \SL\ is not invariant
		under decision-unwinding, as we show later.
		On the contrary, \OGSL\ satisfies such an invariance property.
		This fact allows us to show that this logic has the unbounded tree-model
		property.
		\begin{definition}[Decision-Unwinding]
			\label{def:decunw}
			Let $\GName$ be a \CGS.
			Then, the \emph{decision-un\-winding} of $\GName$ is the \DT\ $\GName[DU]
			\defeq \CGSTuple {\APSet} {\AgSet} {\AcSet[\GName]} {\DecSet[\GName]^{*}}
			{\labFun} {\trnFun} {\epsilon}$ for which there is a surjective function
			$\unwFun : \DecSet[\GName]^{*} \to \StSet[\GName]$ such that \emph{(i)}
			$\unwFun(\epsilon) = \sElm[0\GName]$, \emph{(ii)} $\unwFun(\trnFun(\tElm,
			\decFun)) = \trnFun[\GName](\unwFun(\tElm), \decFun)$, and \emph{(iii)}
			$\labFun(\tElm) = \labFun[\GName](\unwFun(\tElm))$, for all $\tElm \in
			\DecSet[\GName]^{*}$ and $\decFun \in \DecSet[\GName]$.
		\end{definition}
		\noindent
		Note that each \CGS\ $\GName$ has a unique associated decision-unwinding
		$\GName[DU]$.
		
		\cbstart
		We say that a sentence $\varphi$ has the \emph{decision-tree
		model property} if, for each \CGS\ $\GName$, it holds that $\GName \models
		\varphi$ iff $\GName[DU] \models \varphi$.
		By using a standard proof by induction on the structure of \OGSL\ formulas,
		we can show that this logic is invariant under decision-unwinding, i.e.,
		each \OGSL\ sentence has decision-tree model property, and, consequently,
		that it satisfies the unbounded tree-model property.
		\cbend
		For the case of the combined quantification and binding
		prefixes $\qpElm \bpElm \psi$, we can use a technique that allows to build,
		given an elementary dependence map $\spcFun$ satisfying the formula
		on a \CGS\ $\GName$, an elementary dependence map $\spcFun'$
		satisfying the same formula over the \DT\ $\GName[DU]$, and vice versa.
		This construction is based on a step-by-step transformation of the adjoint
		of a dependence maps into another, which is done for each track of the
		original model.
		This means that we do not actually transform the strategy quantifications
		but the equivalent infinite set of action quantifications.
		\begin{theorem}[\OGSL\ Positive Model Properties]
			\label{thm:ogsl(posmodprp)}
			\
			\begin{enumerate}
				\item\label{thm:ogsl(posmodprp:decunw)}
					\OGSL\ is invariant under decision-unwinding;
				\item\label{thm:ogsl(posmodprp:dectremodprp)}
					\OGSL\ has the decision-tree model property.
			\end{enumerate}
		\end{theorem}

		Although this result is a generalization of that proved to hold for \ATLS,
		it actually represents an important demarcation line between \OGSL\ and \SL.
		Indeed, as we show in the following theorem, \SL\ does not satisfy neither
		the tree-model property nor, consequently, the invariance under
		decision-unwinding.
		\begin{theorem}[\SL\ Negative Model Properties]
			\label{thm:sl(negmodprp)}
			\
			\begin{enumerate}
				\item\label{thm:sl(negmodprp:dectremodprp)}
					\hspace{-0.5em}
					\SL\ does not have the decision-tree model property;
				\item\label{thm:sl(negmodprp:decunw)}
					\hspace{-0.5em}
					\SL\ is not invariant under decision-unwinding.
			\end{enumerate}
		\end{theorem}

	\end{paragraph}

	\begin{paragraph}{Bounded tree-model property}

		We now have all tools we need to prove the bounded tree-model property for
		\OGSL, which we recall \SL\ does not satisfy~\cite{MMV10b}.
		Actually, we prove here a stronger property, which we name \emph{bounded
		disjoint satisfiability}.

		To this aim, we first introduce the new concept regarding the satisfiability
		of different instances of the same subsentence of the original
		specification, which intuitively states that these instances can be checked
		on disjoint subtrees of the tree model.
		With more detail, this property asserts that, if two instances use part of
		the same subtree, they are forced to use the same dependence map as well.
		This intrinsic characteristic of \OGSL\ is fundamental to build a unique
		automaton that checks the truth of all subsentences, by simply merging their
		respective automata, without using a projection operation that eliminates
		their proper alphabets, which otherwise can be in conflict.
		In this way, we are able to avoid an exponential blow-up.
		A clearer discussion on this point is reported later in the paper.
		\begin{definition}[\OGSL\ Disjoint Satisfiability]
			\label{def:ogsl(dsjsat)}
			Let $\TName$ be a \CGT, $\varphi \defeq \qpElm \bpElm \psi$ an \OGSL\
			principal sentence, and $\SSet \defeq \set{ \sElm \in \StSet }{ \TName,
			\emptyset, \sElm \models \varphi }$.
			Then, $\TName$ satisfies $\varphi$ \emph{disjointly} over $\SSet$ if there
			exist two functions $\headFun : \SSet \to \SpcSet[\AcSet](\qpElm)$ and
			$\bodyFun : \TrkSet(\epsilon) \to \SpcSet[\AcSet](\qpElm)$ such that, for
			all $\sElm \in \SSet$ and $\asgFun \in \AsgSet(\QPAVSet{\qpElm}, \sElm)$,
			it holds that $\TName, \spcFun(\asgFun), \sElm \models \bpElm \psi$, where
			the elementary dependence maps $\spcFun \in
			\ESpcSet[\StrSet(\sElm)](\qpElm)$ is defined as follows: \emph{(i)}
			$\adj{\spcFun}(\sElm) \defeq \headFun(\sElm)$; \emph{(ii)}
			$\adj{\spcFun}(\trkElm) \defeq \bodyFun(\trkElm' \cdot \trkElm)$, for all
			$\trkElm \in \TrkSet(\sElm)$ with $\card{\trkElm} > 1$, where $\trkElm'
			\in \TrkSet(\epsilon)$ is the unique track such that $\trkElm' \cdot
			\trkElm \in \TrkSet(\epsilon)$.
		\end{definition}

		In the following theorem, we finally describe the crucial step behind our
		automata-theoretic decidability procedure for \OGSL.
		At an high-level, the proof proceeds as follows.
		We start from the satisfiability of the specification $\varphi$ over a \DT\
		$\TName$, whose existence is ensured by
		Item~\ref{thm:ogsl(posmodprp:dectremodprp)} of
		Theorem~\ref{thm:ogsl(posmodprp)} of \OGSL\ positive model properties.
		Then, we construct an intermediate \DT\ $\TName[\sharp]$, called
		\emph{flagged model}, which is used to check the satisfiability of all
		subsentences of $\varphi$ in a disjoint way.
		Intuitively, the flagged model adds a controller agent, named \emph{sharp}
		that decides on which subtree a given subsentence has to be verified.
		Now, by means of Theorem~\ref{thm:ogsl(elm)} on the \OGSL\ elementariness,
		we construct the adjoint functions of the dependence maps used to
		verify the satisfiability of the sentences on $\TName[\sharp]$.
		Then, by applying Corollary~\ref{cor:intspc} and Theorem~\ref{thm:nonintspc}
		of overlapping and non-overlapping dependence maps, respectively, we
		transform the dependence maps over actions, contained in the ranges of the
		adjoint functions, in a bounded version, which preserves the satisfiability
		of the sentences on a bounded pruning $\TName[\sharp]'$ of $\TName[\sharp]$.
		Finally, we remove the additional agent $\sharp$ obtaining the required
		bounded \DT\ $\TName'$.
		Observe that, due to the particular construction of the bounded dependence
		maps, the disjoint \mbox{satisfiability is preserved after the elimination
		of $\sharp$.}
		\begin{theorem}[\OGSL\ Bounded Tree-Model Property]
			\label{thm:ogsl(bndtremodprp)}
			Let $\varphi$ be an \OGSL\ satisfiable sentence and $\PSet \defeq \set{
			((\qpElm, \bpElm), (\psi, i)) \in \LSigSet(\AgSet, \SL \times \{ 0, 1 \})
			}{ \qpElm \bpElm \psi \in \psnt{\varphi} \land i \in \{ 0, 1 \} }$ the set
			of all labeled signatures on $\AgSet$ w.r.t.\ $\SL \times \{ 0, 1 \}$ for
			$\varphi$.
			Then, there exists a $b$-bounded \DT\ $\TName$, with $b = \card{\PSet}
			\cdot \card{\QPVSet(\PSet)} \cdot 2^{\card{\CSet(\PSet)}}$, such that
			$\TName \models \varphi$.
			Moreover, for all $\phi \in \psnt{\varphi}$, it holds that $\TName$
			satisfies $\phi$ disjointly over the set $\set{ \sElm \in \StSet }{
			\TName, \emptyset, \sElm \models \phi }$.
		\end{theorem}

	\end{paragraph}

\end{section}




\begin{section}{Satisfiability Procedure}
	\label{sec:satprc}

	We finally solve the satisfiability problem for \OGSL\ and show that it is
	2\ExpTimeC, as for \ATLS.
	The algorithmic procedures is based on an automata-theoretic approach, which
	reduces the decision problem for the logic to the emptiness problem of a
	suitable universal Co-B\"uchi tree automaton (\UCT, for short)~\cite{GTW02}.
	From an high-level point of view, the automaton construction seems similar to
	what was proposed in literature for \CTLS~\cite{KVW00} and \ATLS~\cite{Sch08}.
	However, our technique is completely new, since it is based on the novel
	notions of elementariness and disjoint satisfiability.

	\begin{paragraph}{Principal sentences}

		To proceed with the satisfiability procedure, we have to introduce a concept
		of encoding for an assignment and the labeling of a \DT.
		\begin{definition}[Assignment-Labeling Encoding]
			\label{def:asglabenc}
			Let $\TName$ be a \DT, $\tElm \in \StSet[\TName]$ one of its states, and
			$\asgFun \in \AsgSet[\TName](\VSet, \tElm)$ an assignment defined on the
			set $\VSet \subseteq \VarSet$.
			A $(\ValSet[ {\AcSet[\TName]} ](\VSet) \times \pow{\APSet})$-labeled
			$\DecSet[\TName]$-tree $\TName' \defeq
			\LTTuple{}{}{\StSet[\TName]}{\uFun}$ is an \emph{assignment-labeling
			encoding} for $\asgFun$ on $\TName$ if $\uFun(\lst{(\trkElm)_{\geq 1}})
			\!=\! (\flip{\asgFun}(\trkElm), \labFun[\TName](\lst{\trkElm}))$, for all
			$\trkElm \in \TrkSet[\TName](\tElm)$.
		\end{definition}
		\noindent
		Observe that there is a unique assignment-labeling encoding for each
		assignment over a given \DT.

		Now, we prove the existence of a \UCT\ $\UName[\bpElm \psi | ^{\AcSet}]$ for
		each \OGSL\ goal $\bpElm \psi$ having no principal subsentences.
		$\UName[\bpElm \psi | ^{\AcSet}]$ recognizes all the assignment-labeling
		encodings $\TName'$ of an a priori given assignment $\asgFun$ over a generic
		\DT\ $\TName$, once the goal is satisfied on $\TName$ under $\asgFun$.
		Intuitively, we start with a \UCW, recognizing all infinite words on the
		alphabet $\pow{\APSet}$ that satisfy the \LTL\ formula $\psi$, obtained by a
		simple variation of the Vardi-Wolper construction~\cite{VW86b}.
		Then, we run it on the encoding tree $\TName'$ by following the directions
		imposed by the assignment in its labeling.
		\begin{lemma}[\OGSL\ Goal Automaton]
			\label{lmm:ogsl(golaut)}
			Let $\bpElm \psi$ an \OGSL\ goal without principal subsentences and
			$\AcSet$ a finite set of actions.
			Then, there exists an \UCT\ $\UName[\bpElm \psi | ^{\AcSet}] \defeq
			\TATuple {\ValSet[\AcSet](\free{\bpElm \psi}) \times \pow{\APSet}}
			{\DecSet} {\QSet[\bpElm \psi]} {\atFun[\bpElm \psi]} {\qElm[0\bpElm\psi]}
			{\aleph_{\bpElm \psi}}$ such that, for all \DT s $\TName$ with
			$\AcSet[\TName] = \AcSet$, states $\tElm \in \StSet[\TName]$, and
			assignments $\asgFun \in \AsgSet[\TName](\free{\bpElm \psi}, \tElm)$, it
			holds that $\TName, \asgFun, \tElm \models \bpElm \psi$ iff $\TName' \in
			\LangSet(\UName[\bpElm \psi | ^{\AcSet}])$, where $\TName'$ is the
			assignment-labeling encoding for $\asgFun$ on $\TName$.
		\end{lemma}

		We now introduce a new concept of encoding regarding the elementary
		dependence maps over strategies.
		\begin{definition}[Elementary Dependence-Labeling Encoding]
			\label{def:elmspclabenc}
			Let $\TName$ be a \DT, $\tElm \in \StSet[\TName]$ one of its states, and
			$\spcFun \in \ESpcSet[ {\StrSet[\TName](\tElm)} ](\qpElm)$ an elementary
			dependence map over strategies for a quantification prefix
			$\qpElm \in \QPSet(\VSet)$ over the set $\VSet \subseteq \VarSet$.
			A $(\SpcSet[ {\AcSet[\TName]} ](\qpElm) \times \pow{\APSet})$-labeled
			$\DirSet$-tree $\TName' \defeq \LTTuple{}{}{\StSet[\TName]}{\uFun}$ is an
			\emph{elementary dependence-labeling encoding} for $\spcFun$ on $\TName$
			if $\uFun(\lst{(\trkElm)_{\geq 1}}) \!=\! (\adj{\spcFun}(\trkElm),
			\labFun[\TName](\lst{\trkElm}))$, for all $\trkElm \!\in\!
			\TrkSet[\TName](\tElm)$.
		\end{definition}
		\noindent
		Observe that also in this case there exists a unique elementary
		dependence-model encoding for each elementary dependence map over
		strategies.

		Finally, in the next lemma, we show how to handle locally the strategy
		quantifications on each state of the model, by simply using a quantification
		over actions modeled by the choice of an action dependence map.
		Intuitively, we guess in the labeling what is the right part of the
		dependence map over strategies for each node of the tree and then verify
		that, for all assignments of universal variables, the corresponding complete
		assignment satisfies the inner formula.
		\begin{lemma}[\OGSL\ Sentence Automaton]
			\label{lmm:ogsl(sntaut)}
			Let $\qpElm \bpElm \psi$ be an \OGSL\ principal sentence without principal
			subsentences and $\AcSet$ a finite set of actions.
			Then, there exists an \UCT\ $\UName[\qpElm \bpElm \psi | ^{\AcSet}]
			\defeq \TATuple {\SpcSet[\AcSet](\qpElm) \times \pow{\APSet}} {\DecSet}
			{\QSet[\qpElm \bpElm \psi]} {\atFun[\qpElm \bpElm \psi]}
			{\qElm[0\qpElm\bpElm\psi]} {\aleph_{\qpElm \bpElm \psi}}$ such that, for
			all \DT s $\TName$ with $\AcSet[\TName] = \AcSet$, states $\tElm \in
			\StSet[\TName]$, and elementary dependence maps over strategies
			$\spcFun \in \ESpcSet[ {\StrSet[\TName](\tElm)} ](\qpElm)$, it holds that
			$\TName, \spcFun(\asgFun), \tElm \emodels \bpElm \psi$, for all $\asgFun
			\in \AsgSet[\TName](\QPAVSet{\qpElm}, \tElm)$, iff $\TName' \in
			\LangSet(\UName[\qpElm \bpElm \psi | ^{\AcSet}])$, where $\TName'$ is
			the elementary dependence-labeling encoding for $\spcFun$ on $\TName$.
		\end{lemma}

	\end{paragraph}

	\begin{paragraph}{Full sentences}

		By summing up all previous results, we are now able to solve the
		satisfiability problem for the full \OGSL\ fragment.

		To construct the automaton for a given \OGSL\ sentence $\varphi$, we first
		consider all \UCT\ $\UName[\phi | ^{\AcSet}]$, for an assigned bounded set
		$\AcSet$, previously described for the principal sentences $\phi \in
		\psnt{\varphi}$, in which the inner subsentences are considered as atomic
		propositions.
		Then, thanks to the disjoint satisfiability property of
		Definition~\ref{def:ogsl(dsjsat)}, we can merge them into a unique \UCT\
		$\UName[\varphi]$ that supplies the dependence map labeling of internal
		components $\UName[\phi | ^{\AcSet}]$, by using the two functions $\headFun$
		and $\bodyFun$ contained into its labeling.
		Moreover, observe that the final automaton runs on a $b$-bounded
		decision tree, where $b$ is obtained from
		Theorem~\ref{thm:ogsl(bndtremodprp)} on the bounded-tree model property.
		\begin{theorem}[\OGSL\ Automaton]
			\label{thm:ogsl(aut)}
			Let $\varphi$ be an \OGSL\ sentence.
			Then, there exists an \UCT\ $\UName[\varphi]$ such that $\varphi$ is
			satisfiable iff $\LangSet(\UName[\varphi]) \neq \emptyset$.
		\end{theorem}

		Finally, by a simple calculation of the size of $\UName[\varphi]$ and the
		complexity of the related emptiness problem, we state in the next theorem
		the precise computational complexity of the satisfiability problem for
		\OGSL.
		\begin{theorem}[\OGSL\ Satisfiability]
			\label{thm:ogsl(sat)}
			The satisfiability problem for \OGSL\ is 2\ExpTimeC.
		\end{theorem}

	\end{paragraph}

\end{section}





	\footnotesize
	\bibliography{References.bib}

	\newpage
	\normalsize




\begin{section}{Mathematical Notation}
	\label{app:mthnot}

	In this short reference appendix, we report the classical mathematical
	notation and some common definitions that are used along the whole work.

	\begin{paragraph}*{Classic objects}

		We consider $\SetN$ as the set of \emph{natural numbers} and $\numcc{m}{n}
		\defeq \set{ k \in \SetN }{ m \leq k \leq n }$, $\numco{m}{n} \defeq \set{ k
		\in \SetN }{ m \leq k < n }$, $\numoc{m}{n} \defeq \set{ k \in \SetN }{ m <
		k \leq n }$, and $\numoo{m}{n} \defeq \set{ k \in \SetN }{ m < k < n }$ as
		its \emph{interval} subsets, with $m \in \SetN$ and $n \in \SetNI \defeq
		\SetN \cup \{ \omega \}$, where $\omega$ is the \emph{numerable infinity},
		i.e., the \emph{least infinite ordinal}.
		Given a \emph{set} $\XSet$ of \emph{objects}, we denote by $\card{\XSet} \in
		\SetNI \cup \{ \infty \}$ the \emph{cardinality} of $\XSet$, i.e., the
		number of its elements, where $\infty$ represents a \emph{more than
		countable} cardinality, and by $\pow{\XSet} \defeq \set{ \YSet }{ \YSet
		\subseteq \XSet }$ the \emph{powerset} of $\XSet$, i.e., the set of all its
		subsets.

	\end{paragraph}

	\begin{paragraph}*{Relations}

		By $\RRel \subseteq \XSet \times \YSet$ we denote a \emph{relation} between
		the \emph{domain} $\dom{\RRel} \defeq \XSet$ and \emph{codomain}
		$\cod{\RRel} \defeq \YSet$, whose \emph{range} is indicated by $\rng{\RRel}
		\defeq \set{ \yElm \in \YSet }{ \exists \xElm \in \XSet .\: (\xElm, \yElm)
		\in \RRel }$.
		We use $\RRel^{-1} \defeq \set{ (\yElm, \xElm) \in \YSet \times \XSet }{
		(\xElm, \yElm) \in \RRel }$ to represent the \emph{inverse} of $\RRel$
		itself.
		Moreover, by $\SRel \cmp \RRel$, with $\RRel \subseteq \XSet \times \YSet$
		and $\SRel \subseteq \YSet \times \ZSet$, we denote the \emph{composition}
		of $\RRel$ with $\SRel$, i.e., the relation $\SRel \cmp \RRel \defeq \set{
		(\xElm, \zElm) \in \XSet \times \ZSet }{ \exists \yElm \in \YSet .\: (\xElm,
		\yElm) \in \RRel \land (\yElm, \zElm) \in \SRel }$.
		We also use $\RRel^{n} \defeq \RRel^{n - 1} \cmp \RRel$, with $n \in
		\numco{1}{\omega}\!$, to indicate the \emph{$n$-iteration} of $\RRel
		\subseteq \XSet \times \YSet$, where $\YSet \subseteq \XSet$ and $\RRel^{0}
		\defeq \set{ (\yElm, \yElm) }{ \yElm \in \YSet }$ is the \emph{identity} on
		$\YSet$.
		With $\RRel^{+} \defeq \bigcup_{n = 1}^{< \omega} \RRel^{n}$ and $\RRel^{*}
		\defeq \RRel^{+} \cup \RRel^{0}$ we denote, respectively, the
		\emph{transitive} and \emph{reflexive-transitive closure} of $\RRel$.
		Finally, for an \emph{equivalence} relation $\RRel \subseteq \XSet \times
		\XSet$ on $\XSet$, we represent with $\class{ \XSet }{ \:\RRel } \defeq
		\set{ [\xElm]_{\RRel} }{ \xElm \in \XSet }$, where $[\xElm]_{\RRel} \defeq
		\set{ \xElm' \in \XSet }{ (\xElm, \xElm') \in \RRel }$, the \emph{quotient}
		set of $\XSet$ w.r.t.\ $\RRel$, i.e., the set of all related equivalence
		\emph{classes} $[\cdot]_{\RRel}$.

	\end{paragraph}

	\begin{paragraph}*{Functions}

		We use the symbol $\YSet^{\XSet} \subseteq \pow{\XSet \times \YSet}$ to
		denote the set of \emph{total functions} $\fFun$ from $\XSet$ to $\YSet$,
		i.e., the relations $\fFun \subseteq \XSet \times \YSet$ such that for all
		$\xElm \in \dom{\fFun}$ there is exactly one element $\yElm \in \cod{\fFun}$
		such that $(\xElm, \yElm) \in \fFun$.
		Often, we write $\fFun : \XSet \to \YSet$ and $\fFun : \XSet \pto \YSet$ to
		indicate, respectively, $\fFun \in \YSet^{\XSet}$ and $\fFun \in
		\bigcup_{\XSet' \subseteq \XSet} \YSet^{\XSet'}$.
		Regarding the latter, note that we consider $\fFun$ as a \emph{partial
		function} from $\XSet$ to $\YSet$, where $\dom{\fFun} \subseteq \XSet$
		contains all and only the elements for which $\fFun$ is defined.
		Given a set $\ZSet$, by $\fFun[\rst \ZSet] \defeq \fFun \cap (\ZSet \times
		\YSet)$ we denote the \emph{restriction} of $\fFun$ to the set $\XSet \cap
		\ZSet$, i.e., the function $\fFun[\rst \ZSet] : \XSet \cap \ZSet \pto \YSet$
		such that, for all $\xElm \in \dom{\fFun} \cap \ZSet$, it holds that
		$\fFun[\rst \ZSet](\xElm) = \fFun(\xElm)$.
		Moreover, with $\emptyfun$ we indicate a generic \emph{empty function},
		i.e., a function with empty domain.
		Note that $\XSet \cap \ZSet = \emptyset$ implies $\fFun[\rst \ZSet] =
		\emptyfun$.
		Finally, for two partial functions $\fFun, \gFun : \XSet \pto \YSet$, we use
		$\fFun \Cup \gFun$ and $\fFun \Cap \gFun$ to represent, respectively, the
		\emph{union} and \emph{intersection} of these functions defined as follows:
		$\dom{\fFun \Cup \gFun} \defeq \dom{\fFun} \cup \dom{\gFun} \setminus \set{
		\xElm \in \dom{\fFun} \cap \dom{\gFun} }{ \fFun(\xElm) \neq \gFun(\xElm) }$,
		$\dom{\fFun \Cap \gFun} \defeq \set{ \xElm \in \dom{\fFun} \cap \dom{\gFun}
		}{ \fFun(\xElm) = \gFun(\xElm) }$, $(\fFun \Cup \gFun)(\xElm) =
		\fFun(\xElm)$ for $\xElm \in \dom{\fFun \Cup \gFun} \cap \dom{\fFun}$,
		$(\fFun \Cup \gFun)(\xElm) = \gFun(\xElm)$ for $\xElm \in \dom{\fFun \Cup
		\gFun} \cap \dom{\gFun}$, and $(\fFun \Cap \gFun)(\xElm) = \fFun(\xElm)$ for
		$\xElm \in \dom{\fFun \Cap \gFun}$.

	\end{paragraph}

	\begin{paragraph}*{Words}

		By $\XSet^{n}$, with $n \in \SetN$, we denote the set of all
		\emph{$n$-tuples} of elements from $\XSet$, by $\XSet^{*} \defeq \bigcup_{n
		= 0}^{< \omega} \XSet^{n}$ the set of \emph{finite words} on the
		\emph{alphabet} $\XSet$, by $\XSet^{+} \defeq \XSet^{*} \setminus \{
		\epsilon \}$ the set of \emph{non-empty words}, and by $\XSet^{\omega}$ the
		set of \emph{infinite words}, where, as usual, $\epsilon \in \XSet^{*}$ is
		the \emph{empty word}.
		The \emph{length} of a word $\wElm \in \XSet^{\infty} \defeq \XSet^{*} \cup
		\XSet^{\omega}$ is represented with $\card{\wElm} \in \SetNI$.
		By $(\wElm)_{i}$ we indicate the \emph{$i$-th letter} of the finite word
		$\wElm \in \XSet^{*}$, with $i \in \numco{0}{\card{\wElm}}$.
		Furthermore, by $\fst{\wElm} \defeq (\wElm)_{0}$ (resp., $\lst{\wElm} \defeq
		(\wElm)_{\card{\wElm} - 1}$), we denote the \emph{first} (resp.,
		\emph{last}) letter of $\wElm$.
		In addition, by $(\wElm)_{\leq i}$ (resp., $(\wElm)_{> i}$), we indicate the
		\emph{prefix} up to (resp., \emph{suffix} after) the letter of index $i$ of
		$\wElm$, i.e., the finite word built by the first $i + 1$ (resp., last
		$\card{\wElm} - i - 1$) letters $(\wElm)_{0}, \ldots, (\wElm)_{i}$ (resp.,
		$(\wElm)_{i + 1}, \ldots, (\wElm)_{\card{\wElm} - 1}$).
		We also set, $(\wElm)_{< 0} \defeq \epsilon$, $(\wElm)_{< i} \defeq
		(\wElm)_{\leq i - 1}$, $(\wElm)_{\geq 0} \defeq \wElm$, and $(\wElm)_{\geq
		i} \defeq (\wElm)_{> i - 1}$, for $i \in \numco{1}{\card{\wElm}}$.
		Mutatis mutandis, the notations of $i$-th letter, first, prefix, and suffix
		apply to infinite words too.
		Finally, by $\pfx{\wElm[1], \wElm[2]} \in \XSet^{\infty}$ we denote the
		\emph{maximal common prefix} of two different words $\wElm[1], \wElm[2] \in
		\XSet^{\infty}$, i.e., the finite word $\wElm \in \XSet^{*}$ for which
		there are two words $\wElm[1|'], \wElm[2|'] \in \XSet^{\infty}$ such that
		$\wElm[1] = \wElm \cdot \wElm[1|']$, $\wElm[2] = \wElm \cdot \wElm[2|']$,
		and $\fst{\wElm[1|']} \neq \fst{\wElm[2|']}$.
		By convention, we set $\pfx{\wElm, \wElm} \defeq \wElm$.

	\end{paragraph}

	\begin{paragraph}*{Trees}

		For a set $\DirSet$ of objects, called \emph{directions}, a
		\emph{$\DirSet$-tree} is a set $\TSet \subseteq \DirSet^{*}$ closed under
		prefix, i.e., if $\tElm \cdot \dElm \in \TSet$, with $\dElm \in \DirSet$,
		then also $\tElm \in \TSet$.
		We say that it is \emph{complete} if it holds that $\tElm \cdot \dElm' \in
		\TSet$ whenever $\tElm \cdot \dElm \in \TSet$, for all $\dElm' < \dElm$,
		where $< \: \subseteq \DirSet \times \DirSet$ is an a priori fixed strict
		total order on the set of directions that is clear from the context.
		Moreover, it is \emph{full} if $\TSet = \DirSet^{*}$.
		The elements of $\TSet$ are called \emph{nodes} and the empty word
		$\epsilon$ is the \emph{root} of $\TSet$.
		For every $\tElm \in \TSet$ and $\dElm \in \DirSet$, the node $\tElm \cdot
		\dElm \in \TSet$ is a \emph{successor} of $\tElm$ in $\TSet$.
		The tree is $b$-\emph{bounded} if the maximal number $b$ of its successor
		nodes is finite, i.e., $b = \max_{\tElm \in \TSet} \card{\set{ \tElm \cdot
		\dElm \in \TSet }{ \dElm \in \DirSet }} < \omega$.
		A \emph{branch} of the tree is an infinite word $\wElm \in \DirSet^{\omega}$
		such that $(\wElm)_{\leq i} \in \TSet$, for all $i \in \SetN$.
		For a finite set $\LabSet$ of objects, called \emph{symbols}, a
		\emph{$\LabSet$-labeled $\DirSet$-tree} is a quadruple
		$\LTDef{\LabSet}{\DirSet}$, where $\TSet$ is a $\DirSet$-tree and $\vFun :
		\TSet \to \LabSet$ is a \emph{labeling function}.
		When $\DirSet$ and $\LabSet$ are clear from the context, we call $\LTStruct$
		simply a (labeled) tree.

	\end{paragraph}

\end{section}




\begin{section}{Proofs of Section~\ref{sec:bndspc}}
	\label{app:bndspc}

	In this appendix, we give the proofs of Theorem~\ref{thm:intspc} and
	Corollary~\ref{cor:intspc} of overlapping dependence maps and
	Theorem~\ref{thm:nonintspc} of non-overlapping dependence maps.
	In particular, to prove the first two results, we need to introduce the
	concept of pivot for a given set of signatures and then show some useful
	related properties.
	Moreover, for the latter result, we define an apposite ad-hoc signature
	dependence, based on a sharp combinatorial construction, in order to maintain
	separated the dependence maps associated to the components of a
	non-overlapping set of signatures.

	\begin{paragraph}{Pivot}

		To proceed with the definitions, we have first to introduce some additional
		notation.
		Let $\ESet$ be a set and $\sigElm \in \SigSet(\ESet)$ a signature.
		Then, $\QPAVSet{\sigElm} \defeq \ESet \setminus \QPEVSet{\sigElm}$
		indicates the set of elements in $\ESet$ associated to universal quantified
		variables.
		Moreover, for an element $\eElm \in \ESet$, we denote by $\QPDepSet(\sigElm,
		\eElm) \defeq \set{\eElm' \in \ESet}{(\eElm', \eElm) \in
		\QPDepSet(\sigElm)}$ the set of elements from which $\eElm$ is functional
		dependent.
		Given another element $\eElm' \in \ESet$, we say that $\eElm$
		\emph{precedes} $\eElm'$ in $\sigElm$, in symbols $\eElm \qpordRel[\sigElm]
		\eElm'$, if $\bFun(\eElm) \qpordRel[\qpElm] \bFun(\eElm')$, where $\sigElm =
		(\qpElm, \bFun)$.
		Observe that this kind of order is, in general, not total, due to the fact
		that $\bFun$ is not necessarily injective.
		Consequently, by $\min_{\qpordRel[\sigElm]} \FSet$, with $\FSet \subseteq
		\ESet$, we denote the set of minimal elements of $\FSet$ w.r.t.\
		$\qpordRel[\sigElm]$.
		Finally, for a given set of signatures $\SSet \subseteq \SigSet(\ESet)$, we
		indicate by $\QPAVSet{\SSet} \defeq \bigcap_{\sigElm \in \SSet}
		\QPAVSet{\sigElm}$ the set of elements that are universal in all signatures
		of $\SSet$, by $\BPColSet(\SSet, \eElm) \defeq \set {\eElm' \in \ESet
		\setminus \QPAVSet{\SSet} }{ (\eElm', \eElm) \in \BPColSet(\SSet)}$ the set
		of existential elements that form a collapsing chain with $\eElm$, and by
		$\BPColSet(\SSet, \sigElm) \defeq \set{ \eElm \in \ESet }{ \exists \eElm'
		\in \QPEVSet{\sigElm} \:.\: (\eElm', \eElm) \in \BPColSet(\SSet) }$ the set
		of elements that form a collapsing chain with at least one existential
		element in $\sigElm$.

		Intuitively, a pivot is an element on which we can extend a partial
		assignment that is shared by a set of dependence maps related to signatures
		via a signature dependence, in order to find a total assignment by an
		iterative procedure.
		Let $\FSet$ the domain of a partial function $\dFun : \ESet \to \DSet$ and
		$\eElm$ an element not yet defined, i.e., $\eElm \in \ESet \setminus \FSet$.
		If, on one hand, $\eElm$ is existential quantified over a signature $\sigElm
		= (\qpElm, \bFun)$ and all the elements from which $\eElm$ depends on that
		signature are in the domain $\FSet$, then the value of $\eElm$ is uniquely
		determined by the related dependence map.
		So, $\eElm$ is a pivot.
		If, on the other hand, $\eElm$ is universal quantified over all signatures
		$\sigElm \in \SSet$ and all elements that form a collapsing chain with
		$\eElm$ are in the domain $\FSet$, then, also in this case we can define
		the value of $\eElm$ being sure to leave the possibility to build a total
		assignment.
		So, also in this case $\eElm$ is a pivot.
		For this reason, pivot plays a fundamental role in the construction of such
		shared assignments.
		The existence of a pivot for a given finite set of signatures $\SSet
		\subseteq \SigSet(\ESet)$ w.r.t.\ a fixed domain $\FSet$ of a partial
		assignment is ensured under the hypothesis that there are no cyclic
		dependences in $\SSet$.
		The existence proof passes through the development of three lemmas
		describing a simple seeking procedure.

		With the previous description and the examples of Section~\ref{sec:bndspc}
		in mind, we now formally describe the properties that an element of the
		support set has to satisfy in order to be a \emph{pivot} for a set of
		signatures w.r.t.\ an a priori given subset of elements.
		\begin{definition}[Pivots]
			\label{def:piv}
			Let $\SSet \subseteq \SigSet(\ESet)$ be a set of signatures on $\ESet$ and
			$\FSet \subset \ESet$ a subset of elements.
			Then, an element $\eElm \in \ESet$ is a \emph{pivot} for $\SSet$ w.r.t.\
			$\FSet$ if $\eElm \not\in \FSet$ and either one of the following items
			holds:
			\begin{enumerate}
				\item\label{def:piv(all)}
					$\eElm \in \QPAVSet{\SSet}$ and $\BPColSet(\SSet, \eElm) \subseteq
					\FSet$;
				\item\label{def:piv(exs)}
					there is a signature $\sigElm \in \SSet$ such that $\eElm \in
					\QPEVSet{\sigElm}$ and $\QPDepSet(\sigElm, \eElm) \subseteq \FSet$.
			\end{enumerate}
		\end{definition}
		\noindent
		Intuitively, Item~\ref{def:piv(all)} asserts that the pivot is universal
		quantified over all signatures and all existential elements that form a
		collapsing chain starting in the pivot itself are already defined.
		On the contrary, Item~\ref{def:piv(exs)} asserts that the pivot is
		existential quantified and, on the relative signature, it depends only on
		already defined elements.

		Before continuing, we provide the auxiliary definition of \emph{minimal
		$\SSet$-chain}.
		\begin{definition}[Minimal $\SSet$-Chain]
			\label{def:minschain}
			Let $\SSet \subseteq \SigSet(\ESet)$ be a set of signatures on $\ESet$ and
			$\FSet \subset \ESet$ a subset of elements.
			A pair $(\vec{\eElm}, \vec{\sigElm}) \in \ESet^{k} \times \SSet^{k}$,
			with $k \in \numco{1}{\omega}$, is a \emph{minimal $\SSet$-chain} w.r.t.\
			$\FSet$ if it is an $\SSet$-chain such that $(\vec{\eElm})_{i} \in
			\min_{(\vec{\sigElm})_{i}} (\ESet \setminus \FSet)$, for all $i \in
			\numco{0}{k}$.
		\end{definition}

		In addition to the definition of pivot, we also give the formal concept of
		\emph{pivot seeker} that is used, in an iterative procedure, to find a pivot
		if this exists.
		\begin{definition}[Pivot Seekers]
			\label{def:pivskr}
			Let $\SSet \subseteq \SigSet(\ESet)$ be a set of signatures on $\ESet$ and
			$\FSet \subset \ESet$ a subset of elements.
			Then, a pair $(\eElm \cdot \vec{\eElm}, \sigElm \cdot \vec{\sigElm}) \in
			\EElm^{k} \times \SSet^{k}$ of sequences of elements and signatures of
			length $k \in \numco{1}{\omega}$ is a \emph{pivot seeker} for $\SSet$
			w.r.t.\ $\FSet$ if the following hold:
			\begin{enumerate}
				\item\label{def:pivskr(min)}
					$\eElm \in \min_{\sigElm} (\ESet \setminus \FSet)$;
				\item\label{def:pivskr(link)}
					$\fst{\vec{\eElm}} \in (\QPEVSet{\sigElm} \cup \BPColSet(\SSet,
					\sigElm)) \setminus \FSet$, if $k > 1$;
				\item\label{def:pivskr(schain)}
					$(\vec{\eElm}, \vec{\sigElm})$ is a minimal $\SSet$-chain, if $k > 1$.
			\end{enumerate}
		\end{definition}
		\noindent
		Intuitively, a pivot seeker is a snapshot of the seeking procedure at a
		certain step.
		Item~\ref{def:pivskr(min)} ensures that the element $\eElm$ we are going to
		consider as a candidate for pivot depends only on the elements defined
		in $\FSet$.
		Item~\ref{def:pivskr(link)} builds a link between the signature $\sigElm$ of
		the present candidate and the head element $\fst{\vec{\eElm}}$ of the
		previous step, in order to maintain information about the dependences that
		are not yet satisfied.
		Finally, Item~\ref{def:pivskr(schain)} is used to ensure that the procedure
		avoids loops by checking pivots on signature already considered.

		As shown through the above mentioned examples, in the case of overlapping
		signatures, we can always find a pivot w.r.t.\ a given set of elements
		already defined, by means of a pivot seeker.

		The following lemma ensures that we can always start the iterative procedure
		over pivot seekers to find a pivot.
		\begin{lemma}[Pivot Seeker Existence]
			\label{lmm:pivskrexs}
			Let $\SSet \subseteq \SigSet(\ESet)$ be a set of signatures on $\ESet$ and
			$\FSet \subset \ESet$ a subset of elements.
			Then, there exists a pivot seeker for $\SSet$ w.r.t.\ $\FSet$ of length
			$1$.
		\end{lemma}
		\begin{proof}
			Let $\sigElm \in \SSet$ be a generic signature and $\eElm \in \ESet$ an
			element such that $\eElm \in \min_{\sigElm} (\ESet \setminus \FSet)$.
			Then, it is immediate to see that the pair $(\eElm, \sigElm) \in \ESet^{1}
			\times \SSet^{1}$ is a pivot seeker for $\SSet$ w.r.t.\ $\FSet$ of length
			$1$, since Item~\ref{def:pivskr(min)} of Definition~\ref{def:pivskr} of
			pivot seekers is verified by construction and
			Items~\ref{def:pivskr(link)} and~\ref{def:pivskr(schain)} are vacuously
			satisfied.
		\end{proof}

		Now, suppose to have a pivot seeker of length not greater than the size of
		the support set $\ESet$ and that no pivot is already found.
		Then, in the case of signatures without cyclic dependences, we can always
		continue the iterative procedure, by extending the previous pivot seeker of
		just one further element.
		\begin{lemma}[Pivot Seeker Extension]
			\label{lmm:pivskrext}
			Let $\SSet \subseteq \SigSet(\ESet)$ be a set of signatures on $\ESet$
			with $\CSet(\SSet) = \emptyset$ and $\FSet \subset \ESet$ a subset of
			elements.
			Moreover, let $(\eElm \cdot \vec{\eElm}, \sigElm \cdot \vec{\sigElm}) \in
			\EElm^{k} \times \SSet^{k}$ be a pivot seeker for $\SSet$ w.r.t.\ $\FSet$
			of length $k \in \numco{1}{\omega}$.
			Then, if $\eElm$ is not a pivot for $\SSet$ w.r.t.\ $\FSet$, there exists
			a pivot seeker for $\SSet$ w.r.t.\ $\FSet$ of length $k + 1$.
		\end{lemma}
		\begin{proof}
			By Item~\ref{def:pivskr(min)} of Definition~\ref{def:pivskr} of pivot
			seekers, we deduce that $\eElm \notin \FSet$ and $\QPDepSet(\sigElm,
			\eElm) \subseteq \FSet$.
			Thus, if $\eElm$ is not a pivot for $\SSet$ w.r.t.\ $\FSet$, by
			Definition~\ref{def:piv} of pivot, we have that $\eElm \not\in
			\QPAVSet{\SSet}$ or $\BPColSet(\SSet, \eElm) \not\subseteq \FSet$ and, in
			both cases, $\eElm \in \QPAVSet{\sigElm}$.
			We now distinguish the two cases.
			\begin{itemize}
				\item
					$\eElm \notin \QPAVSet{\SSet}$.

					There exists a signature $\sigElm' \in \SSet$ such that $\eElm \in
					\QPEVSet{\sigElm'}$.
					So, consider an element $\eElm' \in \min_{\sigElm'} (\ESet \setminus
					\FSet)$.
					We now show that the pair of sequences $(\eElm' \cdot \eElm \cdot
					\vec{\eElm}, \sigElm' \cdot \sigElm \cdot \vec{\sigElm}) \in \ESet^{k
					+ 1} \times \SSet^{k + 1}$ of length $k + 1$ satisfies
					Items~\ref{def:pivskr(min)} and~\ref{def:pivskr(link)} of
					Definition~\ref{def:pivskr}.
					The first item is trivially verified by construction.
					Moreover, $\fst{\eElm \cdot \vec{\eElm}} = \eElm \in
					\QPEVSet{\sigElm'} \setminus \FSet$.
					Hence, the second item holds as well.
				\item
					$\eElm \in \QPAVSet{\SSet}$.

					We necessarily have that $\BPColSet(\SSet, \eElm) \not\subseteq
					\FSet$.
					Thus, there is an element $\eElm' \in \ESet \setminus (\QPAVSet{\SSet}
					\cup \FSet)$ such that $(\eElm', \eElm) \in \BPColSet(\SSet)$.
					Consequently, there exists also a signature $\sigElm' \in \SSet$ such
					that $\eElm' \in \QPEVSet{\sigElm'} \setminus \FSet$.
					So, consider an element $\eElm'' \in \min_{\sigElm'} (\ESet \setminus
					\FSet)$.
					We now show that the pair of sequences $(\eElm'' \cdot \eElm \cdot
					\vec{\eElm}, \sigElm' \cdot \sigElm \cdot \vec{\sigElm}) \in \ESet^{k
					+ 1} \times \SSet^{k + 1}$ of length $k + 1$ satisfies
					Items~\ref{def:pivskr(min)} and~\ref{def:pivskr(link)} of
					Definition~\ref{def:pivskr}.
					The first item is trivially verified by construction.
					Moreover, since $(\eElm', \eElm) \in \BPColSet(\SSet)$, by the
					definition of $\BPColSet(\SSet, \sigElm')$, we have that $\fst{\eElm
					\cdot \vec{\eElm}} = \eElm \in \BPColSet(\SSet, \sigElm') \setminus
					\FSet$.
					Hence, the second item holds as well.
			\end{itemize}

			At this point, it only remains to show that Item~\ref{def:pivskr(schain)}
			of Definition~\ref{def:pivskr} holds, i.e., that $(\eElm \cdot
			\vec{\eElm}, \sigElm \cdot \vec{\sigElm})$ is a minimal $\SSet$-chain
			w.r.t.\ $\FSet$.
			For $k = 1$, we have that Items~\ref{def:schain(dep)}
			and~\ref{def:schain(unq)} of Definition~\ref{def:schain} of $\SSet$-chain
			are vacuously verified.
			Moreover, since $\eElm \in \QPAVSet{\sigElm}$, also
			Item~\ref{def:schain(lst)} of the previous definition holds.
			Finally, the $\SSet$-chain is minimal w.r.t.\ $\FSet$, due to the fact
			that $\eElm \in \min_{\sigElm} (\ESet \setminus \FSet)$.
			Now, suppose that $k > 1$.
			Since $(\vec{\eElm}, \vec{\sigElm})$ is already an $\SSet$-chain, to prove
			Items~\ref{def:schain(dep)} and~\ref{def:schain(unq)} of
			Definition~\ref{def:schain} of $\SSet$-chain, we have only to show that
			$(\eElm, \fst{\vec{\eElm}}) \in \QPDepSet'(\sigElm)$ and $\sigElm \neq
			(\vec{\sigElm})_{i}$, for all $i \in \numco{0}{k - 1}$, respectively.

			By Items~\ref{def:pivskr(min)} and~\ref{def:pivskr(link)} of
			Definition~\ref{def:pivskr}, we have that $\eElm \in \min_{\sigElm}(\ESet
			\setminus \FSet)$ and $\fst{\vec{\eElm}} \in (\QPEVSet{\sigElm} \cup
			\BPColSet(\SSet, \sigElm)) \setminus \FSet$.
			So, two cases arise.
			\begin{itemize}
				\item
					$\fst{\vec{\eElm}} \in \QPEVSet{\sigElm} \setminus \FSet$.

					Since $\eElm \in \QPAVSet{\sigElm} \cap \min_{\sigElm}(\ESet \setminus
					\FSet)$, we can deduce that $(\eElm, \fst{\vec{\eElm}}) \in
					\QPDepSet(\sigElm) \subseteq \QPDepSet'(\sigElm)$.
				\item
					$\fst{\vec{\eElm}} \in \BPColSet(\SSet, \sigElm) \setminus \FSet$.

					By the definition of $\BPColSet(\SSet, \sigElm)$, there exists $\eElm'
					\in \QPEVSet{\sigElm} \setminus \FSet$ such that $(\eElm',
					\fst{\vec{\eElm}}) \in \BPColSet(\SSet)$.
					Now, since $\eElm \in \QPAVSet{\sigElm} \cap \min_{\sigElm}(\ESet
					\setminus \FSet)$, we can deduce that $(\eElm, \eElm') \in
					\QPDepSet(\sigElm)$.
					Thus, by definition of $\QPDepSet'(\sigElm)$, it holds that $(\eElm,
					\fst{\vec{\eElm}}) \in \QPDepSet'(\sigElm)$.
			\end{itemize}

			Finally, suppose by contradiction that there exists $i \in \numco{0}{k -
			1}$ such that $\sigElm = (\vec{\sigElm})_{i}$.
			Two cases can arise.
			\begin{itemize}
				\item
					$i = k - 2$.

					Then, by Item~\ref{def:schain(lst)} of Definition~\ref{def:schain}, we
					have that $(\vec{\eElm})_{i} = \lst{\vec{\eElm}} \in
					\QPAVSet{\lst{\vec{\sigElm}}} = \QPAVSet{(\vec{\sigElm})_{i}}$;
				\item
					$i < k - 2$.

					Then, by Item~\ref{def:schain(dep)} of Definition~\ref{def:schain}, we
					have that $((\vec{\eElm})_{i}, (\vec{\eElm})_{i + 1}) \in
					\QPDepSet'((\vec{\sigElm})_{i})$.
					Consequently, $(\vec{\eElm})_{i} \in \QPAVSet{(\vec{\sigElm})_{i}}$.
			\end{itemize}
			By Definition~\ref{def:minschain} of minimal $\SSet$-chain, since
			$(\vec{\eElm}, \vec{\sigElm})$ is minimal w.r.t.\ $\FSet$, it holds that
			$(\vec{\eElm})_{i} \in \min_{(\vec{\sigElm})_{i}} (\ESet \setminus
			\FSet)$.
			So, $(\vec{\eElm})_{i} \in \QPAVSet{(\vec{\sigElm})_{i}} \cap
			\min_{(\vec{\sigElm})_{i}} (\ESet \setminus \FSet)$.
			Moreover, by Item~\ref{def:pivskr(link)} of Definition~\ref{def:pivskr},
			we have that $(\vec{\eElm})_{0} \in (\QPEVSet{\sigElm} \cup
			\BPColSet(\SSet, \sigElm)) \setminus \FSet =
			(\QPEVSet{(\vec{\sigElm})_{i}} \cup \BPColSet(\SSet, (\vec{\sigElm})_{i}))
			\setminus \FSet$.
			Thus, by applying a reasoning similar to the one used above to prove that
			$(\eElm, \fst{\vec{\eElm}}) \in \QPDepSet'(\sigElm)$, we obtain that
			$((\vec{\eElm})_{i}, (\vec{\eElm})_{0}) \in
			\QPDepSet'((\vec{\sigElm})_{i})$
			Hence, $((\vec{\eElm})_{\leq i}, (\vec{\sigElm})_{\leq i})$ satisfies
			Definition~\ref{def:cycdep} of cyclic dependences.
			So, $((\vec{\eElm})_{\leq i}, (\vec{\sigElm})_{\leq i}) \in \CSet(\SSet)
			\neq \emptyset$, which is a contradiction.
		\end{proof}

		Finally, if we have run the procedure until all elements in $\ESet$ are
		visited, the first one of the last pivot seeker is necessarily a pivot.
		\begin{lemma}[Seeking Termination]
			\label{lmm:skntrm}
			Let $\SSet \subseteq \SigSet(\ESet)$ be a finite set of signatures on
			$\ESet$ with $\CSet(\SSet) = \emptyset$ and $\FSet \subset \ESet$ a subset
			of elements.
			Moreover, let $(\eElm \cdot \vec{\eElm}, \sigElm \cdot \vec{\sigElm}) \in
			\EElm^{k} \times \SSet^{k}$ be a pivot seeker for $\SSet$ w.r.t.\ $\FSet$
			of length $k \defeq \card{\SSet} + 1$.
			Then, $\eElm$ is a pivot for $\SSet$ w.r.t.\ $\FSet$.
		\end{lemma}
		\begin{proof}
			Suppose by contradiction that $\eElm$ is not a pivot for $\SSet$ w.r.t.\
			$\FSet$.
			Then, by Lemma~\ref{lmm:pivskrext} of pivot seeker extension, there exists
			a pivot seeker for $\SSet$ w.r.t.\ $\FSet$ of length $k + 1$, which is
			impossible due to Item~\ref{def:pivskr(schain)} of
			Definition~\ref{def:pivskr} of pivot seekers, since an $\SSet$-chain of
			length $k$ does not exist.
		\end{proof}

		By appropriately combining the above lemmas, we can prove the existence of a
		pivot for a given set of signatures having no cyclic dependences.
		\begin{lemma}[Pivot Existence]
			\label{lmm:pivexs}
			Let $\SSet \subseteq \SigSet(\ESet)$ be a finite set of signatures on
			$\ESet$ with $\CSet(\SSet) = \emptyset$ and $\FSet \subset \ESet$ a subset
			of elements.
			Then, there exists a pivot for $\SSet$ w.r.t.\ $\FSet$.
		\end{lemma}
		\begin{proof}
			By Lemma~\ref{lmm:pivskrexs} of pivot seeker existence, there is a pivot
			seeker of length $1$ for $\SSet$ w.r.t.\ $\FSet$, which can be extended,
			by using Lemma~\ref{lmm:pivskrext} of pivot seeker extension, at most
			$\card{\SSet} < \omega$ times, due to Lemma~\ref{lmm:skntrm} of seeking
			termination, before the reach of a pivot $\eElm$ for $\SSet$ w.r.t.\
			$\FSet$.
		\end{proof}

	\end{paragraph}

	\begin{paragraph}{Big signature dependences}
		\cbstart

		In order to prove Theorem~\ref{thm:nonintspc}, we first introduce big
		signature map $\spcmapFun$.

		\begin{definition}[Big Signature Dependences]
			\label{def:bigsigdep}
			Let $\PSet \subseteq \LSigSet(\ESet)$ be a set of labeled signatures over
			a set $\ESet$, and $\DSet = \PSet \times \QPVSet(\PSet) \times \{0, 1
			\}^{C(\PSet)}$.
			Then, the \emph{big signature dependence} $\spcmapFun \in
			\SpcMapSet[\DSet](\PSet)$ for $\PSet$ over $\DSet$ is defined as follow.
			For all $(\sigElm, \lElm) = ((\qpElm, \bFun), \lElm) \in \PSet$, and
			$\valFun \in \ValSet[\DSet](\QPAVSet{\qpElm})$, we have that:
			\begin{enumerate}
				\item \label{def:bigsigdep(all)}
					$\spcmapFun((\sigElm, \lElm))(\valFun)(\xElm) \defeq \valFun(\xElm)$,
					for all $\xElm \in \QPAVSet{\qpElm}$;
				\item \label{def:bigsigdep(exs)}
					$\spcmapFun((\sigElm, \lElm))(\valFun)(\xElm) \defeq ((\sigElm,
					\lElm), \xElm, \hFun)$, for all $\xElm \in \QPEVSet{\qpElm}$,
					where $\hFun \in \{0, 1 \}^{C(\PSet)}$ is such that, for all
					$(\vec{\eElm}, \vec{\sigElm})$, the following hold:
					\begin{enumerate}
						\item \label{def:bigsigdep(exs:fst)}
							if $\sigElm = \fst{\vec{\sigElm}}$ and $\xElm =
							\bFun(\fst{\vec{\eElm}})$ then $\hFun((\vec{\eElm},
							\vec{\sigElm})) \defeq 1 - \hFun'((\vec{\eElm}, \vec{\sigElm}))$,
							where $\hFun' \in \{0, 1 \}^{C(\PSet)}$ is such that
							$\valFun(\bFun(\lst{\vec{\eElm}})) = ((\sigElm', \lElm'),
							\xElm', \hFun')$, for some $(\sigElm', \lElm') \in \PSet$ and
							$\xElm' \in \QPVSet(\PSet)$;
						\item \label{def:bigsigdep(exs:tail)}
							if there exists $i \in \numco{1}{\card{\vec{\sigElm}}}$ such
							that $\sigElm = (\vec{\sigElm})_{i}$ and $\xElm =
							\bFun((\vec{\eElm})_{i})$, then $\hFun((\vec{\eElm},
							\vec{\sigElm})) \defeq \hFun'((\vec{\eElm}, \vec{\sigElm}))$,
							where $\hFun' \in \{0, 1 \}^{C(\PSet)}$ is such that
							$\valFun(\bFun((\vec{\eElm})_{i})) = ((\sigElm', \lElm'),
							\xElm', \hFun')$, for some $(\sigElm', \lElm') \in \PSet$ and
							$\xElm' \in \QPVSet(\PSet)$;
						\item \label{def:bigsigdep(exs:none)}
							if none of the above cases apply, set $\hFun((\vec{\eElm},
							\vec{\sigElm})) \defeq 0$.
					\end{enumerate}
			\end{enumerate}

		\end{definition}
		Note that Items~\ref{def:bigsigdep(exs:fst)}
		and~\ref{def:bigsigdep(exs:tail)} are mutually exclusive since, by
		definition of cyclic dependence, each signature $(\vec{\sigElm})_{i}$ occurs
		only once in $\vec{\sigElm}$.

		It is easy to see that the previous definition is well formed, i.e., that
		$\spcmapFun$ is actually a labeled signature dependence.
		Indeed the following lemma holds.
		\begin{lemma}
			\label{lmm:bigsigdep}
			Let $\PSet \subseteq \LSigSet(\ESet)$ be a set of labeled signatures over
			a set $\ESet$ and $\DSet = \PSet \times \QPVSet(\PSet) \times \{0, 1
			\}^{C(\PSet)}$.
			Then the big signature dependence $\spcmapFun$ for $\PSet$ over $\DSet$ is
			a labeled signature dependence for $\PSet$ over $\DSet$.
		\end{lemma}
		\begin{proof}
			We have to show that $\wFun(((\qpElm, \bFun), \lElm))$ is a dependence map
			for $\qpElm$ over $\DSet$, for all $(\sigElm, \lElm) \in \PSet$.
			\begin{enumerate}
				\item By Item~\ref{def:bigsigdep(all)} of
					Definition~\ref{def:bigsigdep} it holds that $\wFun((\sigElm,
					\lElm))(\valFun)(\xElm) = \valFun(\xElm)$, for all $\xElm \in
					\QPAVSet{\qpElm}$ and $\valFun \in \ValSet[\DSet](\QPAVSet{\qpElm})$,
					which means $\wFun((\sigElm, \lElm))(\valFun)_{\rst
					\QPAVSet{\qpElm}} = \valFun$, that means that
					Item~\ref{def:qntspc(aqnt)} of Definition~\ref{def:qntspc} holds.
				\item For the Item~\ref{def:qntspc(eqnt)} of
						Definition~\ref{def:qntspc}, let $\valFun[1], \valFun[2] \in
						\ValSet[\DSet](\QPAVSet{\qpElm})$ and $\xElm \in \QPEVSet{\qpElm}$
						such that $(\valFun_{1})_{\rst \QPDepSet(\qpElm,\xElm)} =
						(\valFun_{2})_{\rst \QPDepSet(\qpElm, \xElm)}$.
						We have to prove that $\spcmapFun((\sigElm,
						\lElm))(\valFun[1])(\xElm) = \spcmapFun((\sigElm,
						\lElm))(\valFun[2])(\xElm)$.
						By definition, we have that $\spcmapFun((\sigElm,
						\lElm))(\valFun[1])(\xElm) = ((\sigElm, \lElm), \xElm, \hFun[1])$
						and $\spcmapFun((\sigElm, \lElm))(\valFun[2])(\xElm) = ((\sigElm,
						\lElm), \xElm, \hFun[2])$.
						So, we have only to show that $\hFun[1] = \hFun[2]$.
						To do this, consider a cyclic dependence $(\vec{\eElm},
						\vec{\sigElm}) \in C(\PSet)$ for which there exists $i \in
						\numco{0}{\card{\vec{\sigElm}}}$ such that $\sigElm =
						(\vec{\sigElm})_{i}$ and $\xElm = \bFun((\vec{\eElm})_{i})$.
						Then, we have that $\valFun[1](\ySym) = \valFun[2](\ySym) =
						((\sigElm', \lElm'), \ySym', \hFun')$ for $\ySym =
						\bFun((\vec{\eElm})_{(i-1) \mod \card{\vec{\sigElm}}})$.
						Then, we have the following:
						\begin{itemize}
							\item by Item~\ref{def:bigsigdep(exs:fst)} of
								Definition~\ref{def:bigsigdep}, if $i = 1$ then
								$\hFun[1]((\vec{\eElm}, \vec{\sigElm})) = 1 -
								\hFun[1]'((\vec{\eElm}, \vec{\sigElm})) = \hFun[2]((\vec{\eElm},
								\vec{\sigElm}))$;
							\item by Item~\ref{def:bigsigdep(exs:tail)} of
								Definition~\ref{def:bigsigdep}, if $i \in
								\numoo{1}{\card{\vec{\sigElm}}}$ then $\hFun[1]((\vec{\eElm},
								\vec{\sigElm})) =  \hFun[1]'((\vec{\eElm}, \vec{\sigElm})) =
								\hFun[2]((\vec{\eElm}, \vec{\sigElm}))$.
						\end{itemize}
						On the other side, consider a cyclic dependence $(\vec{\eElm},
						\vec{\sigElm}) \in C(\PSet)$ such that $\sigElm \neq
						(\vec{\sigElm})_{i}$ or $\xElm \neq \bFun((\vec{\eElm})_{i})$, for
						all $i \in \numco{0}{\card{\vec{\sigElm}}}$.
						In this case, by Item~\ref{def:bigsigdep(exs:none)} of
						Definition~\ref{def:bigsigdep}, we have that $\hFun[1]((\vec{\eElm},
						\vec{\sigElm})) = 0 = \hFun[2]((\vec{\eElm}, \vec{\sigElm}))$.
			\end{enumerate}
		\end{proof}

		\cbend
	\end{paragraph}

	\begin{paragraph}{Proofs of theorems}

		We are finally able to show the proofs of the above mentioned results.

		\def\thetheorem{\ref{thm:intspc}}
		\begin{theorem}[Overlapping Dependence Maps]
			Let $\SSet \subseteq \SigSet(\ESet)$ be a finite set of overlapping
			signatures on $\ESet$.
			Then, for all signature dependences $\spcmapFun \in
			\SpcMapSet[\DSet](\SSet)$ for $\SSet$ over a set $\DSet$, it holds that
			$\spcmapFun$ is overlapping.
		\end{theorem}
		\begin{proof}
			By Definition~\ref{def:spcmap} of signature dependence, to prove the
			statement, i.e., that $\cap_{(\qpElm, \bFun) \in \SSet} \set{ \valFun
			\circ \bFun }{ \valFun \in \rng{\spcmapFun(\qpElm, \bFun)} } \neq
			\emptyset$, we show the existence of a function $\dFun \in \DSet^{\ESet}$
			such that, for all signatures $\sigElm = (\qpElm, \bFun) \in \SSet$, there
			is a valuation $\valFun[\sigElm] \!\in\! \rng{\spcmapFun(\sigElm)}$ for
			which $\dFun = \valFun[\sigElm] \cmp \bFun$.

			We build $\dFun$ iteratively by means of a succession of partial functions
			$\dFun_{j} \!:\! \ESet \pto \DSet$, with $j \in \numcc{0}{\card{\ESet}}$,
			satisfying the following invariants:
			\begin{enumerate}
				\item\label{thm:intspc(inv:col)}
					$\dFun[j](\eElm') = \dFun[j](\eElm'')$, for all $(\eElm', \eElm'') \in
					\BPColSet(\SSet) \cap (\dom{\dFun[j]} \times \dom{\dFun[j]})$;
				\item\label{thm:intspc(inv:piv)}
					for all $\eElm \in \dom{\dFun[j]}$, there is $i \in \numco{0}{j}$ such
					that $\eElm$ is a pivot for $\SSet$ w.r.t.\ $\dom{\dFun[i]}$;
				\item\label{thm:intspc(inv:dom)}
					$\dom{\dFun[j]} \subset \dom{\dFun[j + 1]}$, where $j < \card{\ESet}$;
				\item\label{thm:intspc(inv:val)}
					$\dFun[j] = \dFun[j + 1]_{\rst \dom{\dFun[j]}}$, where $j <
					\card{\ESet}$.
			\end{enumerate}

			Before continuing, observe that, since $\QPEVSet{\SSet} = \emptyset$, for
			each element $\eElm \in \ESet \setminus \QPAVSet{\SSet}$, there exists
			exactly one signature $\sigElm[\eElm] = (\qpElm[\eElm], \bFun[\eElm]) \in
			\SSet$ such that $\eElm \in \QPEVSet{\sigElm[\eElm]}$.

			As base case, we simply set $\dFun[0] \defeq \emptyfun$.
			It is immediate to see that the invariants are vacuously satisfied.

			Now, consider the iterative case $j \in \numco{0}{\card{\ESet}}$.
			By Lemma~\ref{lmm:pivexs} of pivot existence, there is a pivot $\eElm[j]
			\in \ESet$ for $\SSet$ w.r.t.\ $\dom{\dFun[j]}$.
			Remind that $\eElm[j] \not\in \dom{\dFun[j]}$.
			At this point, two cases can arise.
			\begin{itemize}
				\item
					$\eElm[j] \in \QPAVSet{\SSet}$.

					If there is an element $\eElm \in \dom{\dFun[j]}$ such that $(\eElm,
					\eElm[j]) \in \BPColSet(\SSet)$ then set $\dFun[j + 1] \defeq
					\dFun[j][ {\eElm[j]} \mapsto \dFun[j](\eElm) ]$.
					By Invariant~\ref{thm:intspc(inv:col)} at step $j$, the choice of such
					an element is irrelevant.
					Otherwise, choose a value $\cElm \in \DSet$, and set $\dFun[j + 1]
					\defeq \dFun[j][ {\eElm[j]} \mapsto \cElm ]$.
					In both cases, all invariants at step $j + 1$ are trivially satisfied
					by construction.
				\item
					$\eElm[j] \not\in \QPAVSet{\SSet}$.

					Consider a valuation $\valFun[j] \in \ValSet[\DSet](\QPAVSet{\qpElm[
					{\eElm[j]} ])}$ such that $\valFun[j](\bFun[ {\eElm[j]} ](\eElm)) =
					\dFun[j](\eElm)$, for all $\eElm \in \dom{\dFun[j]} \cap
					\QPAVSet{\sigElm[ {\eElm[j]} ]}$.
					The existence of such a valuation is ensured by
					Invariant~\ref{thm:intspc(inv:col)} at step $j$, since
					$\dFun[j](\eElm') = \dFun[j](\eElm'')$, for all $\eElm', \eElm'' \in
					\dom{\dFun[j]}$ with $\bFun[ {\eElm[j]} ](\eElm') = \bFun[ {\eElm[j]}
					](\eElm'')$.
					Now, set $\dFun[j + 1] \defeq \dFun[j][ {\eElm[j]} \mapsto
					\spcmapFun(\sigElm[ {\eElm[j]} ])(\valFun[j])(\bFun[ {\eElm[j]}
					](\eElm[j])) ]$.
					It remains to verify the invariants at step $j + 1$.
					Invariants~\ref{thm:intspc(inv:piv)}, \ref{thm:intspc(inv:dom)},
					and~\ref{thm:intspc(inv:val)} are trivially satisfied by construction.
					For Invariant~\ref{thm:intspc(inv:col)}, instead, suppose that there
					exists $(\eElm[j], \eElm) \in \BPColSet(\SSet) \cap (\dom{\dFun[j +
					1]} \times \dom{\dFun[j + 1]})$ with $\eElm[j] \neq \eElm$.
					By Invariant~\ref{thm:intspc(inv:piv)} at step $j$, there is $i \in
					\numco{0}{j}$ such that $\eElm$ is a pivot for $\SSet$ w.r.t.\
					$\dom{\dFun[i]}$, i.e., $\eElm = \eElm[i]$.
					At this point, two subcases can arise, the first of which results to
					be impossible.
					\begin{itemize}
						\item
							$\eElm[i] \in \QPAVSet{\SSet}$.

							By Item~\ref{def:piv(all)} of Definition~\ref{def:piv} of pivot,
							it holds that $\BPColSet(\SSet, \eElm[i]) \subseteq
							\dom{\dFun[i]}$.
							Moreover, since $\eElm[j] \not\in \QPAVSet{\SSet}$ and $(\eElm[j],
							\eElm[i]) \in \BPColSet(\SSet)$, it holds that $\eElm[j] \in
							\BPColSet(\SSet, \eElm[i])$.
							Thus, by a repeated application of
							Invariant~\ref{thm:intspc(inv:dom)} from step $i$ to step $j$, we
							have that $\eElm[j] \in \dom{\dFun[i]} \subset \dom{\dFun[j]}
							\not\ni \eElm[j]$, which is a contradiction.
						\item
							$\eElm[i] \not\in \QPAVSet{\SSet}$.

							Since $\eElm[j], \eElm[i] \not\in \QPAVSet{\SSet}$ and $(\eElm[j],
							\eElm[i]) \in \BPColSet(\SSet)$, it is easy to see that $\sigElm[
							{\eElm[j]} ] = \sigElm[ {\eElm[i]} ]$ and $\bFun[ {\eElm[j]}
							](\eElm[j]) = \bFun[ {\eElm[i]} ](\eElm[i])$.
							Otherwise, we have that $\eElm[j] \in \QPEVSet{\SSet} =
							\emptyset$, which is impossible.
							Hence, it follows that $\dFun[j + 1](\eElm[j]) =
							\spcmapFun(\sigElm[ {\eElm[j]} ])(\valFun[j])(\bFun[ {\eElm[j]}
							](\eElm[j])) = \spcmapFun(\sigElm[ {\eElm[i]}
							])(\valFun[j])(\bFun[ {\eElm[i]} ](\eElm[i]))$.
							Moreover, $\dFun[i + 1](\eElm[i]) = \spcmapFun(\sigElm[ {\eElm[i]}
							])(\valFun[i])(\bFun[ {\eElm[i]} ](\eElm[i]))$.
							Now, it is easy to observe that $\QPDepSet(\qpElm[j], \bFun[
							{\eElm[j]} ](\eElm[j])) \!=\! \QPDepSet(\qpElm[i], \bFun[
							{\eElm[i]} ](\eElm[i]))$, from which we derive that
							$\valFun[j]_{\rst \QPDepSet(\qpElm[j], \bFun[
							{\eElm[j]} ](\eElm[j]))} \!=\! \valFun[i]_{\rst
							\QPDepSet(\qpElm[i], \bFun[ {\eElm[i]} ](\eElm[i]))}$.
							At this point, by Item~\ref{def:qntspc(eqnt)} of
							Definition~\ref{def:qntspc} of dependence maps, it holds
							that $\spcmapFun(\sigElm[ {\eElm[i]} ])(\valFun[j])(\bFun[
							{\eElm[i]} ](\eElm[i])) \!=\! \spcmapFun(\sigElm[ {\eElm[i]}
							])(\valFun[i]) \allowbreak (\bFun[ {\eElm[i]} ](\eElm[i]))$, so,
							$\dFun[j + 1](\eElm[j]) \!=\! \dFun[i + 1](\eElm[i])$.
							Finally, by a repeated application of
							Invariant~\ref{thm:intspc(inv:val)} from step $i + 1$ to step $j$,
							we obtain that $\dFun[i + 1](\eElm[i]) = \dFun[j + 1](\eElm[i])$.
							Hence, $\dFun[j + 1](\eElm[j]) = \dFun[j + 1](\eElm[i])$.
					\end{itemize}
			\end{itemize}

			By a repeated application of Invariant~\ref{thm:intspc(inv:dom)} from step
			$0$ to step $\card{\ESet} - 1$, we have that $\dFun[\card{\ESet}]$ is a
			total function.
			So, we can now prove that $\dFun \defeq \dFun[\card{\ESet}]$ satisfies the
			statement, i.e., $\dFun \in \cap_{(\qpElm, \bFun) \in \SSet} \set{ \valFun
			\circ \bFun }{ \valFun \in \rng{\spcmapFun(\qpElm, \bFun)} }$.

			For each signature $\sigElm = (\qpElm, \bFun) \in \SSet$, consider the
			universal valuation $\valFun[\sigElm | '] \in
			\ValSet[\DSet](\QPAVSet{\qpElm})$ such that $\valFun[\sigElm |
			'](\bFun(\eElm)) = \dFun(\eElm)$, for all $\eElm \in \QPAVSet{\sigElm}$.
			The existence of such a valuation is ensured by
			Invariant~\ref{thm:intspc(inv:col)} at step $\card{\ESet}$.
			Moreover, let $\valFun[\sigElm] \defeq \spcmapFun(\sigElm)(\valFun[\sigElm
			| '])$.
			It remains to prove that $\dFun = \valFun[\sigElm] \cmp \bFun$, by showing
			separately that $\dFun_{\rst \QPAVSet{\sigElm}} = (\valFun[\sigElm] \cmp
			\bFun)_{\rst \QPAVSet{\sigElm}}$ and $\dFun_{\rst \QPEVSet{\sigElm}} =
			(\valFun[\sigElm] \cmp \bFun)_{\rst \QPEVSet{\sigElm}}$ hold.

			On one hand, by Item~\ref{def:qntspc(aqnt)} of
			Definition~\ref{def:qntspc}, for each $\xElm \in \QPAVSet{\qpElm}$, it
			holds that $\valFun[\sigElm | '](\xElm) =
			\spcmapFun(\sigElm)(\valFun[\sigElm | '])(\xElm)$.
			Thus, for each $\eElm \in \QPAVSet{\sigElm}$, we have that
			$\valFun[\sigElm | '](\bFun(\eElm)) = \spcmapFun(\sigElm)(\valFun[\sigElm
			| '])(\bFun(\eElm))$, which implies $\dFun(\eElm) = \valFun[\sigElm |
			'](\bFun(\eElm)) = \spcmapFun(\sigElm)(\valFun[\sigElm | '])(\bFun(\eElm))
			= \valFun[\sigElm](\bFun(\eElm)) = (\valFun[\sigElm] \cmp \bFun)(\eElm)$.
			So, $\dFun_{\rst \QPAVSet{\sigElm}} = (\valFun[\sigElm] \cmp \bFun)_{\rst
			\QPAVSet{\sigElm}}$.

			On the other hand, consider an element $\eElm \in \QPEVSet{\sigElm}$.
			By Invariant~\ref{thm:intspc(inv:piv)} at step $\card{\ESet}$, there is $i
			\in \numco{0}{\card{\ESet}}$ such that $\eElm$ is a pivot for $\SSet$
			w.r.t.\ $\dom{\dFun[i]}$.
			This means that $\eElm[i] = \eElm$ and so $\sigElm[ {\eElm[i]} ] =
			\sigElm$.
			So, by construction, we have that $\dFun[i + 1](\eElm) =
			\spcmapFun(\sigElm)(\valFun[i])(\bFun(\eElm))$.
			Moreover, $\spcmapFun(\sigElm)(\valFun[\sigElm | '])(\bFun(\eElm)) =
			\valFun[\sigElm](\bFun(\eElm)) = (\valFun[\sigElm] \cmp \bFun)(\eElm)$.
			Thus, to prove the required statement, we have only to show that
			$\dFun(\eElm) = \dFun[i + 1](\eElm)$ and
			$\spcmapFun(\sigElm)(\valFun[i])(\bFun(\eElm)) =
			\spcmapFun(\sigElm)(\valFun[\sigElm | '])(\bFun(\eElm))$.
			By a repeated application of Invariants~\ref{thm:intspc(inv:dom)}
			and~\ref{thm:intspc(inv:val)} from step $i$ to step $\card{\ESet} - 1$, we
			obtain that $\dom{\dFun[i]} \subset \dom{\dFun}$, $\dFun[i] = \dFun_{\rst
			\dom{\dFun[i]}}$, and $\dFun[i + 1](\eElm) = \dFun(\eElm)$.
			Thus, by definition of $\valFun[i]$ and $\valFun[\sigElm | ']$, it follows
			that $\valFun[i](\bFun(\eElm')) = \dFun[i](\eElm') = \dFun(\eElm') =
			\valFun[\sigElm | '](\bFun(\eElm'))$, for all $\eElm' \in \dom{\dFun[i]}$.
			At this point, by Item~\ref{def:piv(exs)} of Definition~\ref{def:piv}, it
			holds that $\QPDepSet(\sigElm, \eElm) \subseteq \dom{\dFun[i]}$, which
			implies that $\valFun[i]_{\rst \QPDepSet(\qpElm, \bFun(\eElm))} =
			\valFun[\sigElm | ']_{\rst \QPDepSet(\qpElm, \bFun(\eElm))}$.
			Hence, by Item~\ref{def:qntspc(eqnt)} of Definition~\ref{def:qntspc}, we
			have that $\spcmapFun(\sigElm)(\valFun[i])(\bFun(\eElm)) =
			\spcmapFun(\sigElm)(\valFun[\sigElm | '])(\bFun(\eElm))$.
			So, $\dFun_{\rst \QPEVSet{\sigElm}} = (\valFun[\sigElm] \cmp \bFun)_{\rst
			\QPEVSet{\sigElm}}$.
		\end{proof}

		\def\thecorollary{\ref{cor:intspc}}
		\begin{corollary}[Overlapping Dependence Maps]
			Let $\PSet \subseteq \LSigSet(\ESet, \allowbreak \LSet)$ be a finite set
			of overlapping labeled signatures on $\ESet$ w.r.t.\ $\LSet$.
			Then, for all labeled signature dependences $\spcmapFun \in
			\LSpcMapSet[\DSet](\PSet)$ for $\PSet$ over a set $\DSet$, it holds that
			$\spcmapFun$ is overlapping.
		\end{corollary}
		\begin{proof}
			Consider the set $\PSet' \defeq \set{ (\sigElm, \lElm) \in \PSet }{
			\QPEVSet{\sigElm} \neq \emptyset }$ of all labeled signatures in $\PSet$
			having at least one existential element.
			Since $\PSet$ is overlapping, by Definition~\ref{def:intsig} of
			overlapping signatures, we have that, for all $(\sigElm, \lElm[1]),
			(\sigElm, \lElm[2]) \in \PSet'$, it holds that $\lElm[1] = \lElm[2]$.
			So, let $\SSet \defeq \set{ \sigElm \in \SigSet(\ESet) }{ \exists \lElm
			\in \LSet \:.\: (\sigElm, \lElm) \in \PSet' }$ be the set of first
			components of labeled signatures in $\PSet'$ and $\hFun : \SSet \to
			\PSet'$ the bijective function such that $\hFun(\sigElm) \defeq (\sigElm,
			\lElm)$, for all $\sigElm \in \SSet$, where $\lElm \in \LSet$ is the
			unique label for which $(\sigElm, \lElm) \in \PSet'$ holds.
			Now, since $\SSet$ is overlapping, by Theorem~\ref{thm:intspc} of
			overlapping dependence maps, we have that the signature dependence
			$\spcmapFun \cmp \hFun \in \SpcMapSet[\DSet](\SSet)$ is overlapping as
			well.
			Thus, it is immediate to see that $\spcmapFun_{\rst \PSet'}$ is also
			overlapping, i.e., by Definition~\ref{def:spcmap} of signature
			dependences, there exists $\dFun \in \DSet^{\ESet}$ such that $\dFun \in
			\cap_{((\qpElm, \bFun), \lElm) \in \PSet'} \set{ \valFun \circ \bFun }{
			\valFun \in \rng{\spcmapFun((\qpElm, \bFun), \lElm)}} \neq \emptyset$.
			At this point, consider the labeled signatures $(\sigElm, \lElm) =
			((\qpElm, \bFun), \lElm) \in \PSet \setminus \PSet'$.
			Since $\QPEVSet{\sigElm} = \emptyset$, i.e., $\QPEVSet{\qpElm} =
			\emptyset$, we derive that $\spcmapFun((\sigElm, \lElm)) \in
			\SpcSet[\DSet](\qpElm)$ is the identity dependence map, i.e., it is the
			identity function on $\ValSet[\DSet](\QPVSet(\qpElm))$.
			Thus, $\rng{\spcmapFun((\sigElm, \lElm))} =
			\ValSet[\DSet](\QPVSet(\qpElm))$.
			So, we have that $\dFun \in \cap_{((\qpElm, \bFun), \lElm) \in \PSet}
			\set{ \valFun \circ \bFun }{ \valFun \in \rng{\spcmapFun((\qpElm, \bFun),
			\lElm)}} \neq \emptyset$.
			Hence, again by Definition~\ref{def:spcmap}, it holds that $\spcmapFun$ is
			overlapping.
		\end{proof}

		\def\thetheorem{\ref{thm:nonintspc}}
		\begin{theorem}[Non-Overlapping Dependence Maps]
			Let $\PSet \subseteq \LSigSet(\ESet, \LSet)$ be a set of labeled
			signatures on $\ESet$ w.r.t.\ $\LSet$.
			Then, there exists a labeled signature dependence $\spcmapFun \in
			\LSpcMapSet[\DSet](\PSet)$ for $\PSet$ over $\DSet \!\defeq\! \PSet \times
			\{ 0, 1 \}^{\CSet(\PSet)}$ such that, for all $\PSet' \!\subseteq\!
			\PSet$, it holds that $\spcmapFun_{\rst \PSet'} \in
			\LSpcMapSet[\DSet](\PSet')$ is non-overlapping, if $\PSet'$ is
			non-overlapping.
		\end{theorem}
		\begin{proof}
			\cbstart
			Let $\SSet' \defeq \set{ \sigElm \in \SigSet(\ESet) }{ \exists \lElm \in
			\LSet \:.\: (\sigElm, \lElm) \in \PSet' }$ be the set of signatures that
			occur in some labeled signature in $\PSet'$.

			If $\PSet'$ is non-overlapping, we distinguish the following three cases.
			\begin{enumerate}
				\item
					There exist $(\sigElm, \lElm[1]), (\sigElm, \lElm[2]) \in \PSet'$,
					with $\sigElm = (\qpElm, \bFun)$, such that $\QPEVSet{\sigElm} \neq
					\emptyset$ and $\lElm[1] \neq \lElm[2]$.

					Then, for all valuations $\valFun \in
					\ValSet[\DSet](\QPAVSet{\qpElm})$ and variables $\xElm \in
					\QPEVSet{\qpElm}$, we have that $\spcmapFun((\sigElm,
					\lElm[1]))(\valFun)(\xElm) = ((\sigElm, \lElm[1]), \xElm, \hFun[1])
					\neq ((\sigElm, \lElm[2]), \xElm, \allowbreak \hFun[2]) =
					\spcmapFun((\sigElm, \lElm[1]))(\valFun)(\xElm)$.
					Thus, $\spcmapFun((\sigElm, \lElm[1])) (\valFun)(\xElm) \cmp \bFun
					\neq \spcmapFun((\sigElm, \lElm[2]))(\valFun)(\xElm) \cmp \bFun$, for
					all $\valFun \in \ValSet[\DSet](\QPAVSet{\qpElm})$.
					Hence, $\spcmapFun$ is non-overlapping.
				\item
					$\QPEVSet{\SSet'} \neq \emptyset$.

					Then, there exist $\sigElm' = (\qpElm', \bFun')$, $\sigElm'' =
					(\qpElm'', \bFun'') \in \SSet'$, $\eElm' \in \QPEVSet{\sigElm'}$, and
					$\eElm'' \in \QPEVSet{\sigElm''}$ such that $\sigElm' \neq \sigElm''$
					or $\bFun'(\eElm') \neq \bFun''(\eElm'')$ and, in both cases,
					$(\eElm', \eElm'') \in \BPColSet(\SSet')$.
					By contradiction, let $\dFun \in \cap_{((\qpElm, \bFun), \lElm) \in
					\PSet'}\set{\valFun \cmp \bFun}{\valFun \in \rng{\spcmapFun(((\qpElm,
					\bFun), \lElm))}}$.
					Observe that $\dFun(\eElm') = \dFun(\eElm'')$, for all $(\eElm',
					\eElm'') \in \BPColSet(\SSet')$.
					So, there exist $\valFun' \in \ValSet[\DSet](\QPAVSet{\qpElm'})$ and
					$\valFun'' \in \ValSet[\DSet](\QPAVSet{\qpElm''})$ such that
					$\valFun'(\bFun'(\eElm)) = \dFun(\eElm)$, for all $\eElm \in
					\QPAVSet{\sigElm'}$, and $\valFun''(\bFun''(\eElm)) = \dFun(\eElm)$,
					for all $\eElm \in \QPAVSet{\sigElm''}$.
					Observe that there are $\lElm', \lElm'' \in \LSet$ such that
					$(\sigElm', \lElm'), (\sigElm'', \lElm'') \in \PSet'$.
					So, by the hypothesis of the existence of $\dFun$, we have that
					$\spcmapFun((\sigElm', \lElm'))(\valFun')(\bFun'(\eElm')) =
					\dFun(\eElm') = \dFun(\eElm'') = \spcmapFun((\sigElm'',
					\lElm''))(\valFun'')(\bFun''(\eElm''))$.
					Now, the following cases arise.
					\begin{itemize}
						\item
							$\sigElm' \neq \sigElm''$.

							By Definition~\ref{def:bigsigdep} of big signature dependence, it
							holds that $\spcmapFun((\sigElm',
							\lElm'))(\valFun')(\bFun'(\eElm')) = ((\sigElm', \lElm'),
							\bFun'(\eElm'), \hFun') \neq ((\sigElm'', \lElm''),
							\bFun''(\eElm''), \hFun'') \allowbreak = \spcmapFun((\sigElm'',
							\lElm'')) (\valFun'')(\bFun''(\eElm''))$, which is a
							contradiction.
						\item
							$\sigElm' = \sigElm''$.

							Then, we have that $\bFun'(\eElm') \neq \bFun''(\eElm'')$.
							By Definition~\ref{def:bigsigdep}, it holds that
							$\spcmapFun((\sigElm', \lElm')) \allowbreak
							(\valFun')(\bFun'(\eElm')) = ((\sigElm', \lElm'), \bFun'(\eElm'),
							\hFun') \neq ((\sigElm'', \allowbreak \lElm''), \bFun''(\eElm''),
							\hFun'') = \spcmapFun((\sigElm'', \lElm''))(\valFun'') \allowbreak
							(\bFun''(\eElm''))$, which is a contradiction.
					\end{itemize}
				\item
					$\CSet(\SSet') \neq \emptyset$.

					Then, there exists $(\vec{\eElm}, \vec{\sigElm}) \in C(\SSet')$.
					Let $n \defeq \card{\vec{\sigElm}} - 1$.
					Assume, by contradiction, that there exists $\dFun \in \cap_{((\qpElm,
					\bFun), \lElm) \in \PSet'}\set{\valFun \cmp \bFun}{\valFun \in
					\rng{\spcmapFun(((\qpElm, \bFun), \lElm))}}$.
					Observe again that $\dFun(\eElm') = \dFun(\eElm'')$, for all $(\eElm',
					\eElm'') \in \BPColSet(\SSet')$.
					Now, for all $(\vec{\sigElm})_{i} = (\qpElm[i], \bFun[i]) \in \SSet'$
					there exists $\lElm[i] \in \LSet$ such that $((\vec{\sigElm})_{i},
					\lElm[i]) \in \PSet'$.
					Moreover, let $\valFun[i] \in \ValSet[\DSet](\QPAVSet{\qpElm[i]})$
					such that $\valFun[i](\bFun[i](\eElm)) = \dFun(\eElm)$, for all $\eElm
					\in \QPAVSet{\sigElm[i]}$.
					Then, there exist $n + 1$ functions $\hFun[0], ..., \hFun[n] \in
					\{0,1 \}^{C(\PSet)}$ such that, for all $i \in \numcc{0}{n}$, we have
					that $\dFun((\vec{\eElm})_{i}) = \spcmapFun(((\vec{\sigElm})_{i},
					\lElm[i]))(\valFun[i])(\bFun[i]((\vec{\eElm})_{i})) =
					(((\vec{\sigElm})_{i}, \lElm[i]), \bFun[i]((\vec{\eElm})_{i}),
					\hFun[i])$.
					Observe that, by Item~\ref{def:bigsigdep(exs:tail)} of
					Definition~\ref{def:bigsigdep}, for all $i \in \numco{0}{n}$, it holds
					that $\hFun[i + 1]((\vec{\eElm}, \vec{\sigElm})) =
					\hFun[i]((\vec{\eElm}, \vec{\sigElm}))$ and, in particular,
					$\hFun[0]((\vec{\eElm}, \vec{\sigElm})) = \hFun[n]((\vec{\eElm},
					\vec{\sigElm}))$.
					However, by Item~\ref{def:bigsigdep(exs:fst)} of
					Definition~\ref{def:bigsigdep}, it holds that $\hFun[0]((\vec{\eElm},
					\vec{\sigElm})) = 1 - \hFun[n]((\vec{\eElm}, \vec{\sigElm}))$.
					So, $\hFun[0]((\vec{\eElm}, \vec{\sigElm})) \neq
					\hFun[n]((\vec{\eElm}, \vec{\sigElm}))$, which is a contradiction.
			\end{enumerate}
			\cbend
		\end{proof}

	\end{paragraph}

\end{section}




\begin{section}{Proofs of Section~\ref{sec:modprp}}
	\label{app:modprp}

	In this appendix, we prove Theorem~\ref{thm:sl(negmodprp)} on the negative
	properties for \SL.
	Successively, we introduce
	the concept of flagged model and flagged formulas.
	Finally, we prove Theorem~\ref{thm:ogsl(bndtremodprp)}.


		\def\thetheorem{\ref{thm:sl(negmodprp)}}
		\begin{theorem}[\SL\ Negative Model Properties]
			For \SL, it holds that:
			\begin{enumerate}
				\item
					it is not invariant under decision-unwinding;
				\item
					it does not have the decision-tree model property.
			\end{enumerate}
		\end{theorem}

		\begin{proof}
%
					\emph{[Item (1)].}
					Assume by contradiction that \SL\ is invariant under
					decision-unwinding and consider the two \CGS s $\GName[1] \defeq
					\CGSTuple{\APSet} {\AgSet} {\AcSet} {\StSet} {\labFun}
					{\trnFun[\GName_{1}]} {\sElm[0]}$ and
					$\GName[2] \defeq \CGSTuple{\APSet} {\AgSet} {\AcSet} {\StSet}
					{\labFun} {\trnFun[\GName_{2}]} {\sElm[0]}$, with $\APSet = \{
					\pSym \}$, $\AgSet = \{ \alpha, \beta \}$, $\AcSet = \{ 0, 1 \}$,
					$\StSet = \{ \sSym[0], \sSym[1 | '], \sSym[1 | ''], \sSym[2 |
					'], \sSym[2 | ''], \sSym[3 | '], \sSym[3 | ''] \}$,
					$\labFun(\sElm[2]') = \labFun(\sElm[2]'') = \{\pSym \}$ and
					$\labFun(\sElm) = \emptyset$, for all $\sElm \in \StSet \setminus
					\{\sElm[2]', \sElm[2]''  \}$, and $\trnFun[\GName_{1}]$ and
					$\trnFun[\GName_{2}]$ given as follow.
					If by $\aSym \bSym$ we indicate the decision in which agent $\alpha$
					takes the action $\aSym$ and agent $\beta$ the action $\bSym$,
					then we set $\trnFun[\GName_{1}]$ and $\trnFun[\GName_{2}]$ as follow:
					$\trnFun[\GName_{1}](\sElm[0], 0*) = \trnFun[\GName_{2}](\sElm[0],
					*0)= \sElm[1]'$,
					$\trnFun[\GName_{1}](\sElm[0], 1*)= \trnFun[\GName_{2}](\sElm[0], *1)
					= \sElm[1]''$,
					$\trnFun[\GName_{1}](\sElm[1]', 0*) = \trnFun[\GName_{2}](\sElm[1]',
					0*) = \sElm[2]'$,
					$\trnFun[\GName_{1}](\sElm[1], 1*) = \trnFun[\GName_{2}](\sElm[1], 1*)
					= \sElm[3]'$,
					$\trnFun[\GName_{1}](\sElm[1]'', 0*) = \trnFun[\GName_{2}](\sElm[1]'',
					0*)= \sElm[2]''$,
					$\trnFun[\GName_{1}](\sElm[1]'', 1*) = \trnFun[\GName_{2}](\sElm[1]'',
					1*)= \sElm[3]''$, and
					$\trnFun[\GName_{1}](\sElm, **) = \trnFun[\GName_{2}](\sElm, **) =
					\sElm$, for all $\sElm \in \{\sElm[2]', \sElm[2]'', \sElm[3]',
					\sElm[3]'' \}$.
					Observe that $\GName[1DU] = \GName[2DU]$.

					Then, it is evident that $\GName[1] \models \varphi$ iff $\GName[1DU]
					\models \varphi$ iff $\GName[2DU] \models \varphi$ iff $\GName[2]
					\models \varphi$.
					In particular, the property does have to hold for the \SL\ sentence
					$\varphi = \EExs{\xSym} \EExs{\ySym[\pSym]} \EExs{\ySym[\neg \pSym]}
					((\alpha, \xSym) (\beta, \ySym[\pSym]) \allowbreak (\X \X \pSym))
					\wedge ((\alpha, \xSym) (\beta, \ySym[\neg \pSym]) (\X \X \neg
					\pSym))$.
					It is easy to see that $\GName[1] \not\models \varphi$, while
					$\GName[2] \models \varphi$.
					Thus, \SL\ cannot be invariant under decision-unwinding.

					Indeed, each strategy $\strFun[\xElm]$ of the agent $\alpha$ in
					$\GName[1]$ forces to reach only one state at a time among $\sSym[2 |
					']$, $\sSym[2 | '']$, $\sSym[3 | ']$, and $\sSym[3 | '']$.
					Formally, for each strategy $\strFun[\xElm] \in \StrSet[ {\GName[1]}
					](\sSym[0])$, there is a state $\sElm \in \{ \sSym[2 | '], \sSym[2 |
					''], \sSym[3 | '], \sSym[3 | ''] \}$ such that, for all strategies
					$\strFun[\yElm] \in \StrSet[ {\GName[1]} ](\sSym[0])$, it holds that
					$(\playElm)_{2} = \sElm$, where $\playElm \defeq
					\playFun(\emptyfun[\alpha \mapsto \strFun[\xElm]][ {\beta \mapsto
					\strFun[\yElm]} ], \sSym[0])$.
					Thus, it is impossible to satisfy both the goals $\X \X \pSym$ and $\X
					\X \neg \pSym$ with the same strategy of $\alpha$.

					On the contrary, since $\sSym[0]$ in $\GName[2]$ is owned by the agent
					$\beta$, we may reach both $\sSym[1 | ']$ and $\sSym[1 | '']$ with the
					same strategy $\strFun[\xElm]$ of $\alpha$.
					Thus, if $\strFun[\xElm](\sSym[0] \cdot \sSym[1 | ']) \neq
					\strFun[\xElm](\sSym[0] \cdot \sSym[1 | ''])$, we reach, at the same
					time, either the pair of states $\sSym[2 | ']$ and $\sSym[3 | '']$ or
					$\sSym[2 | ']$ and $\sSym[3 | ']$.
					Formally, there are a strategy $\strFun[\xElm] \in \StrSet[
					{\GName[2]} ](\sSym[0])$, with $\strFun[\xElm](\sSym[0] \cdot \sSym[1
					| ']) \neq \strFun[\xElm](\sSym[0] \cdot \sSym[1 | ''])$, a pair of
					states $(\sElm[\pSym], \sElm[\neg \pSym]) \in \{ (\sSym[2 | '],
					\sSym[3 | '']), (\sSym[2 | ''], \sSym[3 | ']) \}$, and two strategies
					$\strFun[ {\yElm[\pSym]} ], \strFun[ {\yElm[\neg \pSym]} ] \in
					\StrSet[ {\GName[2]} ](\sSym[0])$ such that $(\playElm[\pSym])_{2} =
					\sElm[\pSym]$ and $(\playElm[\neg \pSym])_{2} = \sElm[\neg \pSym]$,
					where $\playElm[\pSym] \defeq \playFun(\emptyfun[\alpha \mapsto
					\strFun[\xElm]][ {\beta \mapsto \strFun[ {\yElm[\pSym]} ]} ],
					\sSym[0])$ and $\playElm[\neg \pSym] \defeq \playFun(\emptyfun[\alpha
					\mapsto \strFun[\xElm]][ {\beta \mapsto \strFun[ {\yElm[\neg \pSym]}
					]} ], \sSym[0])$.
					Hence, we can satisfy both the goals $\X \X \pSym$ and $\X \X \neg
					\pSym$ with the same strategy of $\alpha$.

				\emph{[Item (2)].}
					To prove the statement we have to show that there exists a satisfiable
					sentence that does not have a \DT\ model.
					Consider the \SL\ sentence $\varphi \defeq \varphi_{1} \wedge
					\varphi_{2}$, where $\varphi_{1}$ is the negation of the sentence
					$\varphi$ used in Item (1) and $\varphi_{2} \defeq \AAll{\xSym}
					\AAll{\ySym} (\alpha, \xSym) (\beta, \ySym) \X ((\EExs{\xSym}
					\EExs{\ySym} (\alpha, \xSym) (\beta, \ySym) \X \pSym) \wedge
					(\EExs{\xSym} \EExs{\ySym} (\alpha, \xSym) (\beta, \ySym) \X \neg
					\pSym))$.
					Moreover, note that the sentence $\varphi_{2}$ is equivalent to the
					\CTL\ formula $\A \X ((\E \X \pSym) \wedge (\E \X \neg \pSym))$.
					Then, consider the \CGS\ $\GName \defeq \CGSStruct$
					with $\APSet = \{\pSym\}$, $\AgSet = \{\alpha, \beta\}$, $\AcSet =
					\{0,1 \}$, $\StSet = \{\sElm[0], \sElm[1], \sElm[2], \sElm[3]\}$,
					$\labFun(\sElm[0]) \labFun(\sElm[1]) = \labFun(\sElm[3]) = \emptyset$
					and $\labFun(\sElm[2]) = \{ \pSym \}$, and $\trnFun(\sElm[0],
					**) = \sElm[1]$, $\trnFun(\sElm[1], 0*) = \sElm[1]$, 
					$\trnFun(\sElm[1], 1*) = \sElm[3]$, and $\trnFun(\sElm, **) =
					\sElm$, for all $\sElm \in \{\sElm[2], \sElm[3]\}$.

					It is easy to see that $\GName$ satisfies $\varphi$.
					At this point, let $\TName$ be a \DT\ model of $\varphi_{2}$.
					Then, such a tree has necessarily at least two actions and,
					consequently, two different successors $\tElm[1], \tElm[2] \in
					\DecSet^{*}$ of the	root $\epsilon$, where $\tElm[1], \tElm[2] \in
					\DecSet$ and $\tElm[1](\alpha) = \tElm[2](\alpha)$.
					Moreover, there are two decisions $\decFun[1], \decFun[2] \in \DecSet$
					such that $\pSym \in \labFun(\tElm[1] \cdot \decFun[1])$ and $\pSym
					\not\in \labFun(\tElm[2] \cdot \decFun[2])$.
					Now, let $\strFun[\xSym], \strFun[ {\ySym[\pSym]} ], \strFun[
					{\ySym[\neg \pSym]} ] \in \StrSet(\epsilon)$ be three strategies for
					which the following holds: $\strFun[\xSym](\epsilon) =
					\tElm[1](\alpha)$, $\strFun[ {\ySym[\pSym]} ](\epsilon) =
					\tElm[1](\beta)$, $\strFun[ {\ySym[\neg \pSym]} ](\epsilon) =
					\tElm[2](\beta)$, $\strFun[\xSym](\tElm[1]) = \decFun[1](\alpha)$,
					$\strFun[ {\ySym[\pSym]} ](\tElm[1]) = \decFun[1](\beta)$,
					$\strFun[\xSym](\tElm[2]) = \decFun[2](\alpha)$, and $\strFun[
					{\ySym[\neg \pSym]} ](\tElm[2]) = \decFun[2](\beta)$.
					Then, it is immediate to see that $\TName, \emptyfun[\xSym \mapsto
					\strFun[\xSym]][\ySym[\pSym] \mapsto \strFun[ {\ySym[\pSym]}
					]][\ySym[\neg \pSym] \mapsto \strFun[ {\ySym[\neg \pSym]} ]], \epsilon
					\models ((\alpha, \xSym) (\beta, \ySym[\pSym]) \allowbreak (\X \X
					\pSym)) \wedge ((\alpha, \xSym) (\beta, \ySym[\neg \pSym]) (\X \X \neg
					\pSym))$.
					Thus, we obtain that $\TName \not\models \varphi_{1}$.
					Hence, $\varphi$ does not have a \DT\ model.

		\end{proof}


	\begin{paragraph}{Flagged features}
		A flagged model of a given \CGS\ $\GName$ is obtained adding a
		so-called $\sharp$-agent to the set of agents and flagging every state with
		two flags.
		Intuitively, the $\sharp$-agent takes control of the flag to use in order to
		establish which part of a given formula is checked in the \CGS.
		We start giving first the definition of \emph{plan} and then the concepts of
		\emph{flagged model} and \emph{flagged formulas}.

		\begin{definition}[Plans]
			\label{def:pln}
			A track (resp., path) \emph{plan} in a \CGS\ $\GName$ is a finite (resp.,
			an infinite) sequence of decisions $\plnElm \in \DecSet^{*}$ (resp.,
			$\plnElm \in \DecSet^{\omega}$).
			$\TPlnSet \defeq \DecSet^{*}$ (resp., $\PPlnSet \defeq\DecSet^{\omega}$)
			denotes the set of all track (resp., path) plans.
			Moreover, with each non-trivial track $\trkElm \in \TrkSet$ (resp., path
			$\pthElm \in \PthSet$) it is associated the set $\TPlnSet(\trkElm) \defeq
			\set{ \plnElm \in \DecSet^{\card{\trkElm} - 1} }{ \forall i \in
			\numco{0}{\card{\plnElm}} \,.\, (\trkElm)_{i + 1} = \trnFun((\trkElm)_{i},
			(\plnElm)_{i}) } \subseteq \TPlnSet$ (resp., $\PPlnSet(\pthElm) \defeq
			\set{ \plnElm \in \DecSet^{\omega} }{ \forall i \in \SetN \,.\,
			(\pthElm)_{i + 1} = \trnFun((\pthElm)_{i}, (\plnElm)_{i}) } \subseteq
			\PPlnSet$) of track (resp., path) plans that are \emph{consistent} with
			$\trkElm$ (resp., $\pthElm$).
		\end{definition}

		\begin{definition}[Flagged model]
			Let $\GName = \CGSStruct$ be a \CGS\ with $\card{\AcSet} \geq 2$.
			Let $\sharp \notin \AgSet$ and $\cElm[\sharp] \in \AcSet$.
			Then, the \emph{flagged} \CGS\ is defined as follows:
			\begin{center}
				$\GName[\sharp] = \CGSTuple {\APSet} {\AgSet \cup \{ \sharp \}} {\AcSet}
				{\StSet \times \{ 0, 1 \}} {\labFun[\sharp]} {\trnFun[\sharp]}
				{(\sElm[0], 0)}$
			\end{center}
			where $\labFun[\sharp](\sElm, \iota) \defeq \labFun(\sElm)$, for all
			$\sElm \in \StSet$ and $\iota \in \{ 0, 1 \}$, and
			$\trnFun[\sharp]((\sElm, \iota), \decFun) \defeq (\trnFun(\sElm,
			\decFun_{\rst \AgSet}), \iota')$ with $\iota' = 0$ iff $\decFun(\sharp)
			= \cElm[\sharp]$.
		\end{definition}

		Since $\GName$ and $\GName[\sharp]$ have a different set of agents, an
		agent-closed formula $\varphi$ w.r.t. $\AgSet[\GName]$ is clearly not
		agent-closed w.r.t. $\AgSet[\GName_{\sharp}]$.
		For this reason, we introduce the concept of flagged formulas, that
		represent, in some sense, the agent-closure of formulas.

		\begin{definition}[Flagged formulas]
			Let $\varphi \in \OGSL$.
			The \emph{universal flagged formula} of $\varphi$, in symbol
			$\varphi_{A\sharp}$, is obtained by replacing every principal subsentence
			$\phi \in \psnt{\varphi}$ with the formula $\phi_{A \sharp} \defeq
			\AAll{\xElm[\sharp]} (\sharp, \xElm_{\sharp}) \phi$.
			The existential flagged formula of $\varphi$, in symbol
			$\varphi_{E\sharp}$, is obtained by replacing every principal subsentence
			$\phi \in \psnt{\varphi}$ with the formula $\phi_{E \sharp} \defeq
			\EExs{\xElm[\sharp]} (\sharp, \xElm_{\sharp}) \phi$.
		\end{definition}

		Substantially, these definitions help us to check satisfiability of
		principal subsentences in a separate way.
		The special agent $\sharp$ takes control, over the flagged model, of which
		branch to walk on the satisfiability of some $\phi \in \psnt{\varphi}$.
		Obviously, there is a strict connection between satisfiability of flagged
		formulas over $\GName[\sharp]$ and $\varphi$ over $\GName$.
		Indeed, the following lemma holds.

		\begin{lemma}[Flagged model satisfiability]
			\label{lmm:flagsat}
			Let $\varphi \in \OGSL$ and let $\varphi_{A\sharp}$ and
			$\varphi_{E\sharp}$ the flagged formulas.
			Moreover, let $\GName$ be a $\CGS$ and $\GName[\sharp]$ his relative
			flagged $\CGS$. Then, for all $\sElm \in \StSet$, it holds that:
			\begin{enumerate}
				\item \label{lmm:flagsat(toflag)}
					if $\GName, \emptyset, \sElm \models \varphi$ then  $ \GName[\sharp],
					\emptyset (s, \iota) \models \varphi_{A \sharp}$, for all $\iota \in
					\{0,1\}$;
				\item \label{lmm:flagsat(fromflag)}
					if, for all $\iota \in \{0, 1\}$ it holds that $\GName[\sharp],
					\emptyset, (\sElm, \iota) \models \varphi_{E \sharp}$, then $\GName,
					\emptyset, \sElm \models \varphi$.
			\end{enumerate}
		\end{lemma}

		\begin{proof}
			On the first case, let $\spcFun \in
			\SpcSet_{\StrSet_{\GName}}(\qpElm)$, we  consider $\spcFun_{A \sharp}
			\in \SpcSet_{\StrSet_{\GName[\sharp]}}(\AAll{\xElm[\sharp]} \qpElm)$ such
			that if $\xElm \neq \xElm[\sharp]$ then $\spcFun_{A
			\sharp}(\asgFun)(\xElm) = \spcFun(\asgFun)(\xElm)$, otherwise
			$\spcFun_{A \sharp}(\asgFun)(\xElm) = \asgFun(\xElm[\sharp])$.
			On the second case, let $\spcFun_{E \sharp} \in
			\SpcSet_{\StrSet_{\GName[\sharp]}}(\EExs{\xElm[\sharp]} \qpElm)$, we
			consider $\spcFun \in \SpcSet_{\StrSet_{\GName}}(\qpElm)$ such that
			$\spcFun(\asgFun)(\xElm)) = \spcFun_{E \sharp}(\asgFun)(\xElm))$ (note
			that $\dom{\spcFun(\asgFun)}$ is strictly included in
			$\dom{\spcFun_{E \sharp}(\asgFun)}$).
			Now, given a binding $\bpElm$ and its relative function $\bpFun_{\bpElm}$,
			consider $\bpElm_{\sharp} \defeq (\sharp, \xElm[\sharp]) \bpElm$ and its
			relative function $\bpFun_{\bpElm, \sharp}$.
			We show that in both cases considered above there is some useful relation
			between  $\pthElm_{\bpElm} \defeq \playFun(\spcFun(\asgFun) \circ
			\bpFun_{\bpElm}, \sElm)$ and $\pthElm_{\bpElm,\sharp} \defeq
			\playFun(\spcFun_{\sharp}(\asgFun) \circ \bpFun_{\bpElm,\sharp}, (\sElm,
			\iota))$.
			Indeed, let $\plnElm_{\bpElm}$ the plan such that, for all $i \in \SetN$,
			we have that $(\pthElm_{\bpElm})_{i+1} = \trnFun((\pthElm_{\bpElm})_{i},
			(\plnElm_{\bpElm})_{i})$ and let $\plnElm_{\bpElm,\sharp}$ the plan such
			that, for all $i \in \SetN$, we have that $(\pthElm_{\bpElm,\sharp})_{i+1}
			= \trnFun((\pthElm_{\bpElm,\sharp})_{i},(\plnElm_{\bpElm, \sharp})_{i})$.
			By the definition of play, for each $i \in \SetN$ and $\aElm \in \AgSet$,
			we have that $(\plnElm_{\bpElm})_{i}(\aElm) = (\spcFun(\asgFun) \circ
			\bpFun_{\bpElm})(\aElm)((\pthElm_{\bpElm})_{i})$ and $(\plnElm_{\bpElm,
			\sharp})_{i}(\aElm) = (\spcFun_{\sharp}(\asgFun) \circ \bpFun_{\bpElm,
			\sharp})(\aElm)((\pthElm_{\bpElm, \sharp})_{i})$.
			Clearly, for all $i \in \SetN$, we have that $(\plnElm_{\bpElm})_{i} =
			((\plnElm_{\bpElm,\sharp})_{i})_{\rst \AgSet}$.
			Due to these facts, we can prove by induction that for each $i \in \SetN$
			there exists $\iota \in \{0, 1\}$ such that $(\pthElm_{\bpElm,
			\sharp})_{i} = ((\pthElm_{\bpElm})_{i}, \iota)$.
			The base case is trivial and we omit it here.
			As inductive case, suppose that $(\pthElm_{\bpElm, \sharp})_{i} =
			((\pthElm_{\bpElm})_{i}, \iota)$, for some $i$.
			Then, by definition we have that $(\pthElm_{\bpElm, \sharp})_{i+1} =
			\trnFun[\sharp]((\pthElm_{\bpElm,
			\sharp})_{i},(\plnElm_{\bpElm,\sharp})_{i})$.
			Moreover, by definition of $\trnFun[\sharp]$, we have that
			$(\pthElm_{\bpElm, \sharp})_{i+1} =
			(\trnFun((\pthElm_{\bpElm})_{i},((\plnElm_{\bpElm, \sharp})_{i})_{\rst
			\AgSet}),\iota')$, for some $\iota' \in \{0,1 \}$.
			Since $(\plnElm_{\bpElm})_{i} = ((\plnElm_{\bpElm, \sharp})_{i})_{\rst
			\AgSet}$, we have that $(\pthElm_{\bpElm, \sharp})_{i+1} =
			(\trnFun((\pthElm_{\bpElm})_{i}, (\plnElm_{\bpElm})_{i}), \iota') =
			((\pthElm_{\bpElm})_{i+1}, \iota')$, which is the assert.
			It follows, by definition of $\labFun[\sharp]$, that
			$\labFun((\pthElm_{\bpElm})_{i}) =
			\labFun[\sharp]((\pthElm_{\bpElm,\sharp})_{i})$, for each $i \in \SetN$.
			So, every sentence satisfied on $\pthElm_{\bpElm}$ is satisfied also on
			$\pthElm_{\bpElm,\sharp}$.
			Now we proceed to prove Items~\ref{lmm:flagsat(toflag)}
			and~\ref{lmm:flagsat(fromflag)}, separately.
			%
				\emph{Item~\ref{lmm:flagsat(toflag)}.}
					First, consider the case that $\phi$ is of the form
					$\qpElm \psi$, where $\qpElm$ is a quantification prefix and $\psi$ is
					a boolean composition of goals.
					Since $\GName,\emptyset, \sElm \models \phi$, there exists $\spcFun
					\in \SpcSet_{\StrSet_{\GName}}(\qpElm)$ such that we have
					$\GName,\spcFun(\asgFun),\sElm \models \psi$, for all assignment
					$\asgFun \in \AsgSet_{\GName}(\sElm)$.
					Now, consider $\phi_{A, \sharp} \defeq \AAll{\xElm[\sharp]}(\sharp,
					\xElm[\sharp]) \phi$, which is equivalent to $\AAll{\xElm[\sharp]}
					\qpElm(\sharp, \xElm[\sharp]) \psi$.
					Then, consider $\spcFun_{A \sharp} \in
					\SpcSet_{\StrSet_{\GName[\sharp]}}(\AAll{\xElm[\sharp]} \qpElm)$
					such that $\spcFun_{A \sharp}(\asgFun)(\xElm) =
					\spcFun(\asgFun)(\xElm)$, if $\xElm \neq \xElm[\sharp]$, and
					$\spcFun_{\sharp}(\asgFun)(\xElm) = \asgFun(\xElm[\sharp])$,
					otherwise.
					Clearly, $\spcFun_{A \sharp}$ is build starting from $\spcFun$ as
					described above.
					Then, from the fact that $\GName, \emptyset, \sElm \models \phi$,
					it follows that $\GName[\sharp], \emptyset, (\sElm, \iota) \models
					\phi_{A \sharp}$.
					Now, if we have a formula $\varphi$ embedding some proper principal
					subsentence, then by the induction hypothesis every $\phi \in
					\psnt{\varphi}$ is satisfied by $\GName$ if and only if $\phi_{A,
					\sharp}$ is satisfied by $\GName[\sharp]$.
					By working on the structure of the formula it follows that
					the result holds for $\varphi$ and $\varphi_{A \sharp}$ too, so the
					proof for this Item is done.

				\emph{Item~\ref{lmm:flagsat(fromflag)}.}
					First, consider the case of $\phi$ is of the form $\qpElm \psi$,
					where $\qpElm$ is a quantification prefix and $\psi$ is a
					boolean composition of goals.
					Let $\GName[\sharp], \emptyset,(\sElm, \iota) \models \phi_{E,
					\sharp}$.
					Note that $\phi_{E, \sharp} \defeq \EExs{\xElm[\sharp]}(\sharp,
					\xElm[\sharp]) \qpElm \psi$ is equivalent to $\EExs{\xElm[\sharp]}
					\qpElm(\sharp, \xElm[\sharp]) \psi$, so there exists
					$\spcFun_{E \sharp} \in
					\SpcSet_{\StrSet_{\GName[\sharp]}}(\EExs{\xElm[\sharp]}\qpElm)$ such
					that, for all assignment $\asgFun \in
					\AsgSet_{\GName[\sharp]}(\EExs{\xElm[\sharp]}\qpElm)$, we have that
					$\GName[\sharp], \spcFun_{E \sharp}(\asgFun), (\sElm, \iota)
					\models(\sharp, \xElm[\sharp]) \psi$.
					Then, consider $\spcFun \in \SpcSet_{\StrSet_{\GName}}$ given by
					$\spcFun(\asgFun)(\xElm)) = \spcFun_{\sharp}(\asgFun)(\xElm))$.
					Clearly, $\spcFun$ is build starting from $\spcFun_{E \sharp}$ as
					described above.
					Then, from $\GName[\sharp], \spcFun_{E \sharp}(\asgFun), (\sElm,
					\iota) \models (\sharp, \xElm[\sharp]) \psi$ it follows that $\GName,
					\emptyset, \sElm \models \phi$.
					Now, if we have a formula $\varphi$ embedding some proper principal
					subsentence, then by the induction hypothesis every $\phi \in
					\psnt{\varphi}$ is satisfied by $\GName$ if and only if $\phi_{A,
					\sharp}$ is satisfied by $\GName[\sharp]$.
					By working on the structure of the formula it follows that
					the result holds for $\varphi$ and $\varphi_{E \sharp}$ too, so the
					proof for this Item is done.
		\end{proof}

	\end{paragraph}

	\begin{paragraph}{Proof of Theorem~\ref{thm:ogsl(bndtremodprp)}}

		From now on, by using Item~\ref{thm:ogsl(posmodprp:dectremodprp)} of
		Theorem~\ref{thm:ogsl(posmodprp)}, we can assume to work exclusively on
		\CGT s.
		Let $\SSet_{\phi} \defeq \set{\sElm \in \StSet_{\TName}}{\TName, \emptyset,
		\sElm \models \phi}$ and $\TSet[\phi] \defeq \SSet[\phi] \times \{0, 1\}$.
		By Item \ref{lmm:flagsat(toflag)} of Lemma \ref{lmm:flagsat}, we have that
		$\TName_{\sharp},\emptyset, \tElm \models \phi_{A \sharp}$, for all $\tElm
		\in \TSet[\phi]$.
		Moreover, for all $\tElm \in \TSet[\phi]$, consider a strategy
		$\strFun[\sharp]^{\tElm} \in \StrSet[\TName_{\sharp}](\tElm)$ given by
		$\strFun[\sharp]^{\tElm}(\rho) = \cElm[\sharp]$ iff $\rho = \tElm$.
		Moreover, for all $\phi \in \psnt{\varphi}$, consider the
		function $\AFun_{\phi}: \TrkSet_{\TName_{\sharp}}(\varepsilon) \rightarrow
		2^{(\StSet_{\TName_{\sharp}} \times \TrkSet_{\TName_{\sharp}})}$ given by
		$\AFun_{\phi}(\rho) \defeq \set{(\rho_{i}, \rho')}{i \in
		\numco{0}{\card{\rho}} \land \rho'\in
		\TrkSet_{\TName_{\sharp}}(\emptyset[\sharp \rightarrow
		\strFun^{\rho_{i}}_{\sharp}], \rho_{i}) \land \lst{\rho}=\lst{\rho'}}$.
		Note that $(\lst{\rho}, \lst{\rho}) \in \AFun_{\phi}(\rho)$.
		Indeed: \emph(i) $\lst{\rho} = \rho_{\card{\rho}}$; \emph{(ii)} $\lst{\rho)}
		\in \TrkSet_{\TName_{\sharp}}(\emptyset[\sharp \rightarrow
		\strFun^{\lst{\rho}}_{\sharp}], \lst{\rho})$; and \emph{(iii)} $\lst{\rho}
		= \lst{\lst{\rho}}$.
		Observe that if $(\rho_{i}, \rho') \in \AFun_{\phi}(\rho)$ then $\rho' =
		\rho_{\geq i}$.
		Hence, except for $(\lst{\rho},\lst{\rho})$, there exists at most one pair
		in $\AFun_{\phi}(\rho)$.
		Indeed, by contradiction let $(\rho_{i},\rho_{\geq i})$ and $(\rho_{j},
		\rho_{\geq j})$ both in $\AFun_{\phi}(\rho)$ with $i \lneqq j$ and $j \neq
		\card{\rho}$.
		Then, by the definition of compatible tracks
		$\TrkSet_{\TName_{\sharp}}(\emptyset[\sharp \rightarrow
		\strFun^{\rho}_{\sharp}], \rho_{i})$, there exists a plan $\plnElm \in
		\PlnSet(\rho_{\geq i})$ such that for all $h \in \numco{0} {\card{\rho}-i}$
		we have $\plnElm_{h}(\sharp)=\strFun^{\rho}_{\sharp}((\rho_{\geq i})_{\leq
		h})$.
		Then, by the definition of $\strFun^{\rho}_{\sharp}$, $\plnElm_{h}(\sharp)
		\neq \cElm_{\sharp}$.
		So, by the definition of plan and $\trnFun_{\sharp}$, we have
		that $\rho_{j+1} = (\sElm, 1)$.
		On the other hand, since $(\rho_j, \rho_{\geq j}) \in \AFun_{\phi}(\rho)$,
		then there exists a plan $\plnElm' \in \PlnSet(\rho_{\geq j})$ such that
		$(\plnElm')_{0}(\sharp) = \strFun^{\rho_{\geq j}}_{\sharp}(\rho_{\geq j}) =
		\cElm_{\sharp}$.
		Which implies, by the definition of plan and $\trnFun_{\sharp}$, we have
		that $\rho_{j+1} = (\sElm',0)$, which is in contradiction with the fact that
		the second coordinate of $\rho_{j+1}$ is 1, as shown above.

		This reasoning allows us to build the functions $head_{\phi}$ and
		$body_{\phi}$ for the disjoint satisfiability of $\phi$ over
		$\TName[\sharp]$ on the set $\TSet[\phi]$.
		Indeed, the unique element $(\trkElm[i], \trkElm') \in
		\AFun_{\phi}(\trkElm) \setminus \{ (\lst{\trkElm}, \lst{\trkElm}) \}$
		can be used to define opportunely the elementary dependence map used for
		such disjoint satisfiability.

		\def\thetheorem{\ref{thm:ogsl(bndtremodprp)}}
		\begin{theorem}[\OGSL\ Bounded Tree-Model Property]
			Let $\varphi$ be an \OGSL\ satisfiable sentence and $\PSet \defeq \set{
			((\qpElm, \bpElm), (\psi, i)) \in \LSigSet(\AgSet, \SL \times \{ 0, 1 \})
			}{ \qpElm \bpElm \psi \in \psnt{\varphi} \land i \in \{ 0, 1 \} }$ the set
			of all labeled signatures on $\AgSet$ w.r.t.\ $\SL \times \{ 0, 1 \}$ for
			$\varphi$.
			Then, there exists a $b$-bounded \DT\ $\TName$, with $b = \card{\PSet}
			\cdot \card{\QPVSet(\PSet)} \cdot 2^{\card{\CSet(\PSet)}}$, such that
			$\TName \models \varphi$.
			Moreover, for all $\phi \in \psnt{\varphi}$, it holds that $\TName$
			satisfies $\phi$ disjointly over the set $\set{ \sElm \in \StSet }{
			\TName, \emptyset, \sElm \models \phi }$.
		\end{theorem}

		\begin{proof}
			Since $\varphi$ is satisfiable, then, by
			Item~\ref{thm:ogsl(posmodprp:dectremodprp)} of
			Theorem~\ref{thm:ogsl(posmodprp)}, we have that there exists a \DT\
			$\TName$, such that $\TName \models \varphi$.
			We now prove that there exists a bounded \DT\ $\TName' \defeq
			\CGSTuple{\APSet}{\AgSet}{\AcSet[
			\TName']}{\StSet[\TName']]}{\labFun[\TName']}{ \trnFun [ \TName
			']}{\epsilon}$ with $\AcSet[\TName'] \defeq \numco{0}{n}$	and $n =
			\card{\PSet} \cdot \card{\QPVSet(\PSet)} \cdot 2^{\card{\CSet(\PSet)}}$.
			Since $\TName'$ is a \DT, we have to define only the labeling function
			$\labFun[\TName']$.
			To do this, we need two auxiliary functions $\hFun: \StSet[\TName]
			\times \DecSet[\TName'] \to \DecSet[\TName]$ and $\gFun: \StSet[\TName']
			\to \StSet[\TName]$ that lift correctly the labeling function
			$\labFun[\TName]$ to $\labFun[\TName']$.
			Function $\gFun$ is defined recursively as follows: \emph{(i)}
			$\gFun(\epsilon) \defeq \epsilon$, \emph{(ii)} $\gFun(\tElm' \cdot
			\decFun') \defeq \gFun(\tElm') \cdot \hFun(\gFun(\tElm'), \decFun')$.
			Then, for all $\tElm' \in \StSet[\TName']$, we define
			$\labFun[\TName'](\tElm') \defeq \labFun[\TName](\gFun(\tElm'))$.
			It remains to define the function $\hFun$.
			By Item~\ref{lmm:flagsat(toflag)} of Lemma~\ref{lmm:flagsat}, we have
			that $\TName[\sharp] \models \varphi_{A \sharp}$ and consequently that
			$\TName[\sharp] \models \varphi_{E \sharp}$.
			Moreover, applying the reasoning explained above, $\TName[\sharp]$
			satisfies disjointly $\phi$ over $\SSet[\phi]$, for all $\phi \in
			\psnt{\varphi}$.
			Then, for all $\phi \in \psnt{\varphi}$, we have that there exist a
			function $head_{\phi}: \SSet[\phi] \to \SpcSet[\AcSet_{\TName}](\qpElm)$
			and a function $body_{\phi}: \TrkSet[\TName](\epsilon) \to
			\SpcSet[\AcSet_{\TName}](\qpElm)$ that allow $\TName$ to satisfy $\phi$ in
			a disjoint way over $\SSet[\phi]$.
			Now, by Theorem~\ref{thm:nonintspc}, there exists a signature dependence
			$\spcmapFun \in \LSpcMapSet[\AcSet_{\TName'}](\PSet)$ such that, for all
			$\PSet' \subseteq \PSet$, we have that $\spcmapFun_{\rst \PSet'} \in
			\LSpcMapSet[\AcSet_{\TName'}](\PSet)$ is non-overlapping, if $\PSet$ is
			non-overlapping.
			Moreover, by Corollary~\ref{cor:intspc}, for all $\PSet' \subseteq
			\PSet$, we have that $\spcmapFun_{\rst \PSet'} \in
			\LSpcMapSet[\AcSet_{\TName'}](\PSet)$ is overlapping, if $\PSet$ is
			overlapping.
			At this point, consider the function $\DFun: \DecSet[\TName'] \to
			2^{\PSet}$ that, for all $\decFun' \in \DecSet[\TName']$, is given by
			$\DFun(\decFun') \defeq \set{((\qpElm, \bpElm), (\psi, i)) = \sigElm \in
			\PSet}{\exists \eElm' \in \AcSet[\TName']^{\QPAVSet(\qpElm)}. \decFun' =
			\spcmapFun(\sigElm)(\eElm') \cmp \bpFun}$.
			Note that, for all $\decFun' \in \DecSet[\TName']$, we have that
			$\DFun(\decFun) \subseteq \PSet$ is overlapping.
			Now, consider the functions $\WFun: \StSet[\TName_{\sharp}] \to
			\LSpcMapSet[\AcSet_{\TName}](\PSet)$ such that, for all $\tElm \in
			\StSet[\TName_{\sharp}]$ and $\sigElm = ((\qpElm, \bpElm), (\psi, i)) \in
			\PSet$, is such that
			\begin{center}
				$\WFun(\tElm)(\sigElm) = \left\{
													\begin{array}{ll}
														head_{\phi}(\tElm) &, \tElm \in
														\TSet_{\phi} \\
														body_{\phi}(\trkElm') &,
														\mbox{otherwise}
													\end{array}
													\right.
				$
			\end{center}
			where $\phi = \qpElm \bpElm \psi$ and $\trkElm' \in
			\TrkSet[\TName_{\sharp}](\epsilon)$ is the unique track such that
			$\lst{\trkElm'} = \tElm$.
			Moreover, consider the function $\TFun: \StSet[\TName_{\sharp}] \times
			\DecSet[\TName] \to 2^{\PSet}$ such that, for all $\tElm \in
			\StSet[\TName_{\sharp}]$ and $\decFun \in \DecSet[\TName]$, it is given by
			$\TFun(\tElm, \decFun) \defeq \set{\sigElm = ((\qpElm, \bpElm), (\psi, i))
			\in \PSet}{\exists \eElm \in \AcSet[\TName]^{\QPAVSet{\qpElm}}. \decFun =
			\WFun(\tElm)(\eElm) \cmp \bpFun}$.
			It is easy to see that, for all $\decFun' \in \DecSet[\TName']$ and
			$\tElm \in \StSet[\TName_{\sharp}]$, there exists $\decFun \in
			\DecSet[\TName]$ such that $\DFun(\decFun') \subseteq \TFun(\tElm,
			\decFun)$.
			By Corollary~\ref{cor:intspc}, for all $\tElm \in
			\StSet[\TName_{\sharp}]$, we have that $\WFun(\tElm)_{\rst
			\DFun(\decFun')}$ is overlapping.
			So, by Definition~\ref{def:spcmap}, for all $\tElm \in
			\StSet[\TName_{\sharp}]$ and $\decFun' \in \DecSet[\TName']$, there
			exists $\decFun \in \AcSet[\TName_{\sharp}]^{\AgSet}$ such that $\decFun
			\in \cap_{\sigElm = (\qpElm, \bpFun), (\psi, i) \in
			\DFun(\decFun')}\set{\zFun \circ \bFun}{\zFun \in
			\rng{\WFun(\tElm)(\sigElm)}}$, which implies $\TFun(\tElm, \decFun)
			\supseteq \DFun(\decFun')$.
			Finally, by applying the previous reasoning we obtain the function
			$\hFun$ such that, for all $(\tElm, \decFun') \in \StSet[\TName]
			\times \DecSet[\TName']$, it associates a decision $\hFun(\tElm, \decFun')
			\defeq \decFun \in \DecSet[\TName]$.
			The proof that $\TName' \models \varphi$ proceeds naturally by
			induction and it is omitted here.

		\end{proof}

	\end{paragraph}

\end{section}




\begin{section}{Proofs of Section~\ref{sec:satprc}}
	\label{app:satprc}

	In this appendix, we give the proofs of Lemmas~\ref{lmm:ogsl(golaut)}
	and~\ref{lmm:ogsl(sntaut)} of \OGSL\ goal and sentence automaton and
	Theorems~\ref{thm:ogsl(aut)} and~\ref{thm:ogsl(sat)} of \OGSL\ automaton and
	satisfiability.

	\begin{paragraph}{Alternating tree automata}

		\emph{Nondeterministic tree automata} are a generalization to infinite trees
		of the classical \emph{nondeterministic word automata} on infinite words.
		\emph{Alternating tree automata} are a further generalization of
		nondeterministic tree automata~\cite{MS87}.
		Intuitively, on visiting a node of the input tree, while the latter sends
		exactly one copy of itself to each of the successors of the node, the former
		can send several own copies to the same successor.
		Here we use, in particular, \emph{alternating parity tree automata}, which
		are alternating tree automata along with a \emph{parity acceptance
		condition} (see~\cite{GTW02}, for a survey).

		We now give the formal definition of alternating tree automata.
		\begin{definition}[Alternating Tree Automata]
			\label{def:ata}
			An \emph{alternating tree automaton} (\emph{\ATA}, for short) is a tuple
			$\AName \defeq \ATAStruct$, where $\LabSet$, $\DirSet$, and $\QSet$ are,
			respectively, non-empty finite sets of \emph{input symbols},
			\emph{directions}, and \emph{states}, $\qElm[0] \in \QSet$ is an
			\emph{initial state}, $\aleph$ is an \emph{acceptance condition} to be
			defined later, and $\delta : \QSet \times \LabSet \to \PBoolSet(\DirSet
			\times \QSet)$ is an \emph{alternating transition function} that maps each
			pair of states and input symbols to a positive Boolean combination on the
			set of propositions of the form $(\dElm, \qElm) \in \DirSet \times \QSet$,
			a.k.a. \emph{moves}.
		\end{definition}

		On one side, a \emph{nondeterministic tree automaton} (\emph{\NTA}, for
		short) is a special case of \ATA\ in which each conjunction in the
		transition function $\delta$ has exactly one move $(\dElm, \qElm)$
		associated with each direction $\dElm$.
		This means that, for all states $\qElm \in \QSet$ and symbols $\sigma \in
		\LabSet$, we have that $\atFun(\qElm, \sigma)$ is equivalent to a Boolean
		formula of the form $\bigvee_{i} \bigwedge_{\dElm \in \DirSet} (\dElm,
		\qElm[i, \dElm])$.
		On the other side, a \emph{universal tree automaton} (\emph{\UTA}, for
		short) is a special case of \ATA\ in which all the Boolean combinations that
		appear in $\delta$ are conjunctions of moves.
		Thus, we have that $\atFun(\qElm, \sigma) = \bigwedge_{i} (\dElm[i],
		\qElm[i])$, for all states $\qElm \in \QSet$ and symbols $\sigma \in
		\LabSet$.

		The semantics of the \ATA s is given through the following concept of run.
		\begin{definition}[\ATA\ Run]
			\label{def:ata(run)}
			A \emph{run} of an \ATA\ $\AName = \ATAStruct$ on a $\LabSet$-labeled
			$\DirSet$-tree $\TName = \LTStruct$ is a $(\DirSet \times \QSet)$-tree
			$\RSet$ such that, for all nodes $\xElm \in \RSet$, where $\xElm =
			\prod_{i = 1}^{n} (\dElm[i], \qElm[i])$ and $\yElm \defeq \prod_{i =
			1}^{n} \dElm[i]$ with $n \in \numco{0}{\omega}$, it holds that \emph{(i)}
			$\yElm \in \TSet$ and \emph{(ii)}, there is a set of moves $\SSet
			\subseteq \DirSet \times \QSet$ with $\SSet \models \delta(\qElm[n],
			\vFun(\yElm))$ such that $\xElm \cdot (\dElm, \qElm) \in \RSet$, for all
			$(\dElm, \qElm) \in \SSet$.
		\end{definition}

		In the following, we consider \ATA s along with the \emph{parity acceptance
		condition} (\emph{\APT}, for short) $\aleph \defeq (\FSet_{1}, \ldots,
		\FSet_{k}) \in (\pow{\QSet})^{+}$ with $\FSet_{1} \subseteq \ldots \subseteq
		\FSet_{k} = \QSet$ (see~\cite{KVW00}, for more).
		The number $k$ of sets in the tuple $\aleph$ is called the \emph{index} of
		the automaton.
		We also consider \ATA s with the \emph{co-B\"uchi acceptance condition}
		(\emph{\ACT}, for short) that is the special parity condition with index
		$2$.

		Let $\RSet$ be a run of an \ATA\ $\AName$ on a tree $\TName$ and $\wElm$ one
		of its branches.
		Then, by $\infFun(\wElm) \defeq \set{ \qElm \in \QSet }{ \card{\set{ i \in
		\SetN }{ \exists \dElm \in \DirSet . (w)_{i} = (\dElm, \qElm) }} = \omega }$
		we denote the set of states that occur infinitely often as the second
		component of the letters along the branch $w$.
		Moreover, we say that $w$ satisfies the parity acceptance condition $\aleph
		= (\FSet_{1}, \ldots, \FSet_{k})$ if the least index $i \in \numcc{1}{k}$
		for which $\infFun(w) \cap \FSet_{i} \neq \emptyset$ is even.

		Finally, we can define the concept of language accepted by an \ATA.
		\begin{definition}[\ATA\ Acceptance]
			\label{def:ata(acp)}
			An \ATA\ $\AName = \ATAStruct$ \emph{accepts} a $\LabSet$-labeled
			$\DirSet$-tree $\TName$ iff is there exists a run $\RSet$ of $\AName$ on
			$\TName$ such that all its infinite branches satisfy the acceptance
			condition $\aleph$.
		\end{definition}
		By $\LangSet(\AName)$ we denote the language accepted by the \ATA\ $\AName$,
		i.e., the set of trees $\TName$ accepted by $\AName$.
		Moreover, $\AName$ is said to be \emph{empty} if $\LangSet(\AName) =
		\emptyset$.
		The \emph{emptiness problem} for $\AName$ is to decide whether
		$\LangSet(\AName) = \emptyset$.

	\end{paragraph}

	\begin{paragraph}{Proofs of theorems}

		We are finally able to show the proofs of the above mentioned results.

		\def\thelemma{\ref{lmm:ogsl(golaut)}}
		\begin{lemma}[\OGSL\ Goal Automaton]
			Let $\bpElm \psi$ an \OGSL\ goal without principal subsentences and
			$\AcSet$ a finite set of actions.
			Then, there exists an \UCT\ $\UName[\bpElm \psi | ^{\AcSet}] \defeq
			\TATuple {\ValSet[\AcSet](\free{\bpElm \psi}) \times \pow{\APSet}}
			{\DecSet} {\QSet[\bpElm \psi]} {\atFun[\bpElm \psi]} {\qElm[0\bpElm\psi]}
			{\aleph_{\bpElm \psi}}$ such that, for all \DT s $\TName$ with
			$\AcSet[\TName] = \AcSet$, states $\tElm \in \StSet[\TName]$, and
			assignments $\asgFun \in \AsgSet[\TName](\free{\bpElm \psi}, \tElm)$, it
			holds that $\TName, \asgFun, \tElm \models \bpElm \psi$ iff $\TName' \in
			\LangSet(\UName[\bpElm \psi | ^{\AcSet}])$, where $\TName'$ is the
			assignment-labeling encoding for $\asgFun$ on $\TName$.
		\end{lemma}
		\begin{proof}
			A first step in the construction of the \UCT\ $\UName[\bpElm \psi |
			^{\AcSet}]$, is to consider the \UCW\ $\UName[\psi] \defeq \WATuple
			{\pow{\APSet}} {\QSet[\psi]} {\atFun[\psi]} {\QSet[0\psi]}
			{\aleph_{\psi}}$ obtained by dualizing the \NBW\ resulting from the
			application of the classic Vardi-Wolper construction to the \LTL\ formula
			$\neg \psi$~\cite{VW86b}.
			Observe that $\LangSet(\UName[\psi]) = \LangSet(\psi)$, i.e., this
			automaton recognizes all infinite words on the alphabet $\pow{\APSet}$
			that satisfy the \LTL\ formula $\psi$.
			Then, define the components of $\UName[\bpElm \psi | ^{\AcSet}] \defeq
			\TATuple {\ValSet[\AcSet](\free{\bpElm \psi}) \times \pow{\APSet}}
			{\DecSet} {\QSet[\bpElm \psi]} {\atFun[\bpElm \psi]} {\qElm[0\bpElm\psi]}
			{\aleph_{\bpElm \psi}}$, as follows:
			\begin{itemize}
				\item
					$\QSet[\bpElm \psi] \defeq \{ \qElm[0\bpElm\psi] \} \cup \QSet[\psi]$,
					with $\qElm[0\bpElm\psi] \not\in \QSet[\psi]$;
				\item
					$\atFun[\bpElm \psi](\qElm[0\bpElm\psi], (\valFun, \sigma)) \defeq
					\bigwedge_{\qElm \in \QSet[0\psi]} \atFun[\bpElm \psi](\qElm,
					(\valFun, \sigma))$, for all $(\valFun, \sigma) \in
					\ValSet[\AcSet](\free{\bpElm \psi}) \times \pow{\APSet}$;
				\item
					$\atFun[\bpElm \psi](\qElm, (\valFun, \sigma)) \!\defeq\!
					\bigwedge_{\qElm' \!\in\! \atFun[\psi](\qElm, \sigma)}
					(\valFun \cmp \bndFun[\bpElm], \qElm')$, for all $\qElm \!\in\!
					\QSet[\psi]$ and $(\valFun, \sigma) \in \ValSet[\AcSet](\free{\bpElm
					\psi}) \times \pow{\APSet}$;
				\item
					$\aleph_{\bpElm \psi} \defeq \aleph_{\psi}$.
			\end{itemize}
			Intuitively, the \UCT\ $\UName[\bpElm \psi | ^{\AcSet}]$ simply runs the
			\UCW\ $\UName[\psi]$ on the branch of the encoding individuated by the
			assignment in input.
			Thus, it is easy to see that, for all states $\tElm \in \StSet[\TName]$
			and assignments $\asgFun \in \AsgSet[\TName](\free{\bpElm \psi}, \tElm)$,
			it holds that $\TName, \asgFun, \tElm \models \bpElm \psi$ iff $\TName'
			\in \LangSet(\UName[\bpElm \psi | ^{\AcSet}])$, where $\TName'$ is the
			assignment-labeling encoding for $\asgFun$ on $\TName$.
		\end{proof}

		\def\thelemma{\ref{lmm:ogsl(sntaut)}}
		\begin{lemma}[\OGSL\ Sentence Automaton]
			Let $\qpElm \bpElm \psi$ be an \OGSL\ principal sentence without principal
			subsentences and $\AcSet$ a finite set of actions.
			Then, there exists an \UCT\ $\UName[\qpElm \bpElm \psi | ^{\AcSet}]
			\defeq \TATuple {\SpcSet[\AcSet](\qpElm) \times \pow{\APSet}} {\DecSet}
			{\QSet[\qpElm \bpElm \psi]} {\atFun[\qpElm \bpElm \psi]}
			{\qElm[0\qpElm\bpElm\psi]} {\aleph_{\qpElm \bpElm \psi}}$ such that, for
			all \DT s $\TName$ with $\AcSet[\TName] = \AcSet$, states $\tElm \in
			\StSet[\TName]$, and elementary dependence maps over strategies
			$\spcFun \in \ESpcSet[ {\StrSet[\TName](\tElm)} ](\qpElm)$, it holds that
			$\TName, \spcFun(\asgFun), \tElm \emodels \bpElm \psi$, for all $\asgFun
			\in \AsgSet[\TName](\QPAVSet{\qpElm}, \tElm)$, iff $\TName' \in
			\LangSet(\UName[\qpElm \bpElm \psi | ^{\AcSet}])$, where $\TName'$ is
			the elementary dependence-labeling encoding for $\spcFun$ on $\TName$.
		\end{lemma}
		\begin{proof}
			By Lemma~\ref{lmm:ogsl(golaut)} of \OGSL\ goal automaton, there is an
			\UCT\  $\UName[\bpElm \psi | ^{\AcSet}] \defeq
			\TATuple {\ValSet[\AcSet](\free{\bpElm \psi}) \times \pow{\APSet}}
			{\DecSet} {\QSet[\bpElm \psi]} {\atFun[\bpElm \psi]} {\qElm[0\bpElm\psi]}
			{\aleph_{\bpElm \psi}}$ such that, for all \DT s $\TName$ with
			$\AcSet[\TName] = \AcSet$, states $\tElm \in \StSet[\TName]$, and
			assignments $\asgFun \in \AsgSet[\TName](\free{\bpElm \psi}, \tElm)$, it
			holds that $\TName, \asgFun, \tElm \models \bpElm \psi$ iff $\TName' \in
			\LangSet(\UName[\bpElm \psi | ^{\AcSet}])$, where $\TName'$ is the
			assignment-labeling encoding for $\asgFun$ on $\TName$.

			Now, transform $\UName[\bpElm \psi | ^{\AcSet}]$ into the new \UCT\
			$\UName[\qpElm \bpElm \psi | ^{\AcSet}] \defeq \TATuple
			{\SpcSet[\AcSet](\qpElm) \times \pow{\APSet}} {\DecSet} {\QSet[\qpElm
			\bpElm \psi]} {\atFun[\qpElm \bpElm \psi]} {\qElm[0\qpElm\bpElm\psi]}
			{\aleph_{\qpElm \bpElm \psi}}$, with $\QSet[\qpElm \bpElm \psi] \defeq
			\QSet[\bpElm \psi]$, $\qElm[0\qpElm\bpElm\psi] \defeq \qElm[0\bpElm\psi]$,
			and $\aleph_{\qpElm \bpElm \psi} \defeq \aleph_{\bpElm \psi}$, which is
			used to handle the quantification prefix $\qpElm$ atomically, where the
			transition function is defined as follows: $\atFun[\qpElm \bpElm
			\psi](\qElm, (\spcFun, \sigma)) \defeq \bigwedge_{\valFun \in
			\ValSet[\AcSet](\QPAVSet{\qpElm})} \atFun[\bpElm \psi](\qElm,
			(\spcFun(\valFun), \sigma))$, for all $\qElm \in \QSet[\qpElm \bpElm
			\psi]$ and $(\spcFun, \sigma) \in \SpcSet[\AcSet](\qpElm) \times
			\pow{\APSet}$.
			Intuitively, $\UName[\qpElm \bpElm \psi | ^{\AcSet}]$ reads an action
			dependence map $\spcFun$ on each node of the input tree $\TName'$ labeled
			with a set of atomic propositions $\sigma$ and simulates the execution of
			the transition function $\atFun[\bpElm \psi](\qElm, (\valFun, \sigma))$ of
			$\UName[\bpElm \psi | ^{\AcSet}]$, for each possible valuation $\valFun =
			\spcFun(\valFun')$ on $\free{\bpElm \psi}$ obtained from $\spcFun$ by a
			universal valuation $\valFun' \in \ValSet[\AcSet](\QPAVSet{\qpElm})$.
			It is worth observing that we cannot move the component set
			$\SpcSet[\AcSet](\qpElm)$ from the input alphabet to the states of
			$\UName[\qpElm \bpElm \psi | ^{\AcSet}]$ by making a related guessing of
			the dependence map $\spcFun$ in the transition function, since we have to
			ensure that all states in a given node of the tree $\TName'$, i.e., in
			each track of the original model $\TName$, \mbox{make the same choice for
			$\spcFun$.}

			Finally, it remains to prove that, for all states $\tElm \in
			\StSet[\TName]$ and elementary dependence maps over strategies $\spcFun \in
			\ESpcSet[ {\StrSet[\TName](\tElm)} ](\qpElm)$, it holds that $\TName,
			\spcFun(\asgFun), \tElm \emodels \bpElm \psi$, for all $\asgFun \in
			\AsgSet[\TName](\QPAVSet{\qpElm}, \tElm)$, iff $\TName' \in
			\LangSet(\UName[\qpElm \bpElm \psi | ^{\AcSet}])$, where $\TName'$ is the
			elementary dependence-labeling encoding for $\spcFun$ on $\TName$.

			\emph{[Only if].}
			Suppose that $\TName, \spcFun(\asgFun), \tElm \emodels \bpElm \psi$, for
			all $\asgFun \in \AsgSet[\TName](\QPAVSet{\qpElm}, \tElm)$.
			Since $\psi$ does not contain principal subsentences, we have that
			$\TName, \spcFun(\asgFun), \tElm \models \bpElm \psi$.
			So, due to the property of $\UName[\bpElm \psi | ^{\AcSet}]$,
			it follows that there exists an assignment-labeling encoding
			$\TName[\asgFun | '] \in \LangSet(\UName[\bpElm \psi | ^{\AcSet}])$, which
			implies the existence of a $(\DecSet \times \QSet[\bpElm \psi])$-tree
			$\RSet[\asgFun]$ that is an accepting run for $\UName[\bpElm \psi |
			^{\AcSet}]$ on $\TName[\asgFun | ']$.
			At this point, let $\RSet \defeq \bigcup_{\asgFun \in
			\AsgSet[\TName](\QPAVSet{\qpElm}, \tElm)} \RSet[\asgFun]$ be the union of
			all runs.
			Then, due to the particular definition of the transition function of
			$\UName[\qpElm \bpElm \psi | ^{\AcSet}]$, it is not hard to see that
			$\RSet$ is an accepting run for $\UName[\qpElm \bpElm \psi | ^{\AcSet}]$
			on $\TName'$.
			Hence, $\TName' \in \LangSet(\UName[\qpElm \bpElm \psi | ^{\AcSet}])$.

			\emph{[If].}
			Suppose that $\TName' \in \LangSet(\UName[\qpElm \bpElm
			\psi | ^{\AcSet}])$.
			Then, there exists a $(\DecSet \times \QSet[\qpElm \bpElm \psi])$-tree
			$\RSet$ that is an accepting run for $\UName[\qpElm \bpElm \psi |
			^{\AcSet}]$ on $\TName'$.
			Now, for each $\asgFun \in \AsgSet[\TName](\QPAVSet{\qpElm}, \tElm)$, let
			$\RSet[\asgFun]$ be the run for $\UName[\bpElm \psi | ^{\AcSet}]$ on the
			assignment-state encoding $\TName[\asgFun | ']$ for $\spcFun(\asgFun)$ on
			$\TName$.
			Due to the particular definition of the transition function of
			$\UName[\qpElm \bpElm \psi | ^{\AcSet}]$, it is not hard to see that
			$\RSet[\asgFun] \subseteq \RSet$.
			Thus, since $\RSet$ is accepting, we have that $\RSet[\asgFun]$ is
			accepting as well.
			So, $\TName[\asgFun | '] \in \LangSet(\UName[\bpElm \psi | ^{\AcSet}])$.
			At this point, due to the property of $\UName[\bpElm \psi | ^{\AcSet}]$,
			it follows that $\TName, \spcFun(\asgFun), \tElm \models \bpElm \psi$.
			Since $\psi$ does not contain principal subsentences, we have that
			$\TName, \spcFun(\asgFun), \tElm \!\emodels\! \bpElm \psi$, for all
			$\asgFun \!\in\! \AsgSet[\TName](\QPAVSet{\qpElm}, \tElm)$.
		\end{proof}

		\def\thetheorem{\ref{thm:ogsl(aut)}}
		\begin{theorem}[\OGSL\ Automaton]
			Let $\varphi$ be an \OGSL\ sentence.
			Then, there exists an \UCT\ $\UName[\varphi]$ such that $\varphi$ is
			satisfiable iff $\LangSet(\UName[\varphi]) \neq \emptyset$.
		\end{theorem}
		\begin{proof}
			By Theorem~\ref{thm:ogsl(bndtremodprp)} of \OGSL\ bounded tree-model
			property, if an \OGSL\ sentence $\varphi$ is satisfiable, it is
			satisfiable in a disjoint way on a $b$-bounded \DT\ with $b \defeq
			\card{\PSet} \cdot \card{\QPVSet(\PSet)} \cdot 2^{\card{\CSet(\PSet)}}$,
			where $\PSet \defeq \set{ ((\qpElm, \bpElm), (\psi, i)) \in
			\LSigSet(\AgSet, \SL \times \{ 0, 1 \}) }{ \qpElm \bpElm \psi \in
			\psnt{\varphi} \land i \in \{ 0, 1 \} }$ is the set of all labeled
			signatures on $\AgSet$ w.r.t.\ $\SL \times \{ 0, 1 \}$.
			Thus, we can build an automaton that accepts only $b$-bounded tree
			encodings.
			To do this, in the following, we assume $\AcSet \defeq \numco{0}{b}$.

			Consider each principal subsentence $\phi \in \psnt{\varphi}$ of $\varphi$
			as a sentence with atomic propositions in $\APSet \cup \psnt{\varphi}$
			having no inner principal subsentence.
			This means that these subsentences are considered as fresh atomic propositions.
			Now, let $\UName[\phi | ^{\AcSet}] \defeq \TATuple
			{\SpcSet[\AcSet](\qpElm) \times \pow{\APSet \cup \psnt{\varphi}}}
			{\DecSet} {\QSet[\phi]} {\atFun[\phi]} {\qElm[0\phi]} {\aleph_{\phi}}$ be
			the \UCT s built in Lemma~\ref{lmm:ogsl(sntaut)}.
			Moreover, set $\MSet \defeq \set{ \mFun \in \psnt{\varphi} \to
			\bigcup_{\qpElm \in \QPSet(\VSet), \VSet \subseteq \VarSet}
			\SpcSet[\AcSet](\qpElm) }{ \forall \phi = \qpElm \bpElm \psi \in
			\psnt{\varphi} \:.\: \mFun(\phi) \in \SpcSet[\AcSet](\qpElm) }$.
			Then, we define the components of the \UCT\ $\UName[\varphi] \defeq
			\TATuple {\MSet \times \MSet \times \pow{\APSet \cup \psnt{\varphi}}}
			{\DecSet} {\QSet} {\atFun} {\qElm[0]} {\aleph}$, as follows:
			\begin{itemize}
				\item
					$\QSet \defeq \{ \qElm[0], \qElm[c] \} \cup \bigcup_{\phi \in
					\psnt{\varphi}} \{ \phi \} \times \QSet[\phi]$;
				\item
					$\atFun(\qElm[0], (\mFun[h], \mFun[b], \sigma)) \defeq
					\atFun(\qElm[c], (\mFun[h], \mFun[b], \sigma))$, if $\sigma
					\models \varphi$, and $\atFun(\qElm[0], (\mFun[h], \mFun[b],
					\sigma)) \defeq \Ff$, otherwise, where $\varphi$ is considered here as
					a Boolean formula on $\APSet \cup \psnt{\varphi}$;
				\item
					$\atFun(\qElm[c], (\mFun[h], \mFun[b], \sigma)) \defeq
					\bigwedge_{\decFun \in \DecSet} (\decFun, \qElm[c]) \wedge
					\bigwedge_{\phi \in \sigma \cap \psnt{\varphi}} \allowbreak
					\atFun[\phi](\qElm[0\phi], (\mFun[h](\phi), \sigma)) [(\decFun, \qElm)
					/ (\decFun, (\phi, \qElm))]$;
				\item
					$\atFun((\phi, \qElm), (\mFun[h], \mFun[b], \sigma)) \defeq
					\atFun[\phi](\qElm, (\mFun[b](\phi), \sigma)) [(\decFun, \qElm') /
					\allowbreak (\decFun, (\phi, \qElm'))]$;
				\item
					$\aleph \defeq \bigcup_{\phi \in \psnt{\varphi}} \{ \phi \} \times
					\aleph_{\phi}$.
			\end{itemize}
			Intuitively, $\UName[\varphi]$ checks whether there are principal
			subsentences $\phi$ of $\varphi$ contained into the labeling, for all
			nodes of the input tree, by means of the checking state $\qElm[c]$.
			In the affirmative case, it runs the related automata $\UName[\phi |
			^{\AcSet}]$ by supplying them, as dependence maps on actions, the heading
			part $\mFun[h]$, when it starts, and the body part $\mFun[b]$,
			otherwise.
			In this way, it checks that the disjoint satisfiability is verified.

			We now prove that the above construction is correct.

			\emph{[Only if].}
			Suppose that $\varphi$ is satisfiable.
			Then, by Theorem~\ref{thm:ogsl(bndtremodprp)} there exists a $b$-bounded
			\DT\ $\TName$ such that $\TName \models \varphi$.
			In particular, w.l.o.g., assume that $\AcSet[\TName] = \AcSet$.
			Moreover, for all $\phi = \qpElm \bpElm \psi \in \psnt{\varphi}$, it holds
			that $\TName$ satisfies $\phi$ disjointly over the set $\SSet[\phi] \defeq
			\set{ \tElm \in \StSet[\TName] }{ \TName, \emptyset, \tElm \models \phi
			}$.
			This means that, by Definition~\ref{def:ogsl(dsjsat)} of \OGSL\ disjoint
			satisfiability, there exist two functions $\headFun[\phi] : \SSet[\phi]
			\to \SpcSet[\AcSet](\qpElm)$ and $\bodyFun[\phi] :
			\TrkSet[\TName](\epsilon) \to \SpcSet[\AcSet](\qpElm)$ such that, for all
			$\tElm \in \SSet[\phi]$ and $\asgFun \in \AsgSet[\TName](\QPAVSet{\qpElm},
			\tElm)$, it holds that $\TName, \spcFun[\phi, \tElm](\asgFun), \tElm
			\models \bpElm \psi$, where the elementary dependence map $\spcFun[\phi,
			\tElm] \in \ESpcSet[ {\StrSet[\TName](\tElm)} ](\qpElm)$ is defined as
			follows: \emph{(i)} $\adj{\spcFun[\phi, \tElm]}(\tElm) \defeq
			\headFun[\phi](\tElm)$; \emph{(ii)} $\adj{\spcFun[\phi, \tElm]}(\trkElm)
			\defeq \bodyFun[\phi](\trkElm' \cdot \trkElm)$, for all $\trkElm \in
			\TrkSet[\TName](\tElm)$ with $\card{\trkElm} > 1$, where $\trkElm' \in
			\TrkSet[\TName](\epsilon)$ is the unique track such that $\trkElm' \cdot
			\trkElm \in \TrkSet[\TName](\epsilon)$.

			Now, let $\TName[\varphi]$ be the \DT\ over $\APSet \cup
			\psnt{\varphi}$ with $\AcSet[ {\TName[\varphi]} ] = \AcSet$ such that
			\emph{(i)} $\labFun[ {\TName[\varphi]} ](\tElm) \cap \APSet =
			\labFun[\TName](\tElm)$ and \emph{(ii)} $\phi \in \labFun[
			{\TName[\varphi]} ](\tElm)$ iff $\tElm \in \SSet[\phi]$, for all $\tElm
			\in \StSet[ {\TName[\varphi]} ] = \StSet[\TName]$ and $\phi \in
			\psnt{\varphi}$.

			By Lemma~\ref{lmm:ogsl(sntaut)}, we have that $\TName[\phi, \tElm | '] \in
			\LangSet(\UName[\phi | ^{\AcSet}])$, where $\TName[\phi, \tElm | ']$ is
			the elementary dependence-labeling encoding for $\spcFun[\phi, \tElm]$ on
			$\TName[\varphi]$.
			Thus, there is a $(\DecSet \times \QSet[\phi])$-tree $\RSet[\phi, \tElm]$
			that is an accepting run for $\UName[\phi | ^{\AcSet}]$ on $\TName[\phi,
			\tElm | ']$.
			So, let $\RSet[\phi, \tElm | ']$ be the $(\DecSet \times \QSet)$-tree
			defined as follows: $\RSet[\phi, \tElm | '] \defeq \set{ (\tElm \cdot
			\tElm', (\phi, \qElm)) }{ (\tElm', \qElm) \in \RSet[\phi, \tElm] }$.

			At this point, let $\RSet \defeq \RSet[c] \cup \bigcup_{\phi \in
			\psnt{\varphi}, \tElm \in \SSet[\phi]} \RSet[\phi, \tElm | ']$ be the
			$(\DecSet \times \QSet)$-tree, where $\RSet[c] \defeq \{ \epsilon \} \cup
			\set{ (\tElm, \qElm[c]) }{ \tElm \in \StSet[\TName] \land \tElm \neq
			\epsilon }$, and $\TName' \defeq \LTTuple{}{}{\StSet[\TName]}{\uFun}$ one
			of the $(\MSet \times \MSet \times \pow{\APSet \cup
			\psnt{\varphi}})$-labeled $\DecSet$-tree satisfying the following
			property: for all $\tElm \in \StSet[\TName]$ and $\phi \in
			\psnt{\varphi}$, it holds that $\uFun(\tElm) = (\mFun[h], \mFun[b],
			\sigma)$, where \emph{(i)} $\sigma \cap \APSet = \labFun[\TName](\tElm)$,
			\emph{(ii)} $\phi \in \sigma$ iff $\tElm \in \SSet[\phi]$, \emph{(iii)}
			$\mFun[h](\phi) = \headFun[\phi](\tElm)$, if $\tElm \in \SSet[\phi]$, and
			\emph{(iv)} $\mFun[b](\phi) = \bodyFun[\phi](\trkElm[\tElm])$ with
			$\trkElm[\tElm] \in \TrkSet[\TName](\epsilon)$ the unique track such that
			$\lst{\trkElm[\tElm]} = \tElm$.
			Moreover, since $\TName \models \varphi$, we have that $\labFun[
			{\TName[\varphi]} ](\epsilon) \models \varphi$, where, in the last
			expression, $\varphi$ is considered as a Boolean formula on $\APSet \cup
			\psnt{\varphi}$.
			Then, it is easy to prove that $\RSet$ is an accepting run for
			$\UName[\varphi]$ on $\TName'$, i.e., $\TName' \in
			\LangSet(\UName[\varphi])$.
			Hence, $\LangSet(\UName[\varphi]) \neq \emptyset$.

			\emph{[If].}
			Suppose that there is an $(\MSet \times \MSet \times
			\pow{\APSet \cup \psnt{\varphi}})$-labeled $\DecSet$-tree $\TName' \defeq
			\LTTuple{}{}{\DecSet^{*}}{\uFun}$ such that $\TName' \in
			\LangSet(\UName[\varphi])$ and let the $(\DecSet \times \QSet)$-tree
			$\RSet$ be the accepting run for $\UName[\varphi]$ on $\TName'$.
			Moreover, let $\TName$ be the \DT\ over $\APSet \cup \psnt{\varphi}$ with
			$\AcSet[\TName] = \AcSet$ such that, for all $\tElm \in \StSet[\TName]$,
			it holds that $\uFun(\tElm) = (\mFun[h], \mFun[b],
			\labFun[\TName](\tElm))$, for some $\mFun[h], \mFun[b] \in \MSet$.

			Now, for all $\phi = \qpElm \bpElm \psi \in \psnt{\varphi}$, we make the
			following further assumptions:
			\begin{itemize}
				\item
					$\SSet[\phi] \defeq \set{ \tElm \in \StSet[\TName] }{ \exists
					\mFun[h], \mFun[b] \in \MSet, \sigma \in \pow{\APSet \cup
					\psnt{\varphi}} \:.\: \allowbreak \uFun(\tElm) = (\mFun[h], \mFun[b],
					\sigma) \land \phi \in \sigma }$;
				\item
					let $\RSet[\phi, \tElm]$ be the $(\DecSet \times \QSet[\phi])$-tree
					such that $\RSet[\phi, \tElm] \defeq \{ \epsilon \} \cup \set{
					(\tElm', \qElm) }{ (\tElm \cdot \tElm', (\phi, \qElm)) \in \RSet }$,
					for all $\tElm \in \SSet[\phi]$;
				\item
					let $\TName[\phi, \tElm | ']$ be the elementary dependence-labeling
					encoding for $\spcFun[\phi, \tElm] \in \ESpcSet[
					{\StrSet[\TName](\tElm)} ](\qpElm)$ on $\TName$, for all $\tElm \in
					\SSet[\phi]$, where $\adj{\spcFun[\phi, \tElm]}(\tElm) \defeq
					\mFun[h](\phi)$, with $\uFun(\tElm) = (\mFun[h], \mFun[b], \sigma)$
					for some $\mFun[b] \in \MSet$ and $\sigma \in \pow{\APSet \cup
					\psnt{\varphi}}$, and $\adj{\spcFun[\phi, \tElm]}(\trkElm) \defeq
					\mFun[b](\phi)$, with $\uFun(\lst{\trkElm}) = (\mFun[h], \mFun[b],
					\sigma)$ for some $\mFun[h] \in \MSet$ and $\sigma \in \pow{\APSet
					\cup \psnt{\varphi}}$, for all $\trkElm \in \TrkSet[\TName](\tElm)$
					with $\card{\trkElm} > 1$.
			\end{itemize}

			Since $\RSet$ is an accepting run, it is easy to prove that
			$\RSet[\phi, \tElm]$ is an accepting run for $\UName[\phi | ^{\AcSet}]$ on
			$\TName[\phi, \tElm | ']$.
			Thus, $\TName[\phi, \tElm | '] \in \LangSet(\UName[\phi | ^{\AcSet}])$.
			So, by Lemma~\ref{lmm:ogsl(sntaut)}, it holds that $\TName, \spcFun[\phi,
			\tElm](\asgFun), \tElm \models \bpElm \psi$, for all $\tElm \in
			\SSet[\phi]$ and $\asgFun \!\in \!\AsgSet[\TName](\QPAVSet{\qpElm}, \tElm)$,
			which means that $\SSet[\phi] \!=\! \set{ \tElm \in \StSet[\TName] }{ \TName,
			\emptyset, \tElm \models \phi }$.

			Finally, since $\labFun[ {\TName[\varphi]} ](\epsilon) \models \varphi$,
			we have that $\TName \models \varphi$, where, in the first expression,
			$\varphi$ is considered as a Boolean formula on $\APSet \cup
			\psnt{\varphi}$.
		\end{proof}

		\def\thetheorem{\ref{thm:ogsl(sat)}}
		\begin{theorem}[\OGSL\ Satisfiability]
			The satisfiability problem for \OGSL\ is 2\ExpTimeC.
		\end{theorem}
		\begin{proof}
			By Theorem~\ref{thm:ogsl(aut)} of \OGSL\ automaton, to verify whether an
			\OGSL\ sentence $\varphi$ is satisfiable we can calculate the emptiness of
			the \UPT\ $\UName[\varphi]$.
			This automaton is obtained by merging all \UCT s $\UName[\phi |
			^{\AcSet}]$, with $\phi = \qpElm \bpElm \psi \in \psnt{\varphi}$, which
			in turn are based on the \UCT s $\UName[\bpElm \psi | ^{\AcSet}]$ that
			embed the \UCW s $\UName[\psi]$.
			By a simple calculation, it is easy to see that $\UName[\varphi]$ has
			$2^{\AOmicron{\card{\varphi}}}$ states.

			Now, by using a well-known nondeterminization procedure for \APT
			s~\cite{MS95}, we obtain an equivalent \NPT\ $\NName[\varphi]$ with
			$2^{2^{\AOmicron{\card{\varphi}}}}$ states and index
			$2^{\AOmicron{\card{\varphi}}}$.

			The emptiness problem for such a kind of automaton with $n$ states and
			index $h$ is solvable in time $\AOmicron{n^{h}}$.
			Thus, we get that the time complexity of checking whether $\varphi$ is
			satisfiable is $2^{2^{\AOmicron{\card{\varphi}}}}$.
			Hence, the membership of the satisfiability problem for \OGSL\ in
			2\ExpTime\ directly follows.
			Finally the thesis is proved, by getting the relative lower bound from
			the same problem for \CTLS
		\end{proof}

	\end{paragraph}

\end{section}


\end{document}